\documentclass[apjl,iop]{emulateapj}

\usepackage{natbib}
\usepackage{graphicx}
\usepackage{txfonts}
\begin{document}

\title{Model comparison for the density structure across solar coronal waveguides}
\shorttitle{Cross-field density structure of solar coronal waveguides}
\shortauthors{Arregui, Soler, \& Asensio Ramos}

\author{I. Arregui\altaffilmark{1,2},  R. Soler\altaffilmark{3,4}, and A. Asensio Ramos\altaffilmark{1,2}}
\altaffiltext{1}{Instituto de Astrof\'{\i}sica de Canarias, V\'{\i}a L\'actea s/n, E-38205 La Laguna, Tenerife, Spain}
\altaffiltext{2}{Departamento de Astrof\'{\i}sica, Universidad de La Laguna, E-38206 La Laguna, Tenerife, Spain}
\altaffiltext{3}{Solar Physics Group, Departament de F'\'{\i}sica, Universitat de les Illes Balears, E-07122 Palma de Mallorca, Spain}
\altaffiltext{4}{Institute of Applied Computing \& Community Code (IAC$^3$), Universitat de les Illes Balears, E-07122 Palma de Mallorca, Spain}
 \email{iarregui@iac.es}

\begin{abstract}
The spatial variation of physical quantities, such as the mass density, across solar atmospheric waveguides governs the timescales and spatial scales for wave damping and energy dissipation. The direct measurement of the spatial distribution of density, however, is difficult and indirect seismology inversion methods have been suggested as an alternative. We applied Bayesian inference,  model comparison, and model-averaging  techniques to the inference of the cross-field density structuring in solar magnetic waveguides using information on periods and damping times for resonantly damped magnetohydrodynamic (MHD) transverse kink oscillations. Three commonly employed alternative profiles were used to model the variation of the mass density across the waveguide boundary. Parameter inference enabled us to obtain information on physical quantities such as the Alfv\'en travel time, the density contrast, and the transverse inhomogeneity length scale. The inference results from alternative density models were compared and their differences quantified. Then, the relative plausibility of the considered models was assessed by performing model comparison. Our results indicate that the evidence in favor of any of the three models is minimal, unless the oscillations are strongly damped. In such a circumstance, the application of model-averaging techniques enables the computation of an evidence-weighted inference that takes into account the plausibility of each model in the calculation of a combined inversion for the unknown physical parameters.
\end{abstract}

\keywords{magnetohydrodynamics (MHD) --- methods: statistical --- Sun: corona --- Sun: oscillations}
 
\section{Introduction}

The dissipation of magnetic wave energy is considered a relevant process in the heating of the plasma in the solar atmosphere \citep{parnell12,arregui15}. High resolution observations show that wave activity is all pervasive in the solar atmosphere \citep[see e.g.][]{nakariakov05,demoortel12,jess15}, but quantifying the importance of magnetohydrodynamic (MHD) waves in plasma heating processes  remains a difficult task. The significance of currently considered physical processes leading to wave energy transport and dissipation such as resonant absorption \citep{goossens91},  phase mixing \citep{heyvaerts83}, or Alfv\'en wave turbulence \citep{vanballegooijen11} relies on the variation of physical properties across the magnetic field. In particular, the cross-field variation of density and magnetic field strength determines the timescales and spatial scales for the damping of transverse waves \citep{hollweg88,goossens11},  how fast energy is transferred to small length scales \citep{heyvaerts83,soler15b}, the timing for the onset of dissipative effects \citep{lee86}, and the fraction of the wave energy available to be converted into heat \citep{goossens13, vandoorsselaere14}. 

Seismology inversion techniques offer an indirect method to obtain information on the physical properties of magnetic and plasma structures and their spatial variation. The method of MHD seismology, first suggested by \cite{uchida70}, \cite{rosenberg70}, and \cite{roberts84}, combines information on observed and theoretical properties of MHD waves to obtain information on difficult-to-measure physical parameters. Early implementations of the method have proven to be useful in the determination of e.g., the magnetic field strength in transversely oscillating coronal loops \citep{nakariakov01,vandoorsselaere08}, the coronal density scale height \citep{andries05b}, the magnetic flux tube expansion \citep{verth08b}, or the Alfv\'en velocity in prominence threads \citep{lin09}, see \cite{demoortel05}, \cite{nakariakov05}, \cite{demoortel12}, and \cite{arregui12b} for reviews.

A particular application of MHD seismology aims at obtaining information on the spatial variation of the mass density across solar coronal waveguides. An early example can be found in \cite{goossens02a} and involves the comparison between the theoretically predicted damping ratio of resonantly damped transverse kink waves in radially inhomogeneous waveguides and the observed damping ratio of coronal loop oscillations. By assuming a density contrast between the interior of the loop and the external corona, estimates for the transverse inhomogeneity length scale can be obtained. Recent expansions of this idea have considered the Bayesian inversion of both density contrast and transverse inhomogeneity length scale by making use of different damping regimes \citep{arregui13b} or the computation of their marginal posteriors from observed damping ratios \citep{arregui14}.  In these studies, the density structuring was arbitrarily prescribed. 

In two recent papers, \cite{soler13} and \cite{soler14a} have considered the impact of the assumed cross-field density profile on theoretically predicted periods and damping times for resonantly damped transverse oscillations and the  ensuing influence on the inferred physical parameters through inversion. \cite{soler13} describe the basic theory behind the application of the Frobenius method implementation to compute resonantly damped oscillations in nonuniform waveguides. The seismology analysis by \cite{soler14a} for parameters such as the Alfv\'en speed, the density contrast, and the transverse inhomogeneity length scale shows  that significantly different inversion results are obtained depending on the adopted density model.

In our study, a comparative analysis between the alternative density models employed by \cite{soler14a} is presented. The aim is to quantify to what extent they result in distinct parameter inference results and to evaluate which one among the presented models better explains observed data for period and damping times. This is done by considering a Bayesian approach to the solution of the inference problem and by computing the relative plausibility between models. We additionally present a model-averaging procedure to compute a combined inversion result, weighted with the evidence for each model in view of observed data.

The layout of the paper is as follows. Section~\ref{models} describes the models considered in our study of transverse oscillations in one-dimensional flux tubes with alternative forms for the cross-field variation of the density. In Section~\ref{forward} the solutions to the forward problem for transverse oscillations and the classic inversion results are discussed. Parameter inference results in the Bayesian framework are described in Section~\ref{inference}. Then, the plausibility between the considered alternative models is discussed in Section~\ref{comparison}.
Evidence-weighted posterior distributions for the unknown parameters are obtained in Section~\ref{averaging}.  In Section~\ref{conclusions}, our conclusions are presented.

\section{Cross-field density models}\label{models}

The simplest models that still capture global properties of MHD kink waves in solar flux tubes, such as the oscillatory period and the damping time by resonant absorption, consider the waveguides as one-dimensional density enhancements in cylindrically symmetric models with the magnetic field pointing along the $z$-direction. Under the zero plasma-$\beta$ approximation, the density distribution is arbitrarily imposed to model the waveguide as a region of enhanced mass density that varies in the radial direction to connect the internal higher-density medium to the relatively lower external density region. The transition between the internal, $\rho_{\rm i}$, and the external, $\rho_{\rm e}$, densities occurs at a nonuniform layer of thickness $l/R$, with $R$ the radius of the waveguide. At a given location inside the nonuniform density layer the global kink mode frequency matches the local Alfv\'en frequency. This produces the time decay of the wave amplitude for standing kink waves \citep{hollweg88,goossens02a,ruderman02}. This resonant damping process also operates for propagating kink waves, resulting in the spatial decay of the wave amplitude as the wave propagates guided by the plasma structure \citep{terradas10,pascoe12}.

\begin{figure}
\epsscale{1.00}
\plotone{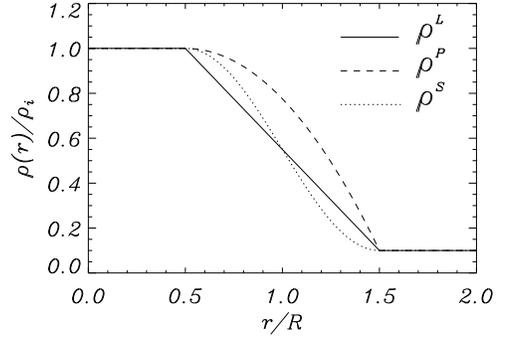}
\caption{Spatial variation of the equilibrium mass density across the magnetic flux tube for waveguide models with a linear profile (solid line), a sinusoidal profile (dotted line), and a parabolic profile (dashed line). The ratio of internal to external density is $\rho_{\rm i}/\rho_{\rm e}=10$ and the thickness of the nonuniform layer is $l/R=1$.}
              \label{figmodels}
    \end{figure}

Initial theoretical investigations on the damping of transverse kink waves by resonant absorption focused on the consideration of thin tube and thin boundary models. This is the so-called TTTB approximation, which assumes that the wavelength is long and that the thickness of the nonuniform transitional layer is short in comparison to the radius of the waveguide ($R/\lambda\ll1$, $l/R\ll1$) \citep[see, e.g.,][]{hollweg88,goossens95a,goossens02a,ruderman02}. The analysis was then extended to include fully nonuniform layers in both one-dimensional \citep{vandoorsselaere04a,soler13} and two-dimensional \citep{arregui05} density models. 

Following \cite{soler14a}, we consider three alternative models for the density variation across the magnetic field. We use $M^S$, $M^L$, and $M^P$ to name the sinusoidal, linear, and parabolic density models used in this work. By adopting the same superscript convention for the radial variation of the density at the nonuniform layer, we have

\begin{equation}\label{rhos}
\rho^S(r)=\frac{\rho_{\rm i}}{2}\left[\left(1+\frac{\rho_{\rm e}}{\rho_{\rm i}}\right)-\left(1-\frac{\rho_{\rm e}}{\rho_{\rm i}}\right)\sin\left(\frac{\pi}{l}(r-R)\right)\right],
\end{equation}
for the sinusoidal model,

\begin{equation}\label{rhop}
\rho^L(r)=\rho_{\rm i}-\frac{\rho_{\rm i}-\rho_{\rm e}}{l}\left(r-R+\frac{l}{2}\right),
\end{equation}
for the linear model, and

\begin{equation}\label{rhol}
\rho^P(r)=\rho_{\rm i}-\frac{\rho_{\rm i}-\rho_{\rm e}}{l^2}\left(r-R+\frac{l}{2}\right)^2.
\end{equation}
for the parabolic model. None of the specifically assumed profiles is expected to be an accurate quantitative representation of the real density
variation across solar flux tubes, but they provide a means to model alternative cases that, by means of the model comparison technique presented in this paper, can offer information about the plausibility of different profiles. Figure~\ref{figmodels} shows the spatial variation of the density distribution for the three considered profiles for fixed values of the density contrast and the thickness of the nonuniform layer. 

\section{Forward and inverse solutions}\label{forward}

The basic theory for the time/spatial damping of standing/propagating kink waves due to resonant absorption has been developed by a number of studies \citep{hollweg88,goossens95a,goossens11,terradas10, verth10,soler11a,soler11c,soler11d}.  Numerical simulations have confirmed the obtained damping properties, by analyzing the spatial and temporal characteristics of the mode coupling process  in coronal loop models and arbitrary inhomogeneous structures \citep{terradas06a,terradas06c,terradas08b,pascoe10,pascoe11,pascoe12}.  
For standing waves, an analytical expression for the period of the fundamental kink mode can be obtained under the thin tube approximation 
 \citep[see, e.g.,][]{goossens08a}

\begin{equation}\label{period}
P=\tau_{\rm Ai}\sqrt{2}  \left(\frac{\zeta+1}{\zeta}\right)^{1/2}.
\end{equation}
Here $\tau_{\rm A,i}$ is the internal Alfv\'en travel time and $\zeta=\rho_{\rm i}/\rho_{\rm e}$ the density contrast. This solution to the forward problem expresses a relation between the observable quantity (period) and  two physical parameters to be determined (Alfv\'en travel time and density contrast). The period is independent of the transverse inhomogeneity length scale, $l/R$, because of the adopted thin tube approximation. It becomes dependent on this parameter when this assumption is relaxed \citep{vandoorsselaere04a,arregui05}. The error on the period associated with the use of the TTTB approximation was studied by \cite{soler14a}

Because of resonant absorption, time damping occurs and  the amplitude of the radial velocity component decays, while the azimuthal velocity component becomes larger in the vicinity of the resonant position. The classic theory for resonant Alfv\'en waves under the thin boundary approximation  provides us with an analytical expression for the damping time, $\tau_{\rm d}$, as a function of the relevant physical parameters. In units of the oscillatory period, this expression is of the form \citep[see, e.g.,][]{hollweg88,goossens02a,ruderman02}

\begin{equation}\label{damping}
\frac{\tau_{\rm d}}{P}= F \left(\frac{R}{l}\right)\left(\frac{\zeta+1}{\zeta-1}\right).
\end{equation}
The numerical factor $F$ depends on the assumed density profile at the nonuniform layer. The factor is $2/\pi$ for the sinusoidal profile, $4/\pi^2$ for the linear profile, and $4\sqrt{2}/\pi^2$ for the parabolic profile. The numerical values for these factors do not seem to differ to a great extent, but the analytical inversions performed by \cite{soler14a} indicate that significant differences may arise between the resulting one-dimensional inversion curves when fully numerical results to the forward problem are used. It must be noted that  analogous expressions to the ones given in Eqs.~(\ref{period}) and (\ref{damping}) can be obtained in the case of spatial damping of propagating kink waves \citep{terradas10}, upon replacement of the period by the wavelength and the damping time by the damping length. The reason is that resonant damping does not make any distinction with respect to the standing or propagating character of the wave and, under the thin tube and thin boundary approximations, the influence of the particular density profile at the nonuniform transitional layer is solely contained in the factor $F$. In both cases, the dynamics corresponds to a surface Alfv\'en wave \citep{goossens09,goossens12a}. For standing kink waves, the observational consequence is  the attenuation of the amplitude in time. For propagating kink waves, resonant absorption produces the attenuation of wave amplitude in space.

\begin{figure}
\epsscale{1.00}
\plotone{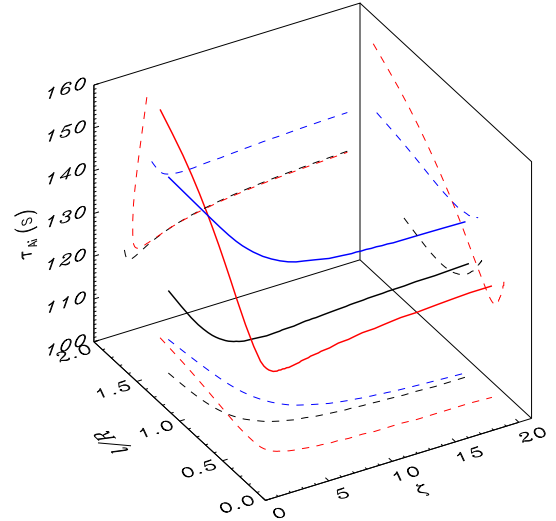}
        \caption{Classic seismological inversion result of the event with strong damping ($P=185$ s and $\tau_{\rm d}=200$ s) with the one-dimensional solution curves in the three-dimensional parameter space corresponding to the sinusoidal (black), linear (red), and parabolic (blue) density models. The dashed lines are the projections of the inversion curves to the various planes.}
              \label{figsoler}
    \end{figure}
    
Once the solution to the forward problem is obtained by solving Eqs.~(\ref{period}) and (\ref{damping}) for a range of values of the physical parameters,  their inference can be performed using the analytical inversion technique developed by \cite{goossens08a}. In the inversion procedure, the three quantities ($\tau_{\rm Ai}$, $\zeta$, and $l/R$) are the seismic quantities of interest to be inferred from observed period and damping time. As we only have two observables and three unknowns, there are in principle an infinite number of solutions that can equally well explain the observed data, as first pointed out by \cite{arregui07a}. These solutions follow a one-dimensional curve in the three-dimensional parameter space of unknowns. The curve was first obtained numerically by \cite{arregui07a} and analytically by \cite{goossens08a} for a particular density model in the context of standing kink waves and by \cite{goossens12b} in the context of propagating kink waves. The method is applicable to both coronal loop and prominence fine-structure oscillations \citep[see reviews by][]{goossens08b, arregui12a,arregui12b}. Solution curves for the three density models under consideration were obtained by \cite{soler14a}. Although the inversion solution is unable to fully constrain the three parameters of interest, the seismic variables are found to be constrained to intervals that can be calculated analytically, under the thin tube and thin boundary approximations. Their expressions can be found in \cite{goossens08a} and \cite{goossens08b}.

The forward solutions for period and damping time, outside the thin tube and thin boundary approximations, have been computed for particular cross-field density models by \cite{vandoorsselaere04a} in one-dimensional equilibrium models and by \cite{arregui05} in two-dimensional equilibrium models. The two observables are then functions of the three unknown parameters, $\tau_{\rm Ai}$, $\zeta$, and $l/R$. 
\cite{soler14a} computed numerical solutions for the three density models described above and compared the differences between he numerical  forward solutions and those computed under the TTTB approximations and also between the forward solutions obtained with the three alternative density models. Furthermore, \cite{soler14a} applied the analytical inversion procedure to forward solutions obtained using the three density models. 

An example inversion solution (see Figure~\ref{figsoler}), shows the one-dimensional solution curve that links those valid combinations of ($\tau_{\rm Ai}$, $\zeta$, $l/R$) that reproduce a particular observed period and damping time.  As mentioned in that study,  the impact of the assumed density profile at the nonuniform layer is strong both in the shape of the one-dimensional inversion solution curves in the three-dimensional parameter space and on the obtained valid intervals of the seismic variables. Thus, the inferred intervals of the seismic variables are directly affected by the specific choice of density variation. From the obtained results, \cite{soler14a} conclude that the particular choice for the density model at the nonuniform layer can have a strong impact on the seismologically inferred parameters.

In our study, we will employ both analytical approximations (given by Eqs.~[\ref{period}] and [\ref{damping}]) and the numerical solutions, as given by \cite{soler14a}, to the forward problem. The inversion procedure, however, will differ, and, in addition, model comparison and averaging techniques are implemented. The classic approach to the inversion problem deals with finding the mathematical solution for the one-dimensional inversion curve in the three-dimensional parameter space by imposing the exact matching of the theoretically predicted period and damping time values with those observed. The procedure being mathematically correct, we recall that parameter inference has to be pursued under circumstances in which information from observations is incomplete and uncertain. For this reason, the inference has to be carried out by following a probabilistic approach. We will apply three levels of Bayesian inference to the problem of determining the cross-field density profile from observed values of period and damping time. The first level involves the inversion of the three parameters of interest, the internal Alfv\'en travel time, and the two parameters that define the cross-field density profile, namely, the density contrast and the transverse inhomogeneity length scale, assuming that each of the models described above is true. The second level deals with the comparison between the three assumed models, to ascertain which one is more plausible in view of given data. The third level uses all the information on the data and the three models under consideration to produce parameter inversions in which the resulting inference is an average of the particular inferences for each model, weighted with the plausibility for each model in view of data.

\section{Parameter inference}\label{inference}

Figure~\ref{figsoler} indicates that the classic inversion leads to well-differentiated results for the inference of the three unknowns depending on the assumed cross-field density model. Our first aim is to evaluate in a quantitative manner how different the results are when a fully Bayesian approach is employed, which, in addition, enables us to correctly propagate uncertainty from measured wave properties to inferred parameters. 

We have performed the inference for the unknown physical parameters, $\tau_{\rm Ai}$, $\zeta$, and $l/R$ by first generating theoretical predictions for period and damping time by resonant absorption. The latter two parameters completely define the cross-sectional density profile, once one of the three alternative models (linear, parabolic, sinusoidal) is selected. The forward predictions by each model are computed for different combinations of the equilibrium parameters both under the TTTB approximations, using the algebraic expressions ~(\ref{period}) and (\ref{damping}), and by employing the Frobenius method developed by \cite{soler13} and applicable to any arbitrary density profile at the nonuniform layer.  In the following, we will refer to them as the TTTB and the numerical forward solutions, respectively. Those predictions are then compared to observed data for period and damping time in the inversion. 

Let us gather the three parameters to be inferred in the vector of unknowns $\mbox{\boldmath$\theta$}=(\tau_{\rm Ai}, \zeta, l/R)$ and the two observable quantities in $d=(P,\tau_{\rm d})$. The inference is based on the use of Bayes' theorem \citep{bayes63}, which provides us with the probability density function of the unknown parameters conditional on the observed data and the model $M$, $p(\mbox{\boldmath$\theta$} | d, M)$, as a combination of the likelihood function, $p(d|\mbox{\boldmath$\theta$},M)$, and the prior probability density function for the unknown parameters, $p(\mbox{\boldmath$\theta$}|M)$. Their relationship is given by the expression

\begin{equation}\label{bayes}
p(\mbox{\boldmath$\theta$} | d,M)=\frac{p(d | \mbox{\boldmath$\theta$},M)p(\mbox{\boldmath$\theta$}|M)}{\int p(d|\mbox{\boldmath$\theta$}, M)p(\mbox{\boldmath$\theta$}|M)d\mbox{\boldmath$\theta$}},
\end{equation}
which assumes that model $M$ under consideration is true. The denominator is the so-called evidence,  an integral of the likelihood over the prior distribution that normalizes the likelihood and turns it into a probability. This can be ignored when performing parameter inference, because it is independent of the vector of parameters, but plays the central role in model comparison, as will become apparent in Section~\ref{comparison}. Both prior and likelihood represent probabilities that are directly assigned, whilst the posterior is computed. This posterior contains all the information that can be gathered concerning the vector of unknowns, conditional on observed data. The magnitude of the posterior probability density function is a measure of the degree of belief on the values the parameter vector can take on.

We proceed by first assigning the direct probabilities to be used in Bayes' theorem for the  likelihood function and the prior distribution. We adopt a Gaussian likelihood function that relates the observed period and damping time, ($P$, $\tau_{\rm d}$), and the predictions of the model, ($P^M(\mbox{\boldmath$\theta$})$, $\tau_{\rm d}^M(\mbox{\boldmath$\theta$})$). Under the TTTB solutions they are given in Eqs.~(\ref{period}) and (\ref{damping}), where the particular value of $F$ corresponding to each density model has to be used. Outside the TTTB approximation they are numerically computed. Then, the observed values and the theoretical/numerical predictions can be compared by adopting a likelihood of the form

\begin{eqnarray}\label{like}
p(P, \tau_{\rm d}|\mbox{\boldmath$\theta$}, M)=&&(2\pi\sigma_P\sigma_{\tau_{\rm d}})^{-1} \nonumber\\
&&\times \exp \left\{\frac{\left[P-P^{M}(\mbox{\boldmath$\theta$})\right]^2}{2 \sigma^2_P} + \frac{\left[\tau_{\rm d}-\tau_{\rm d}^{M}(\mbox{\boldmath$\theta$})\right]^2}{2 \sigma^2_{\tau_{\rm d}}}\right\},
\end{eqnarray}
with $\sigma_P$ and $\sigma_{\tau_{\rm d}}$ the uncertainties associated with the measured period and damping time, respectively. Concerning the prior information, we adopt uniform prior distributions for the unknown parameters over given ranges, so that all the values inside those ranges are equally probable a priori. We therefore assign

\begin{equation}\label{prior}
p(\theta_i)=\frac{1}{\theta^{max}_i-\theta^{min}_i} \mbox{\hspace{0.4cm}} \mbox{for}  \mbox{\hspace{0.4cm}} \theta^{min}_i\leq\theta_i\leq\theta^{max}_i  \mbox{\hspace{0.4cm}}  i = 1,2,3
\end{equation}
and zero otherwise, where $\theta_1$, $\theta_2$, and $\theta_3$ correspond to $\tau_{\rm Ai}$, $\zeta$, and $l/R$, respectively.  Unless otherwise stated, the following ranges have been taken: $\tau_{\rm Ai}\in[1,500]$ s, 
$\zeta\in[1.1,10]$, and  $l/R\in[0.01,2.0]$. Application of Bayes' rule to expressions (\ref{like}) and (\ref{prior}) provides us with the full posterior, $p(\mbox{\boldmath $\theta$}| d, M)$.

In our analysis, rather than sampling the posterior using a Markov Chain Monte Carlo technique as in \cite{arregui11b}, we make use of the basic definition of marginal posteriors, by performing the required integrals over the parameter space. This was first done by \cite{arregui14} to infer the density contrast and transverse inhomogeneity length scale from measured damping ratio values. As shown by these authors, although the classic one-dimensional curves in the parameter space point to an infinite number of equally valid solutions, not all of them are equally probable. To obtain the degree of belief in terms of marginal posteriors, the following integrals have to be computed:

\begin{equation}
\begin{array}{ll}
p(\tau_{\rm Ai} | \{P, \tau_{\rm d}\}, M)=\int p(\{\tau_{\rm Ai},\zeta,l/R\} | \{P, \tau_{\rm d}\}, M) \ d\zeta\  d(l/R), \label{m1}\\\\
p(\zeta | \{P, \tau_{\rm d}\}, M)=\int p(\{\tau_{\rm Ai},\zeta,l/R\} | \{P, \tau_{\rm d}\}, M) \ d\tau_{\rm Ai}\  d(l/R),  \\\\
\mbox{and}\\\\
p(l/R | \{P, \tau_{\rm d}\}, M)=\int p(\{\tau_{\rm Ai},\zeta,l/R\} | \{P, \tau_{\rm d}\}, M) \ d\tau_{\rm Ai} \ d\zeta. \label{m3} \\
 \end{array}
\end{equation}
The marginal posteriors encode all the information for each model parameter available in the priors and the data. Their computation correctly propagates uncertainty from data to inferred parameters. The analysis is applied to two cases from a set of observations by \cite{ofman02b}, one representing observations with moderate damping ($P=272$ s,  $\tau_{\rm d}=849$ s, $\tau_{\rm d}/P=3.12$) and another representing observations with strong damping ($P=185$ s, $\tau_{\rm d}=200$ s, $\tau_{\rm d}/P=1.08$). A timescale of 30 s is taken as  the uncertainty on measured period and damping time, in line with current observational capabilities. 

\begin{figure*}
   \centering
   \includegraphics[width=0.3\textwidth]{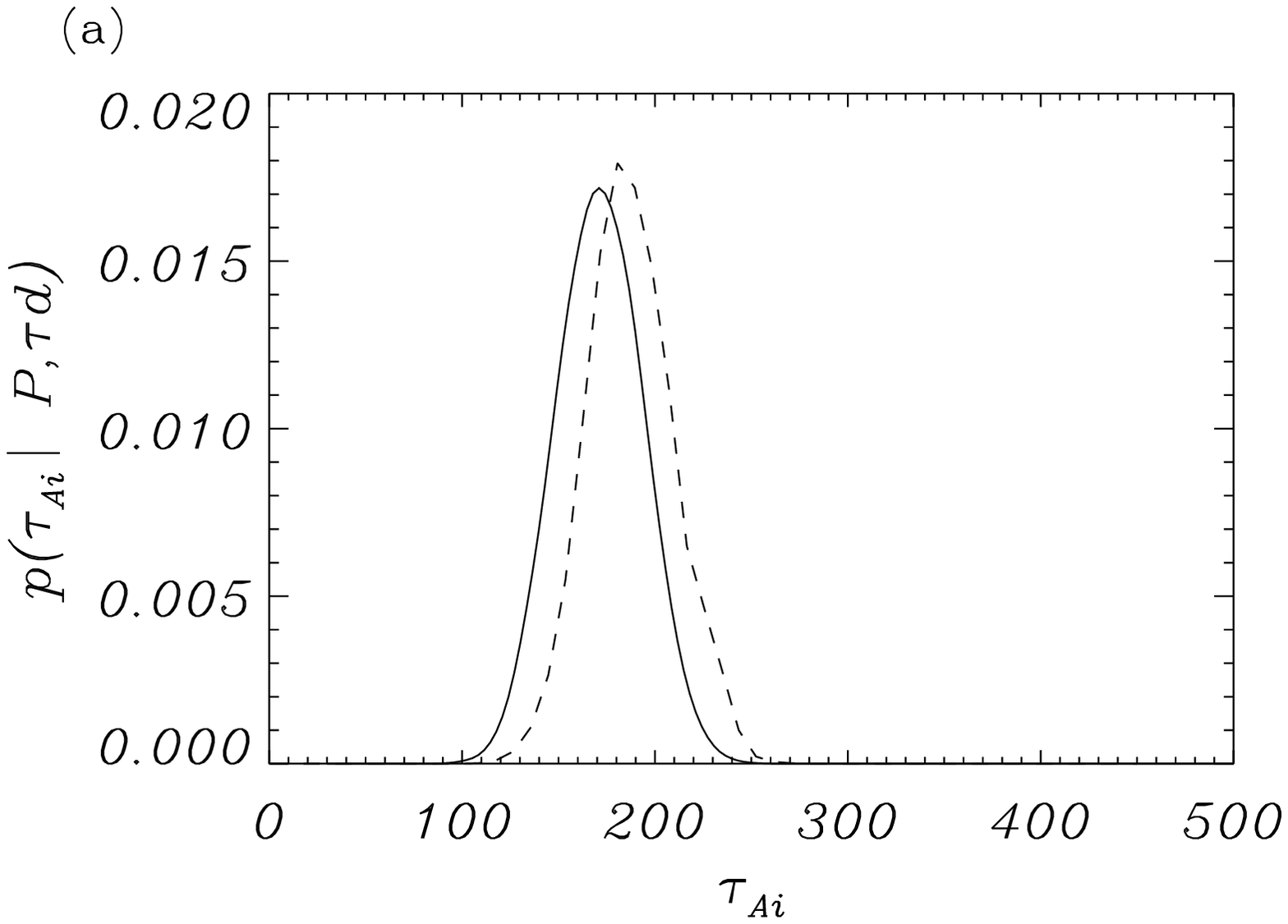}
   \includegraphics[width=0.3\textwidth]{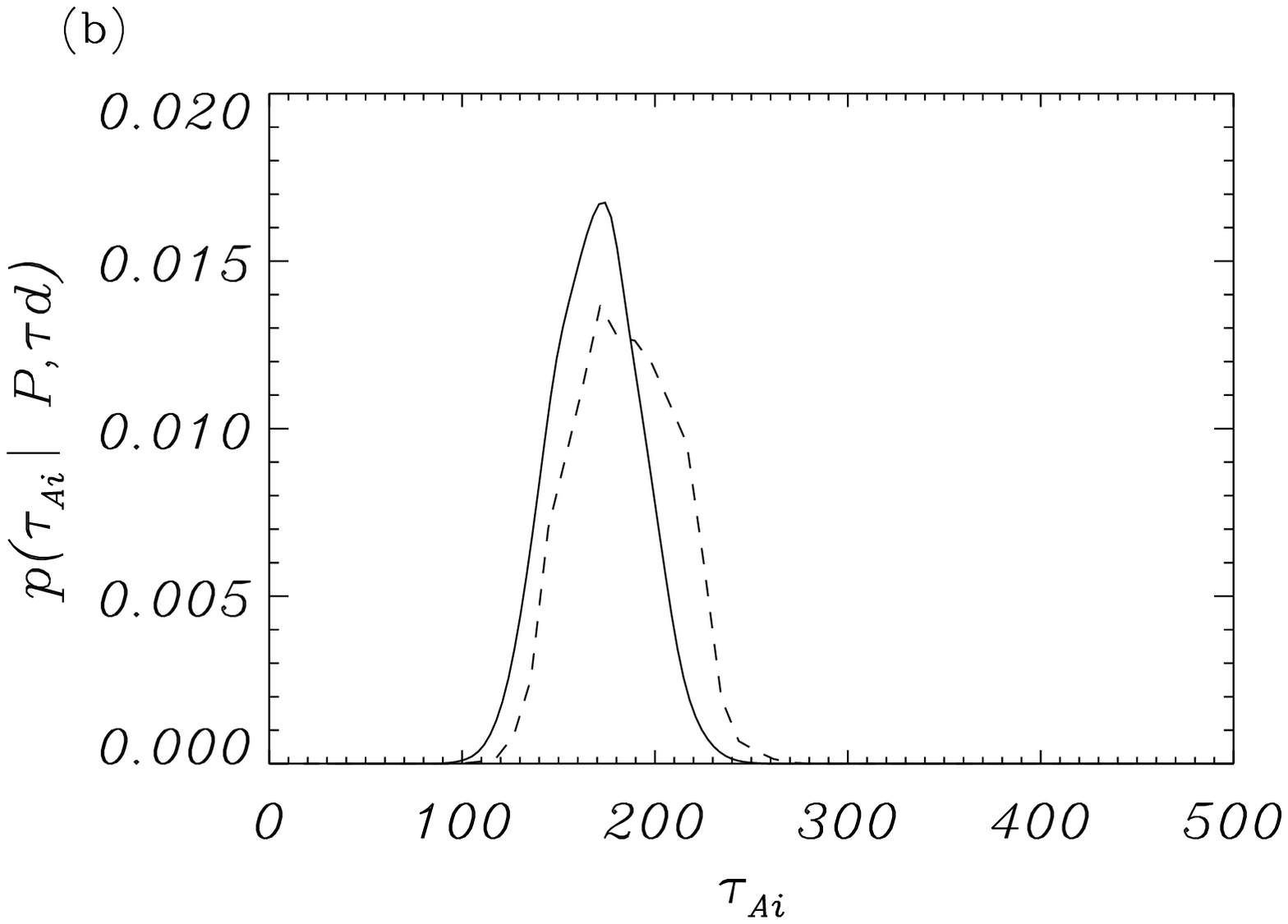}
   \includegraphics[width=0.3\textwidth]{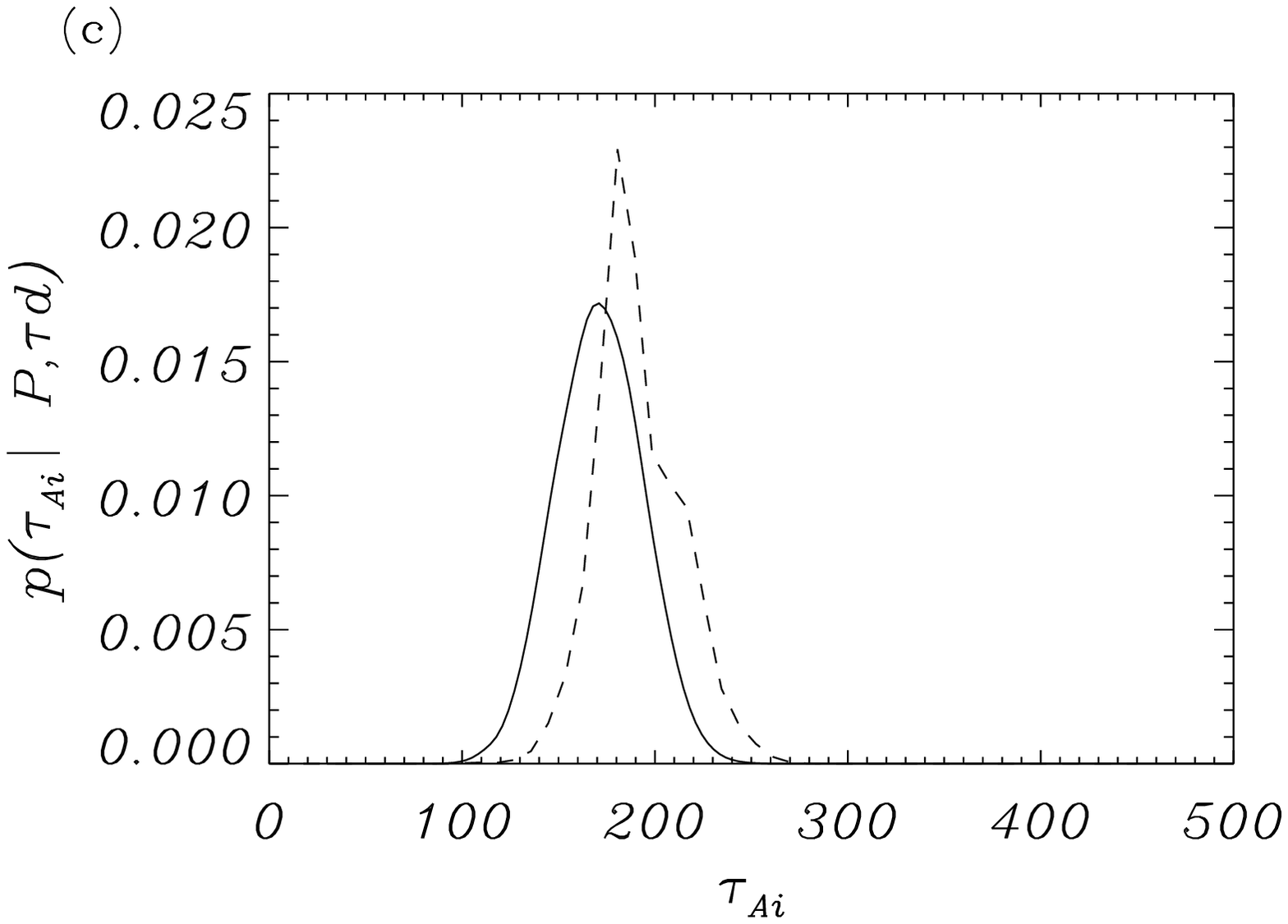}\\
   \includegraphics[width=0.3\textwidth]{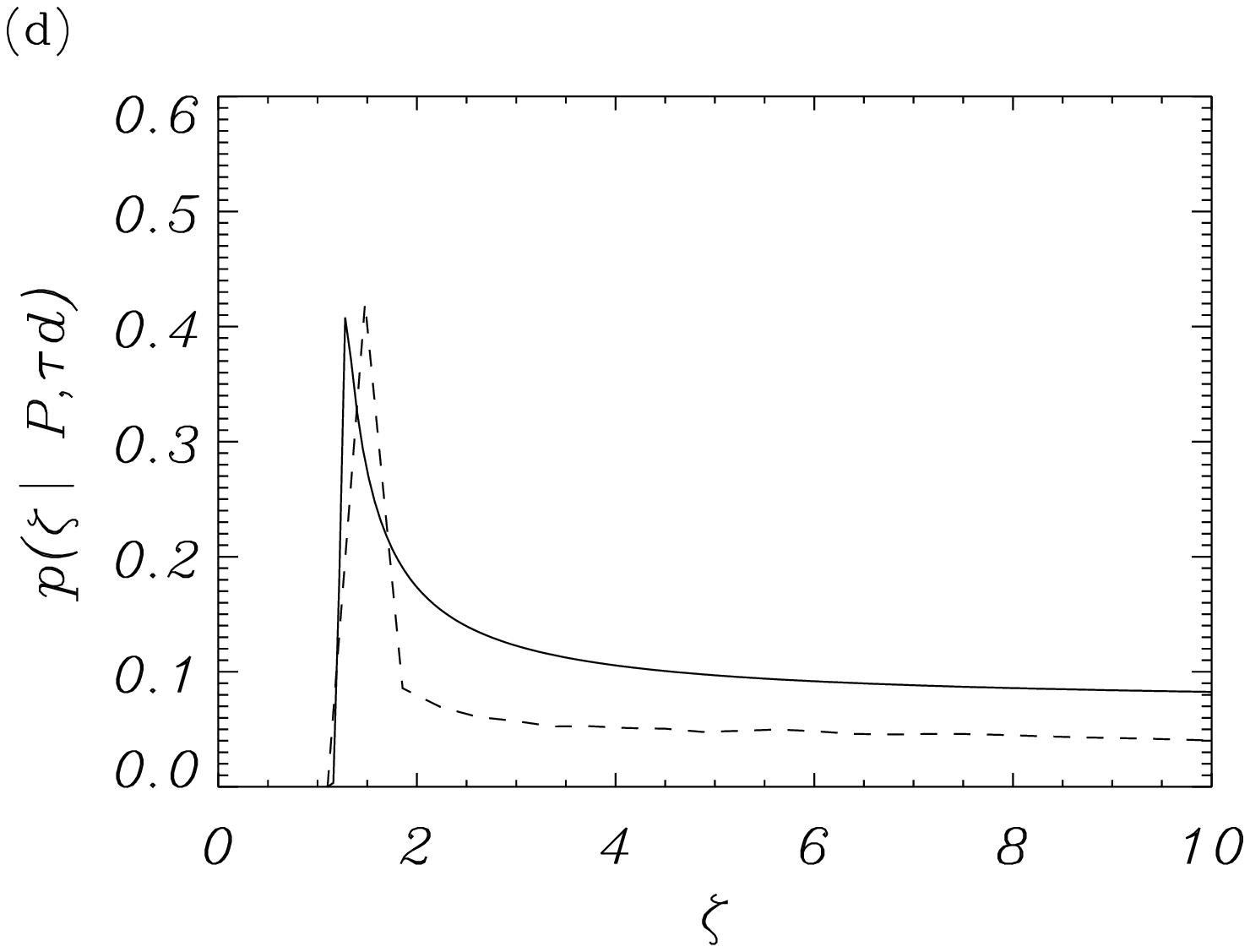}
   \includegraphics[width=0.3\textwidth]{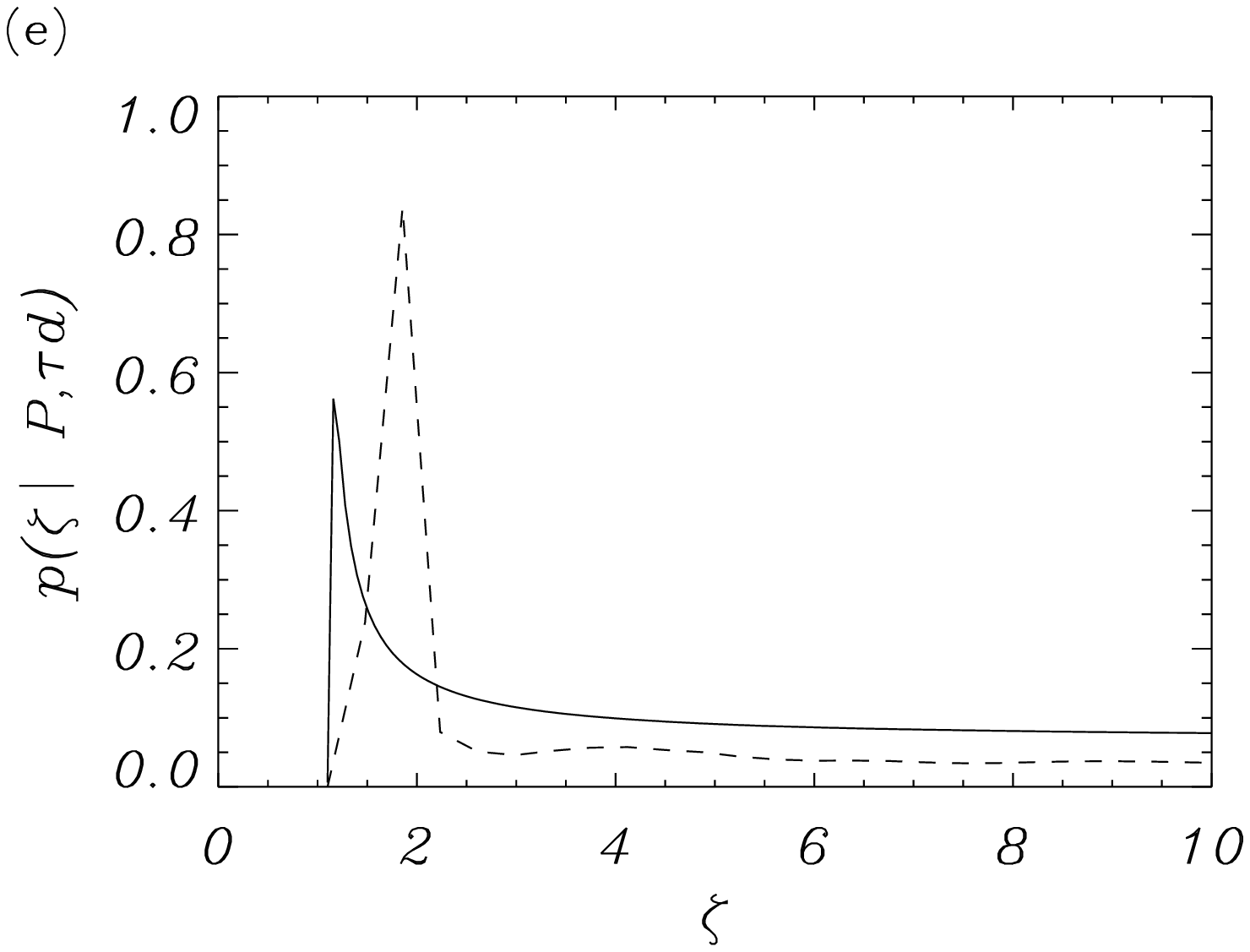}
   \includegraphics[width=0.3\textwidth]{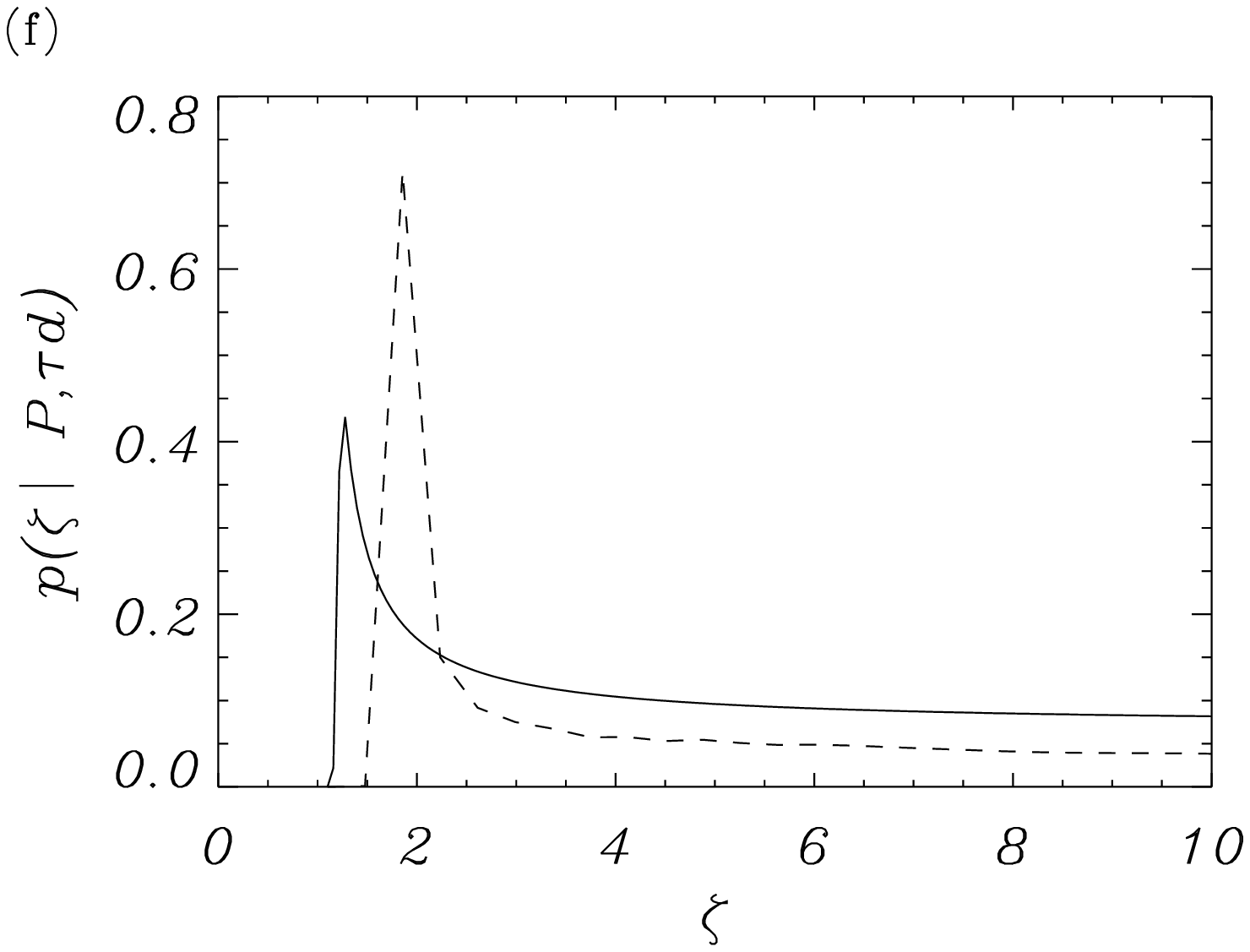}\\
   \includegraphics[width=0.3\textwidth]{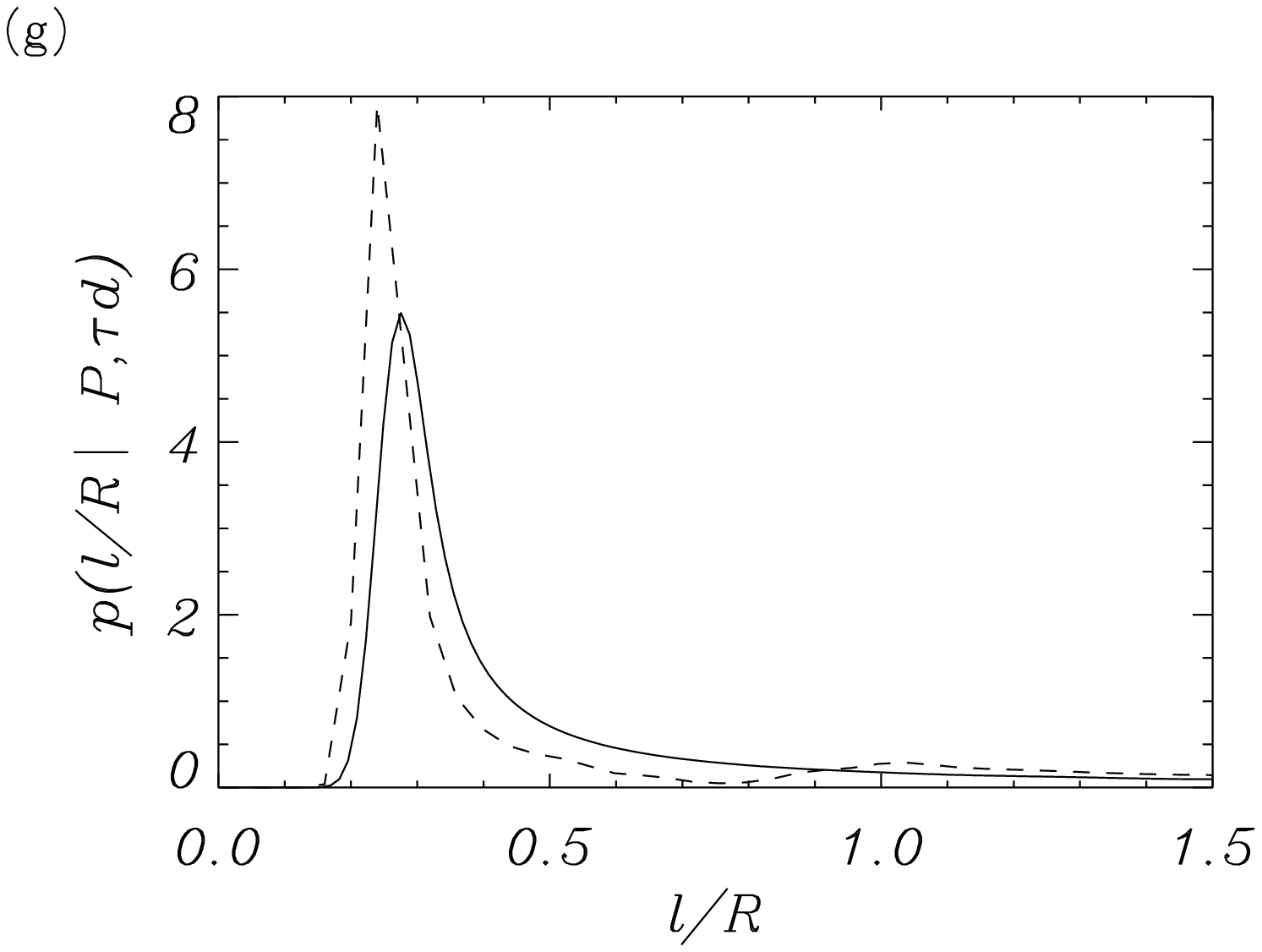}
   \includegraphics[width=0.3\textwidth]{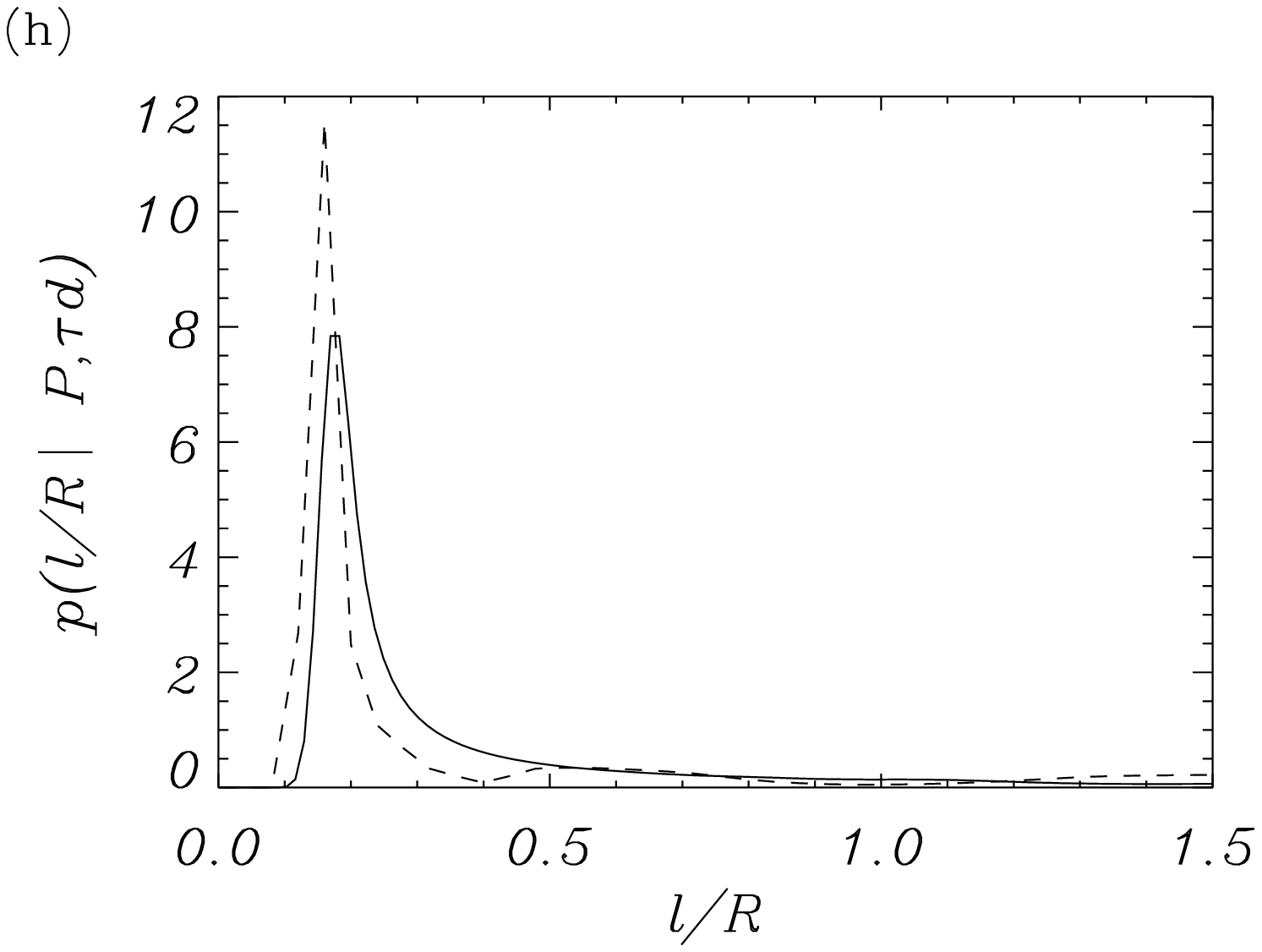}
   \includegraphics[width=0.3\textwidth]{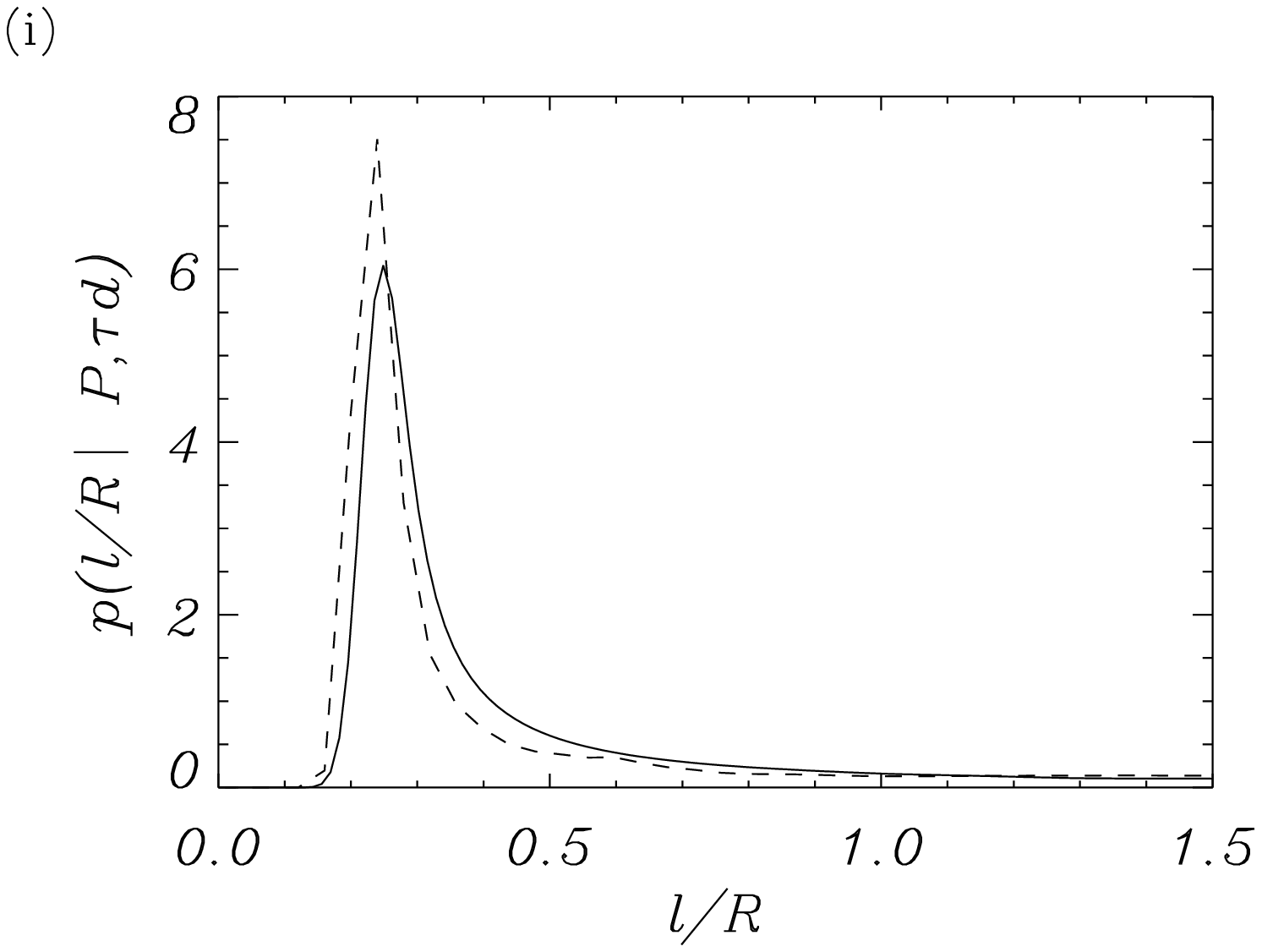}
   \caption{Marginal posteriors for the three parameters of interest ($\tau_{\rm Ai}$, $\zeta$, $l/R$) computed using expressions~(\ref{m1}) for the moderate damping case ($P=272$ s; $\tau_{\rm d}=849$ s; $\sigma_{\rm P}=\sigma_{\rm \tau_{\rm d}}= 30$ s). Left column: sinusoidal density profile; middle: linear profile; right: parabolic profile. In each panel, solid lines represent the inference performed using the TTTB forward solutions, while dashed lines are for the inference performed using the numerical forward solutions. Summary data for the posteriors with the median and errors at the 68\% credible region are shown in Table~\ref{inf1table}.}
              \label{inf1f1}%
    \end{figure*}
    
\begin{figure*}
   \centering
   \includegraphics[width=0.3\textwidth]{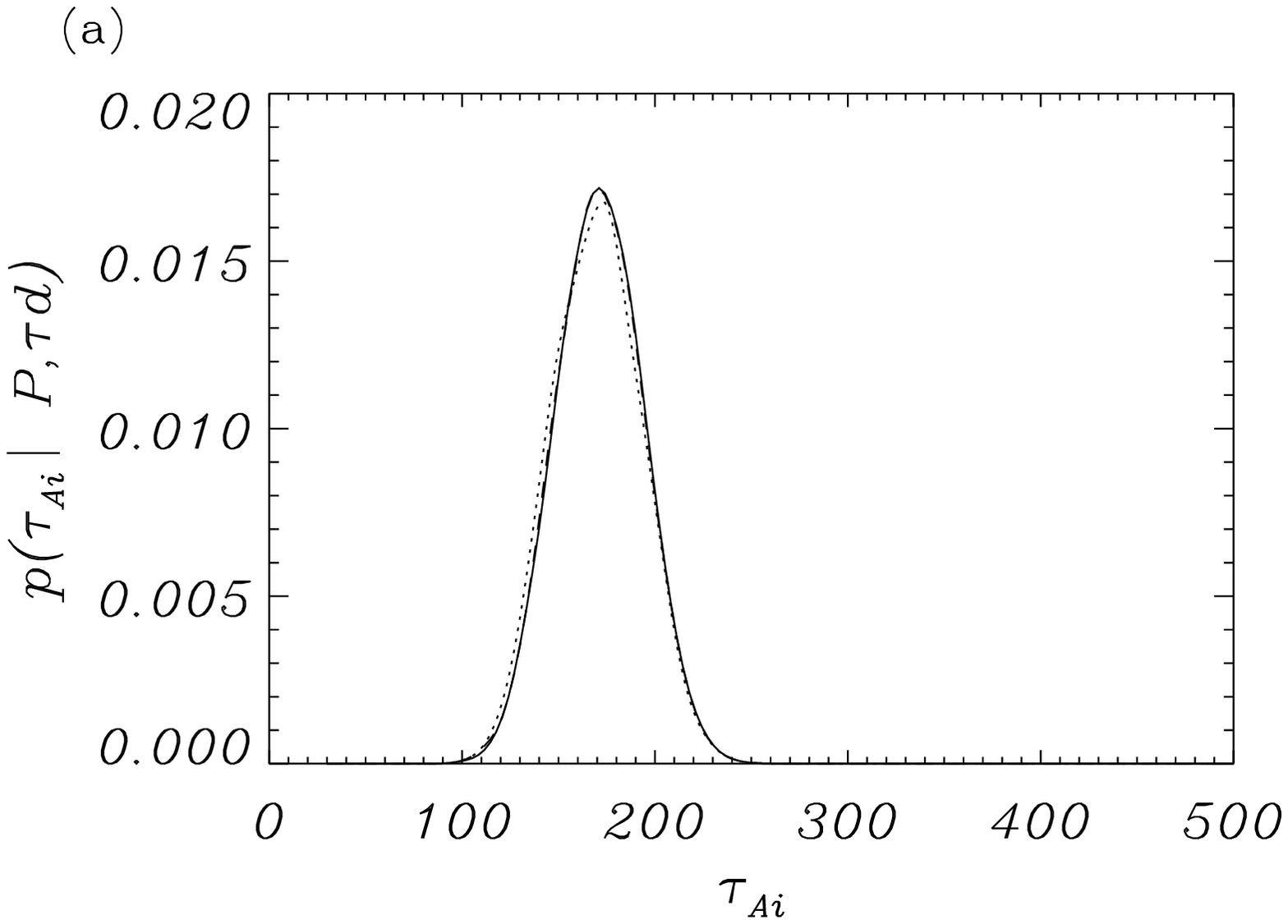}
   \includegraphics[width=0.3\textwidth]{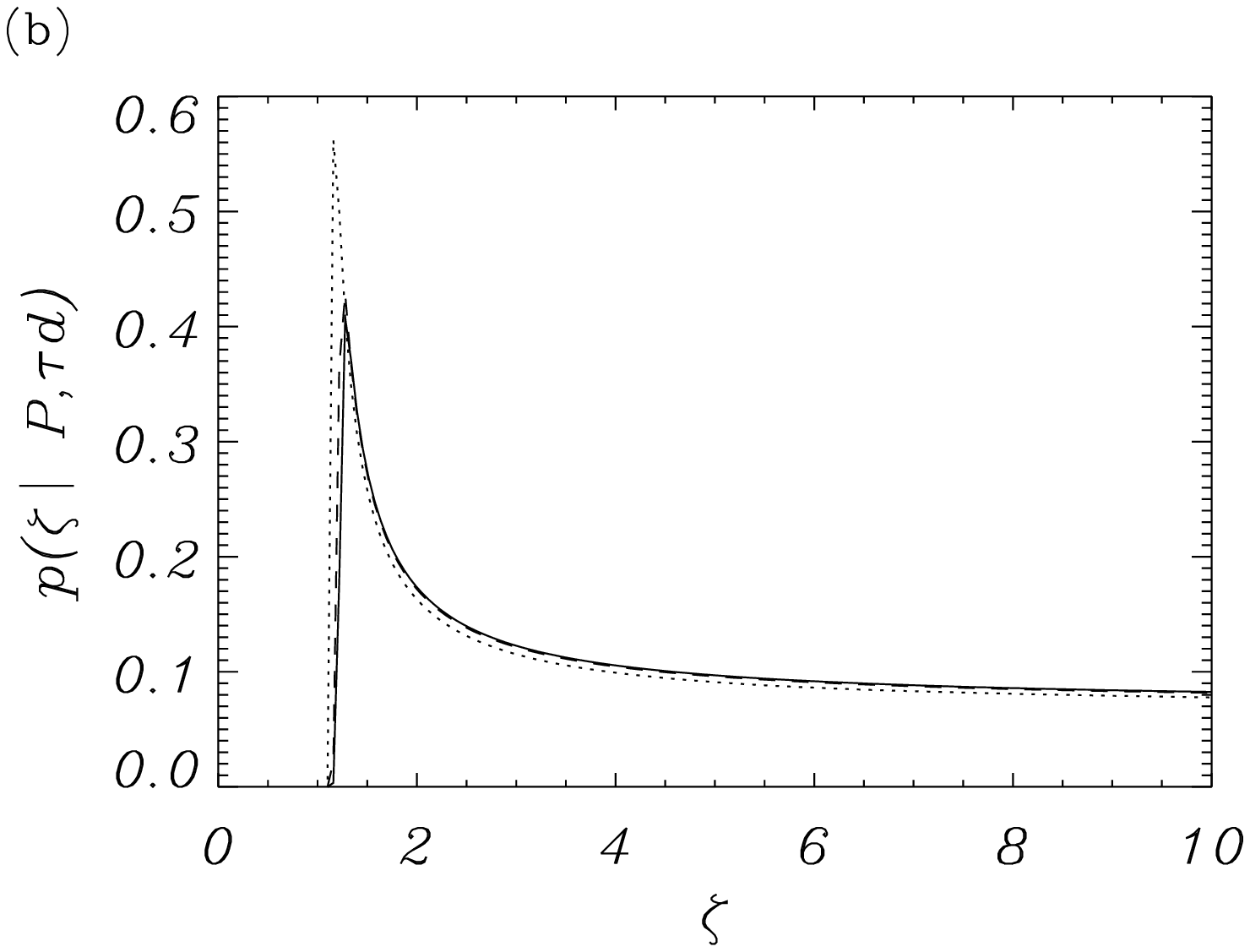}
   \includegraphics[width=0.3\textwidth]{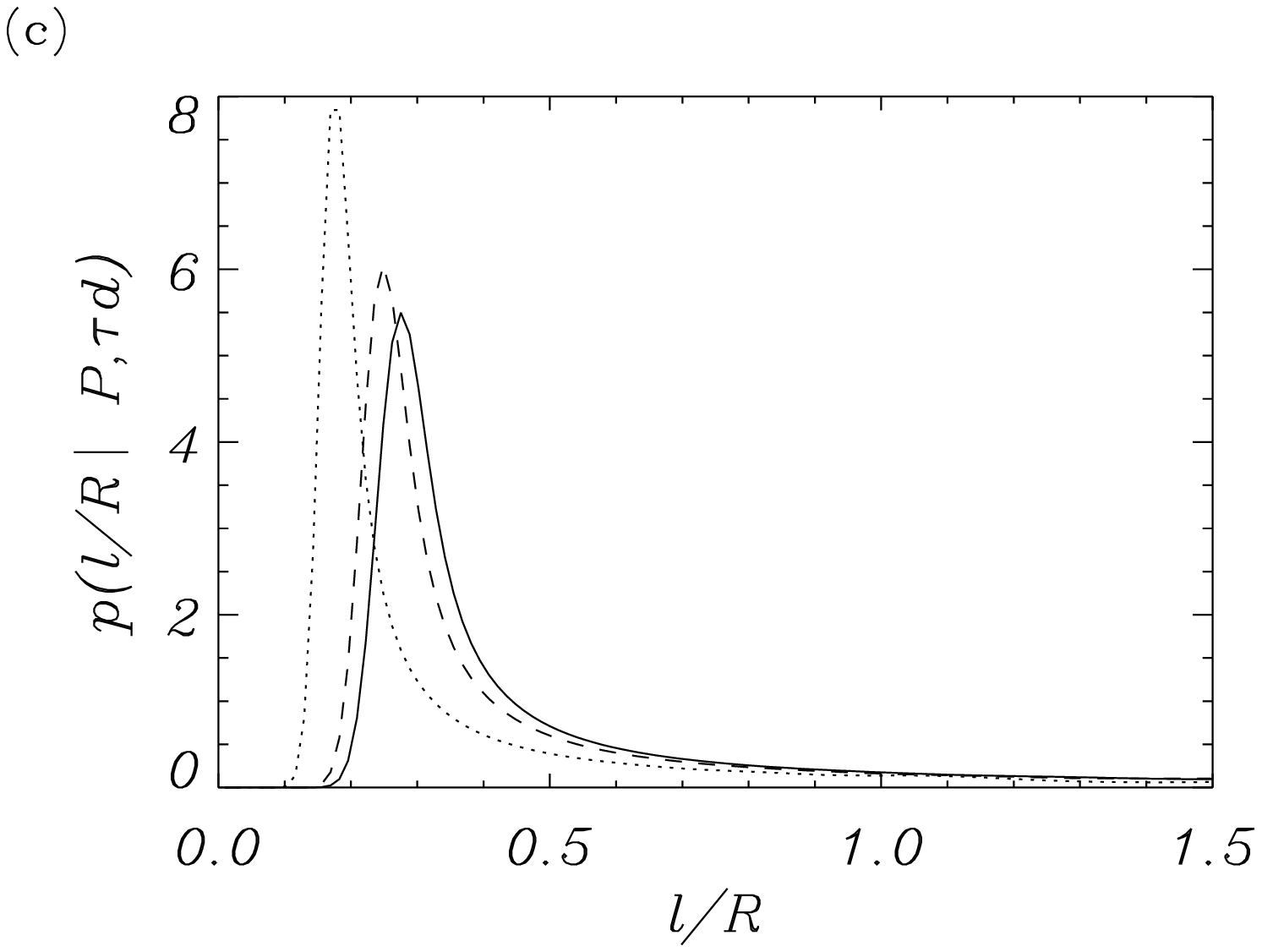}\\
   \includegraphics[width=0.3\textwidth]{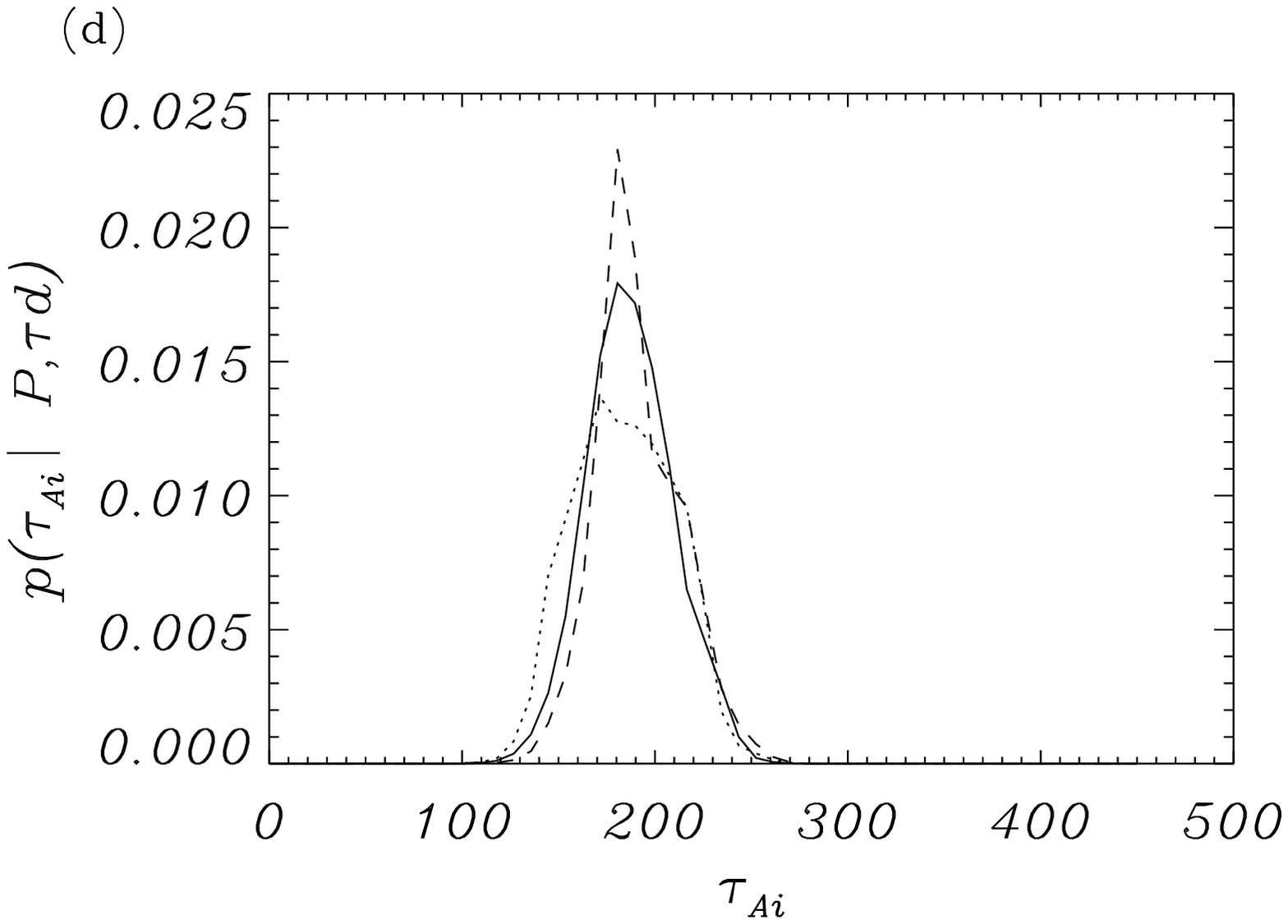}
   \includegraphics[width=0.3\textwidth]{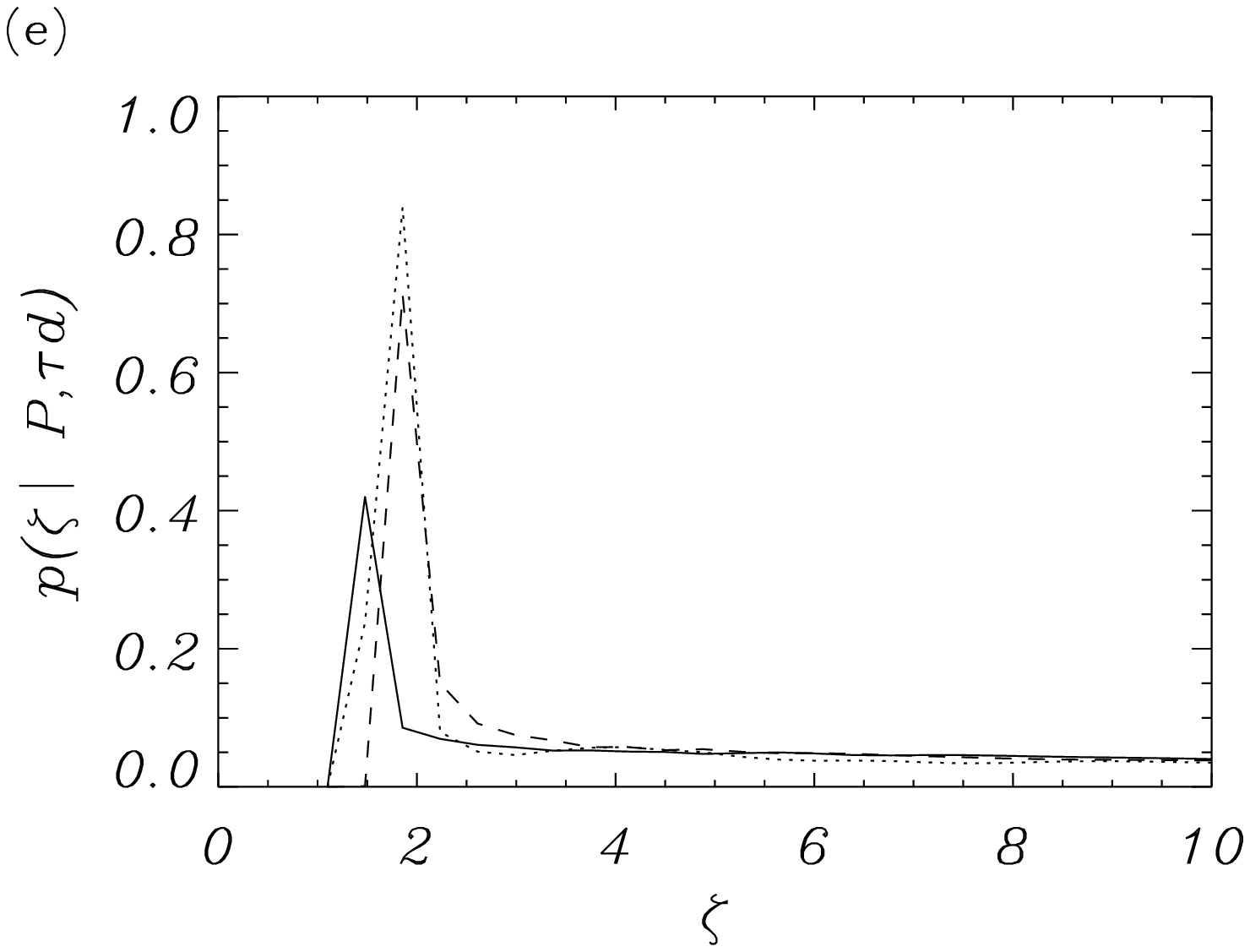}
   \includegraphics[width=0.3\textwidth]{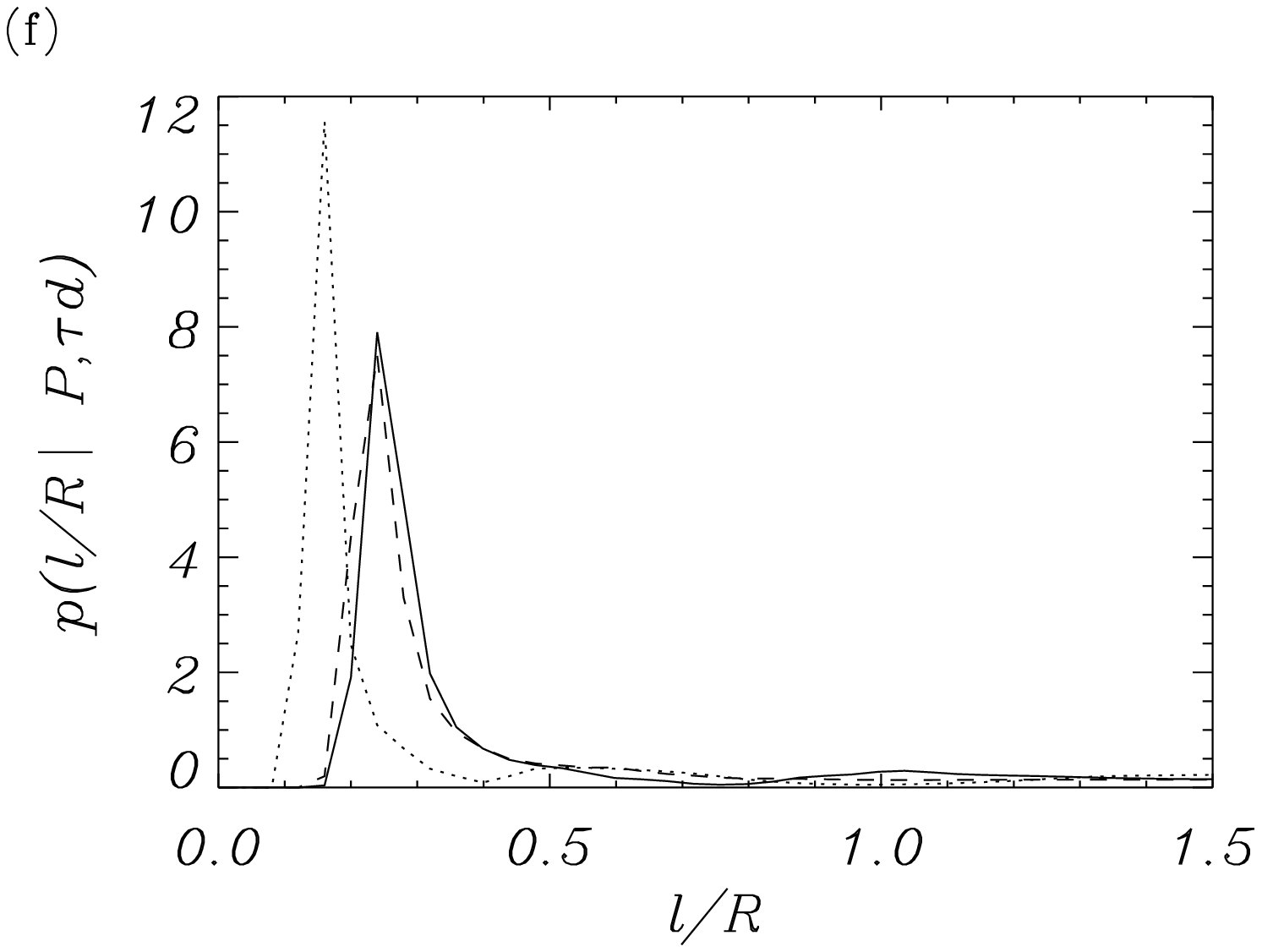}
   \caption{Marginal posteriors shown in Fig.~\ref{inf1f1} now displayed for direct comparison between results for the three alternative density models: sinusoidal (solid), linear (dotted), parabolic (dashed). Top row: TTTB results; bottom row: numerical results.}
              \label{inf1f2}%
    \end{figure*}

\begin{figure*}
   \centering
   \includegraphics[width=0.3\textwidth]{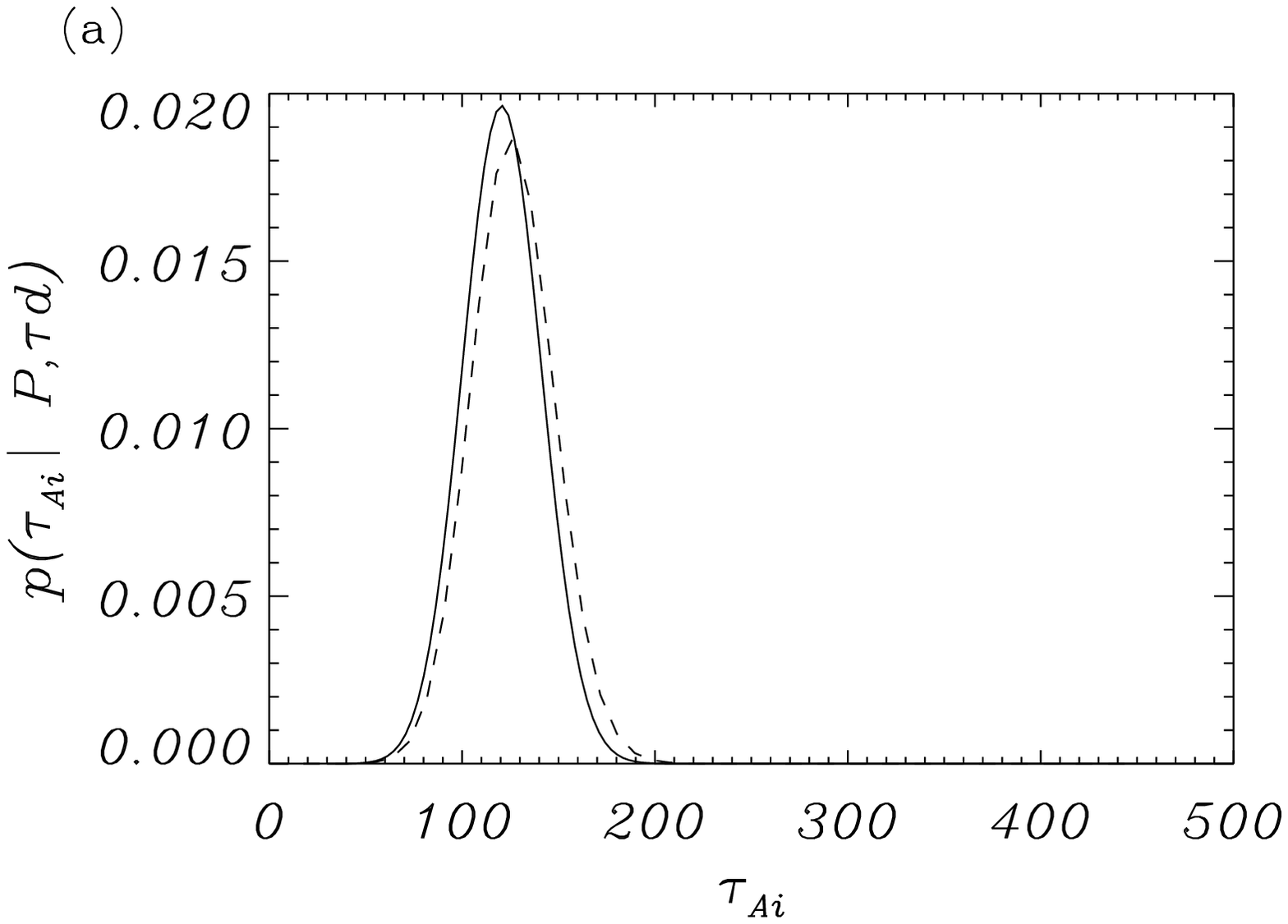}
   \includegraphics[width=0.3\textwidth]{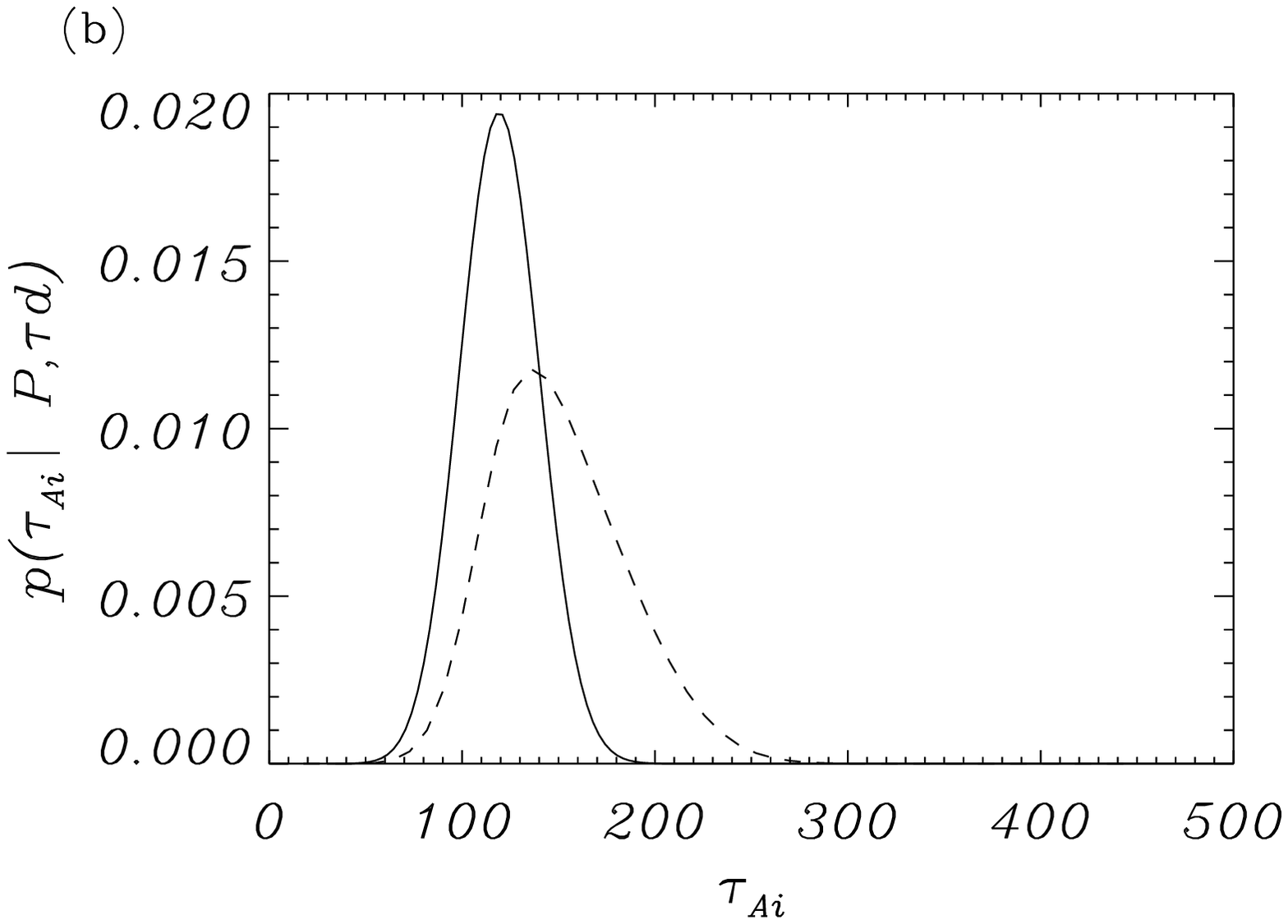}
   \includegraphics[width=0.3\textwidth]{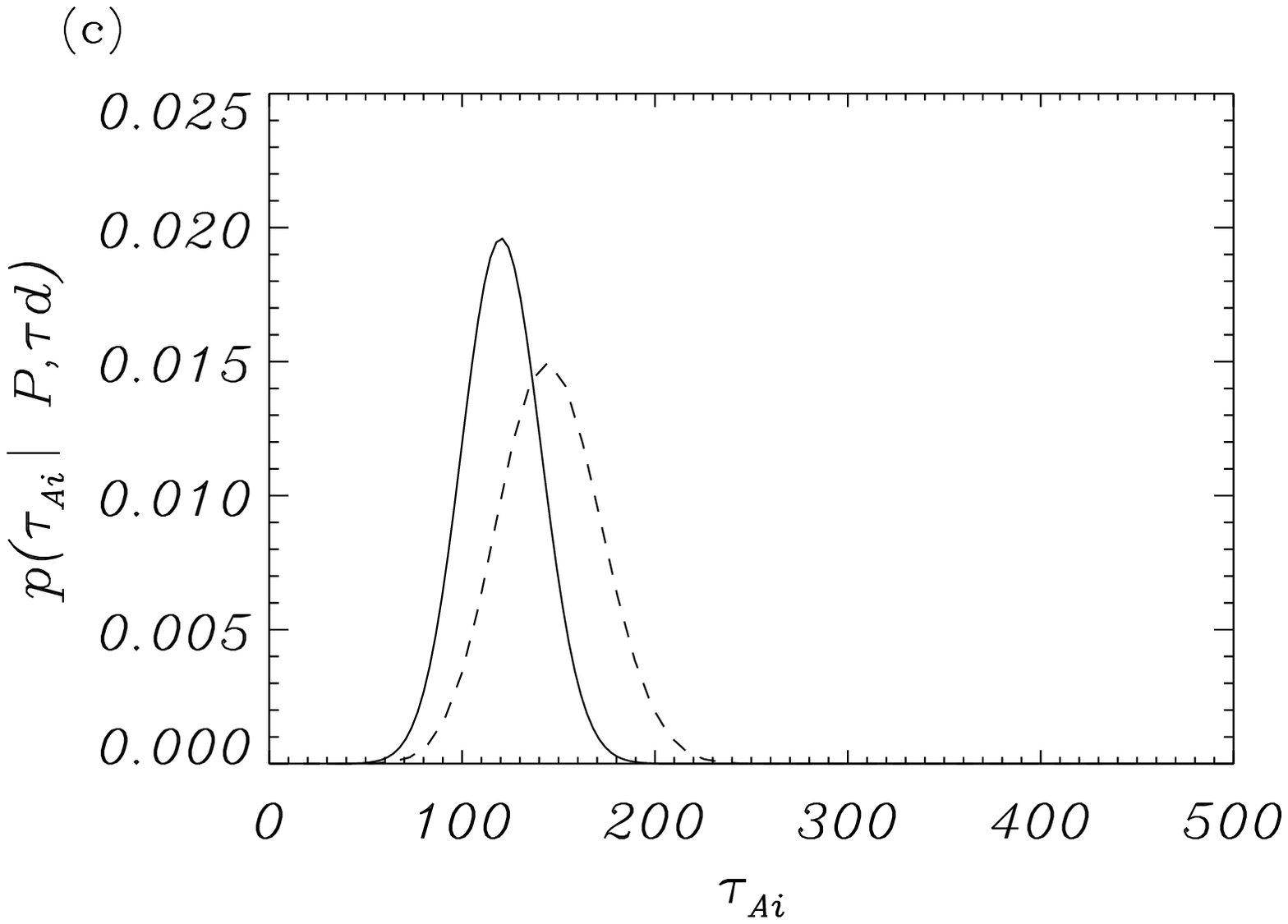}\\
   \includegraphics[width=0.3\textwidth]{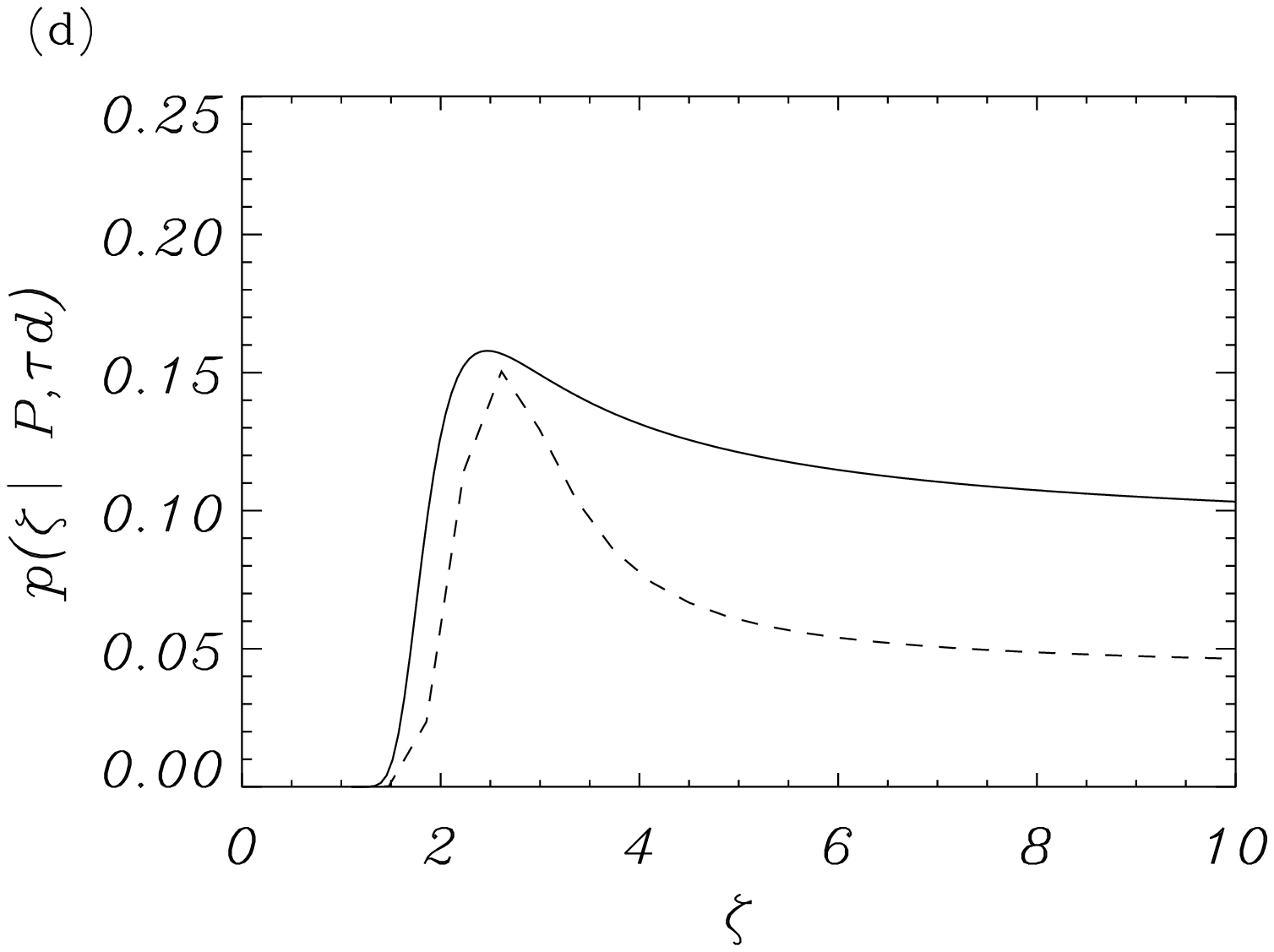}
   \includegraphics[width=0.3\textwidth]{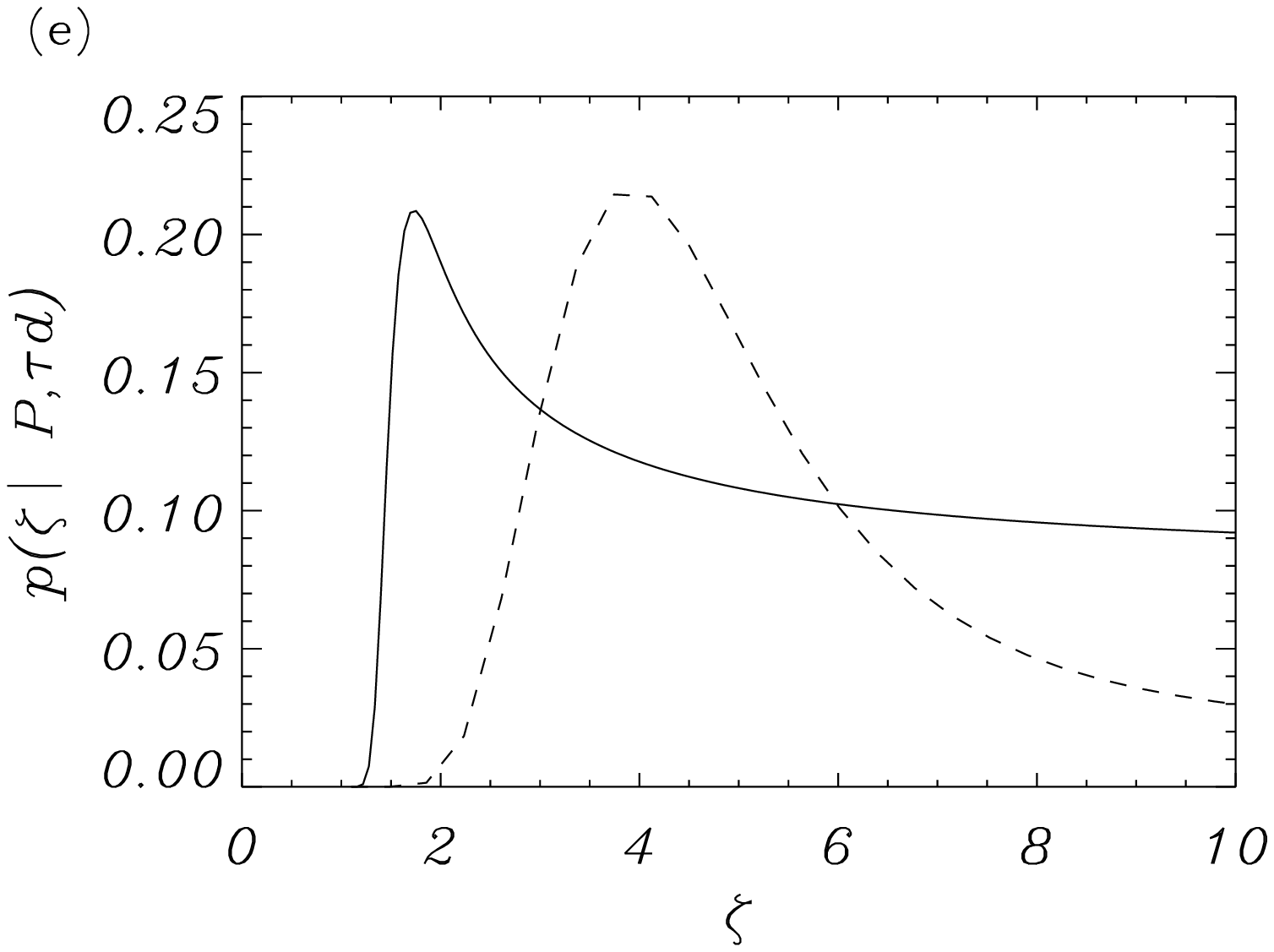}
   \includegraphics[width=0.3\textwidth]{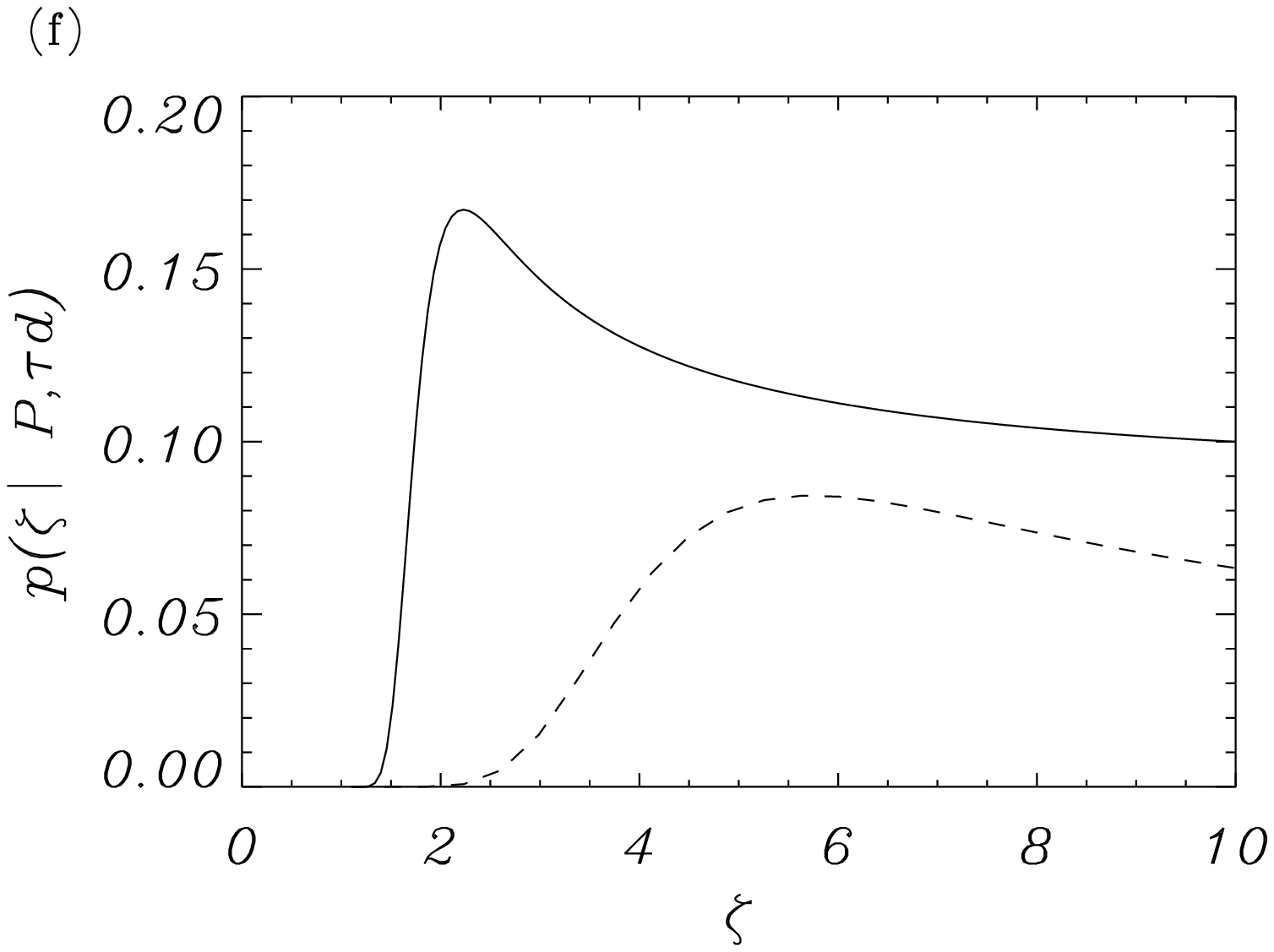}\\
   \includegraphics[width=0.3\textwidth]{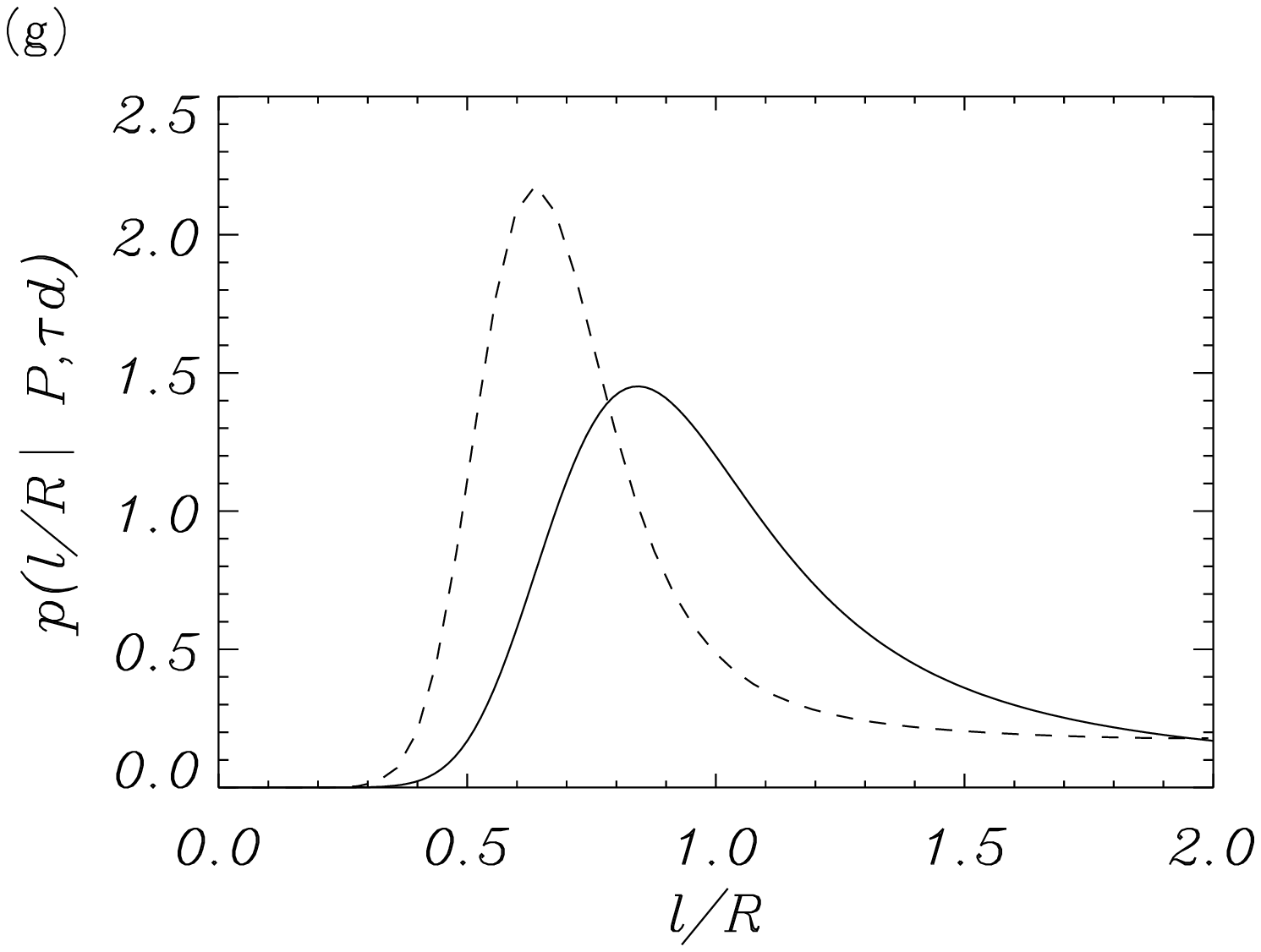}
   \includegraphics[width=0.3\textwidth]{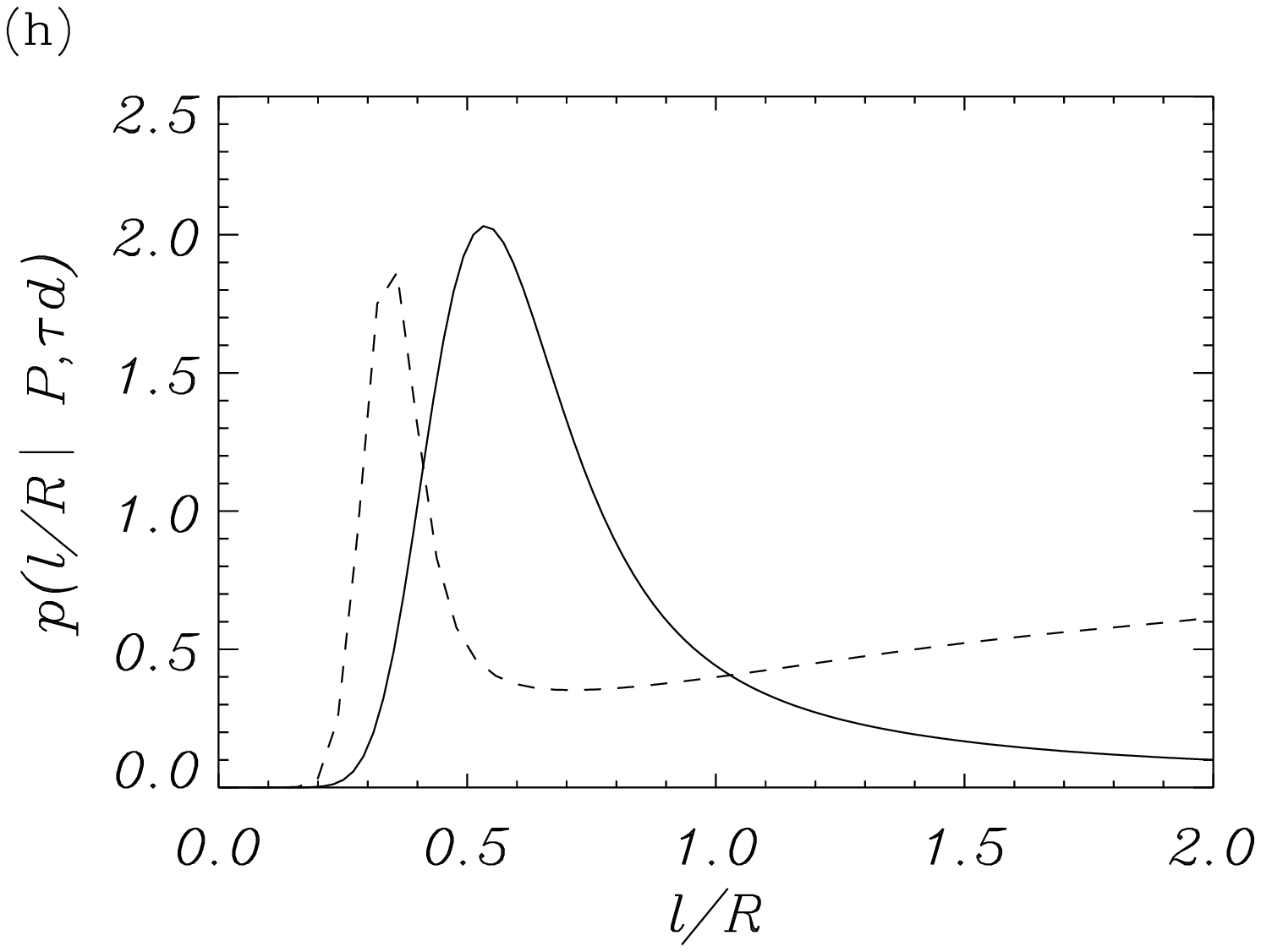}
   \includegraphics[width=0.3\textwidth]{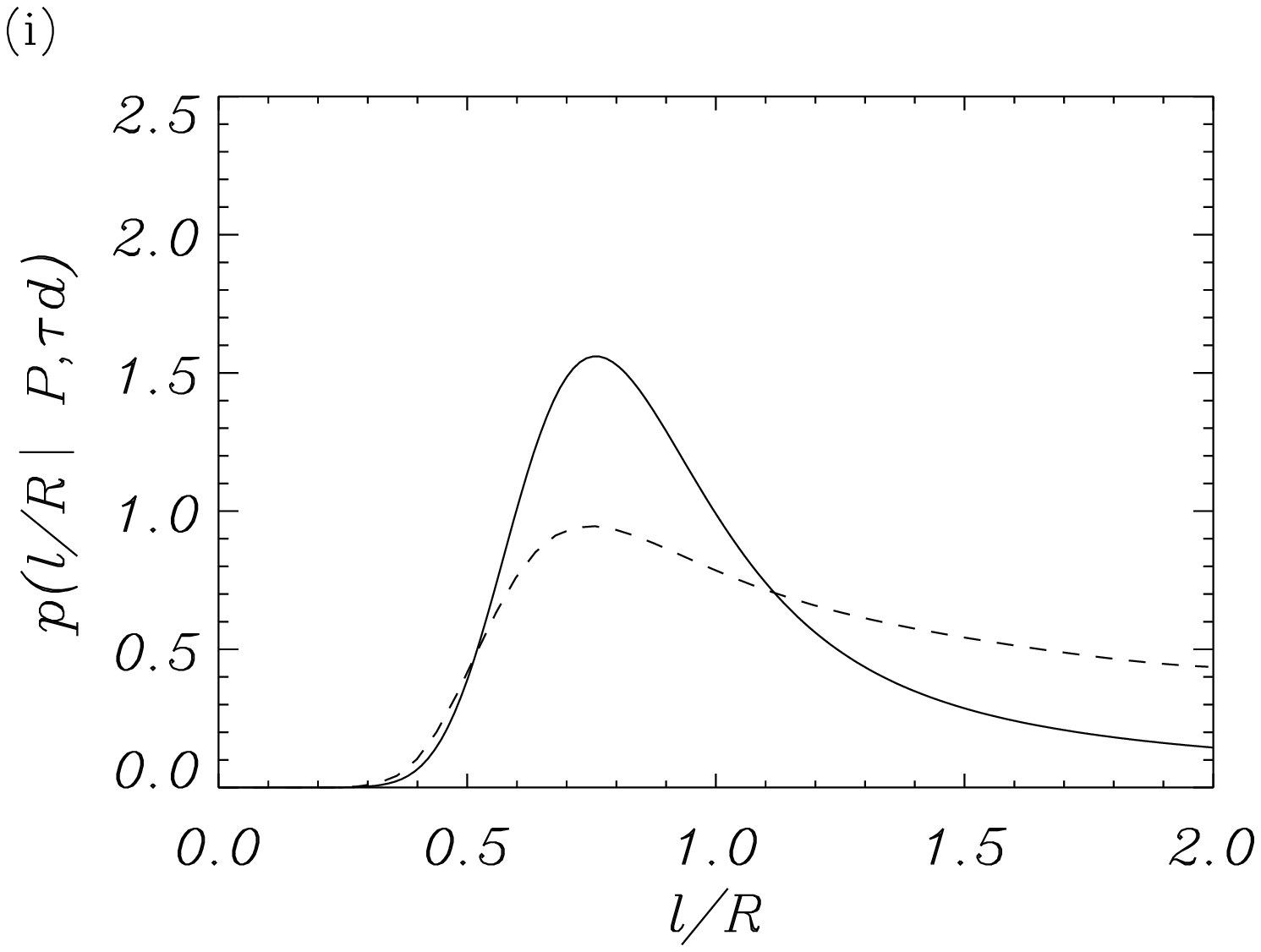}
   \caption{Marginal posteriors for the three parameters of interest ($\tau_{\rm Ai}$, $\zeta$, $l/R$) computed using expressions~(\ref{m1}) for the strong damping case ($P=185$ s; $\tau_{\rm d}=200$ s; $\sigma_{\rm P}=\sigma_{\rm \tau_{\rm d}}= 30$ s). Left column: sinusoidal density profile; middle: linear profile; right: parabolic profile. On each panel, solid lines represent the inference performed using the TTTB forward solutions, while dashed lines are for the inference performed using the numerical forward solutions. Summary data for the posteriors with the median and errors at the 68\% credible region are shown in Table~\ref{inf2table}.}
              \label{inf2f1}%
    \end{figure*}

\begin{figure*}
   \centering
   \includegraphics[width=0.3\textwidth]{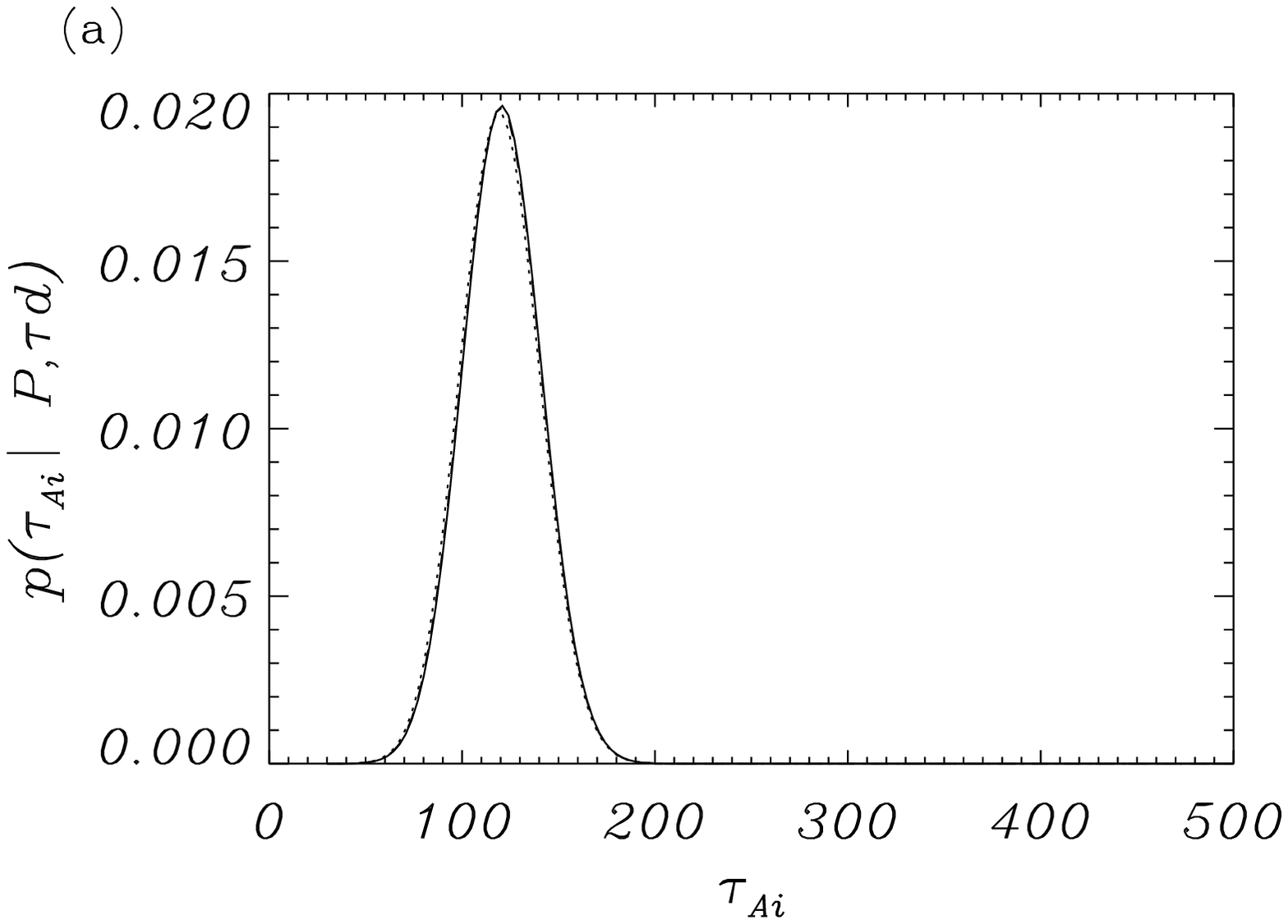}
   \includegraphics[width=0.3\textwidth]{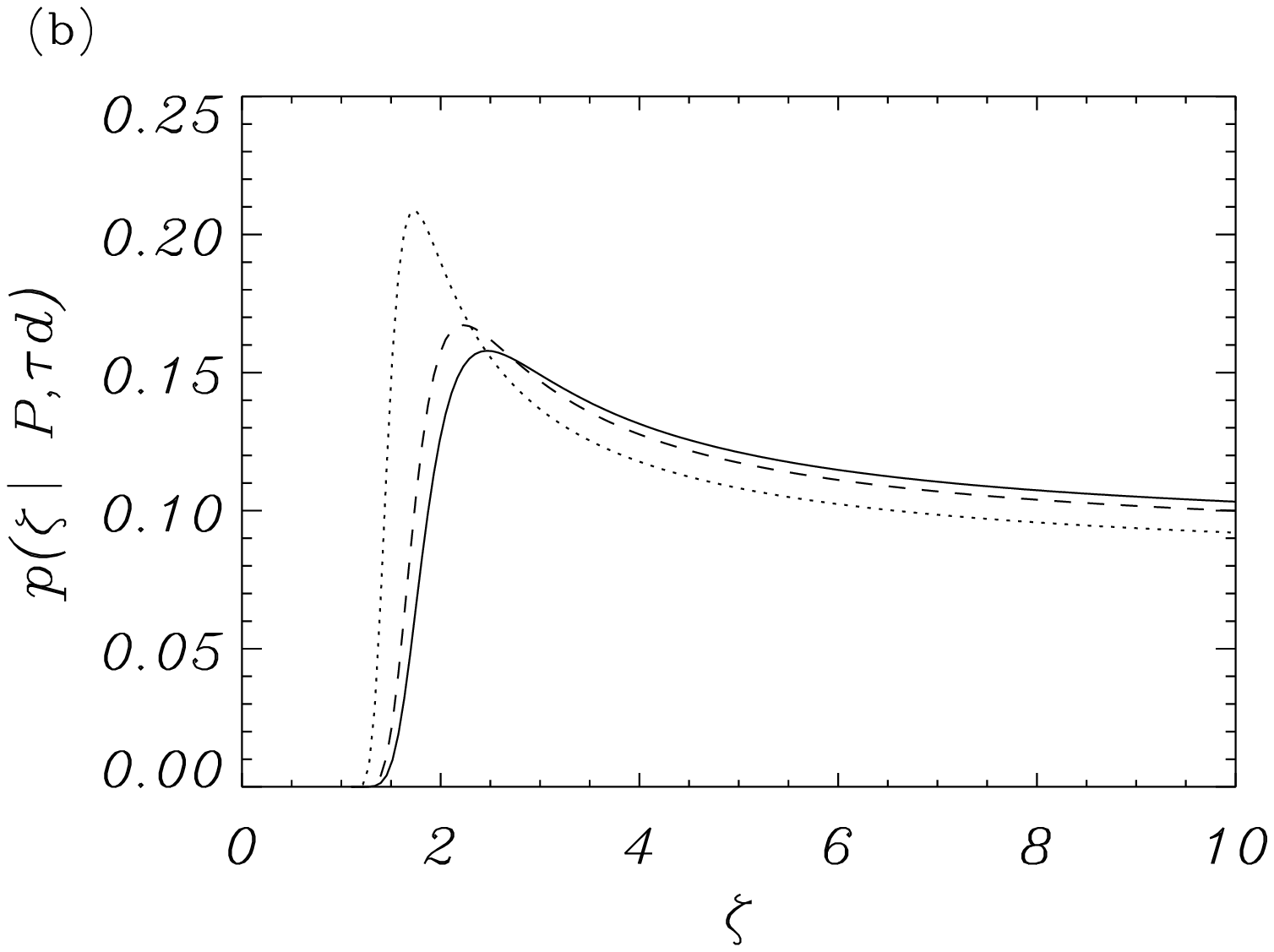}
   \includegraphics[width=0.3\textwidth]{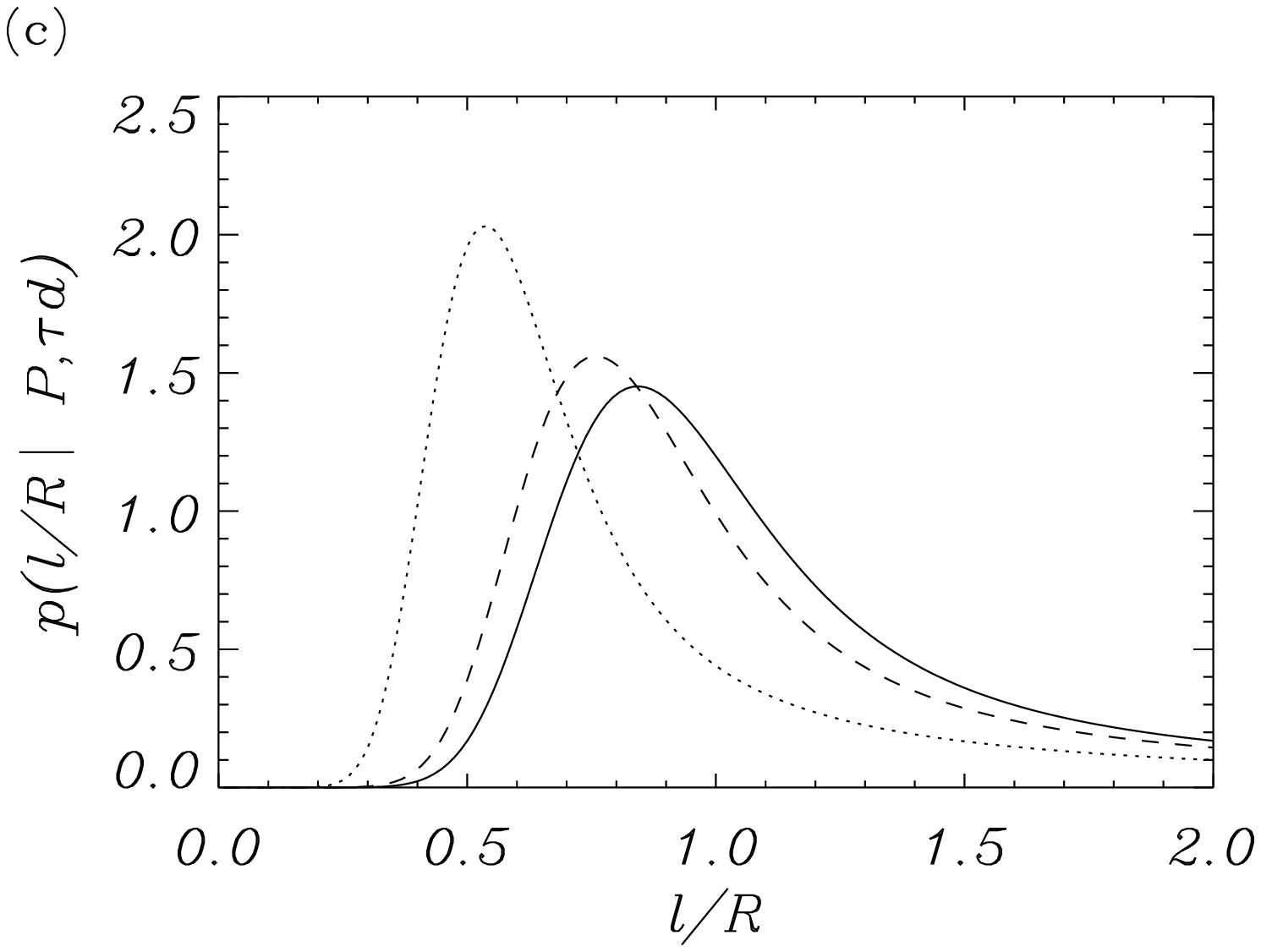}\\
   \includegraphics[width=0.3\textwidth]{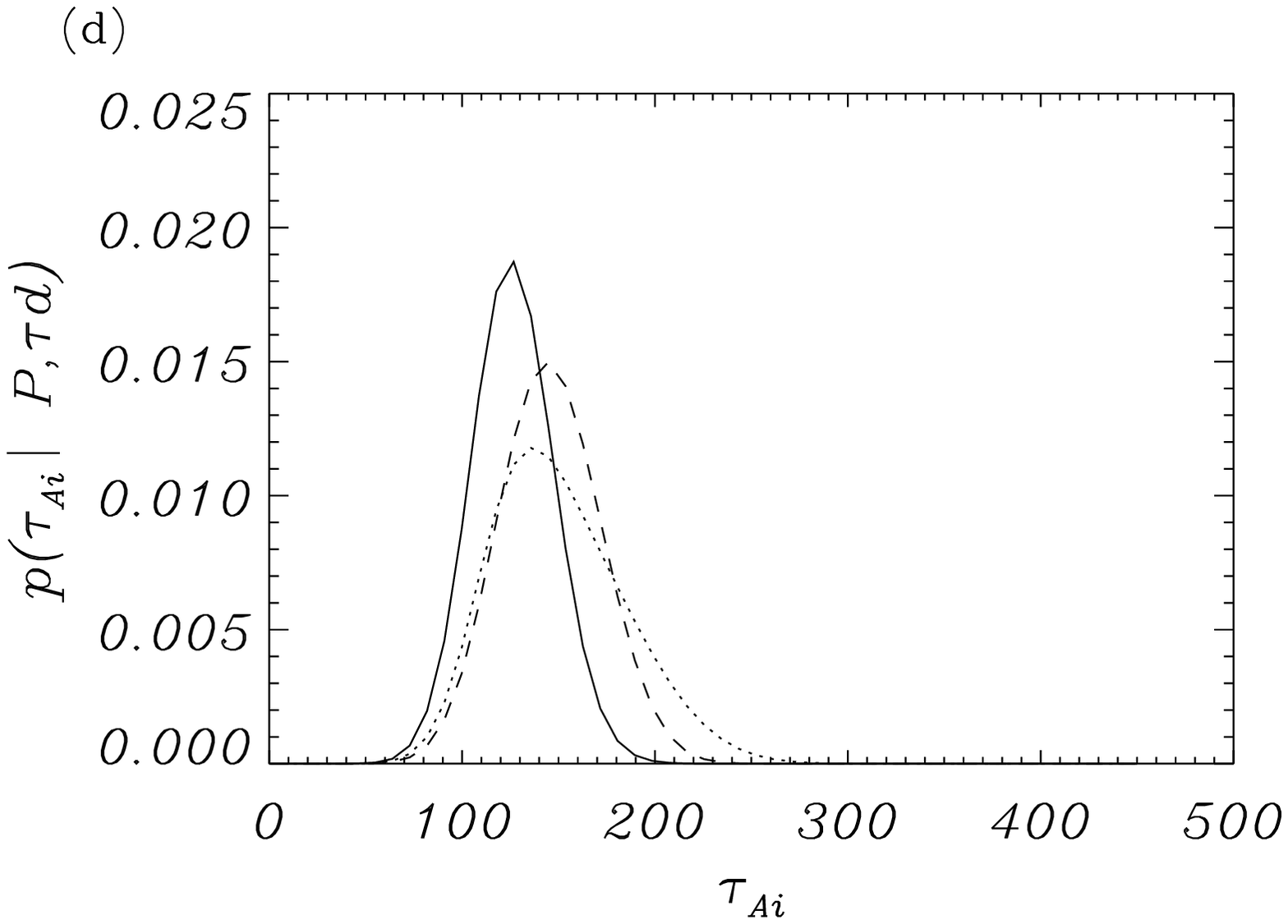}
   \includegraphics[width=0.3\textwidth]{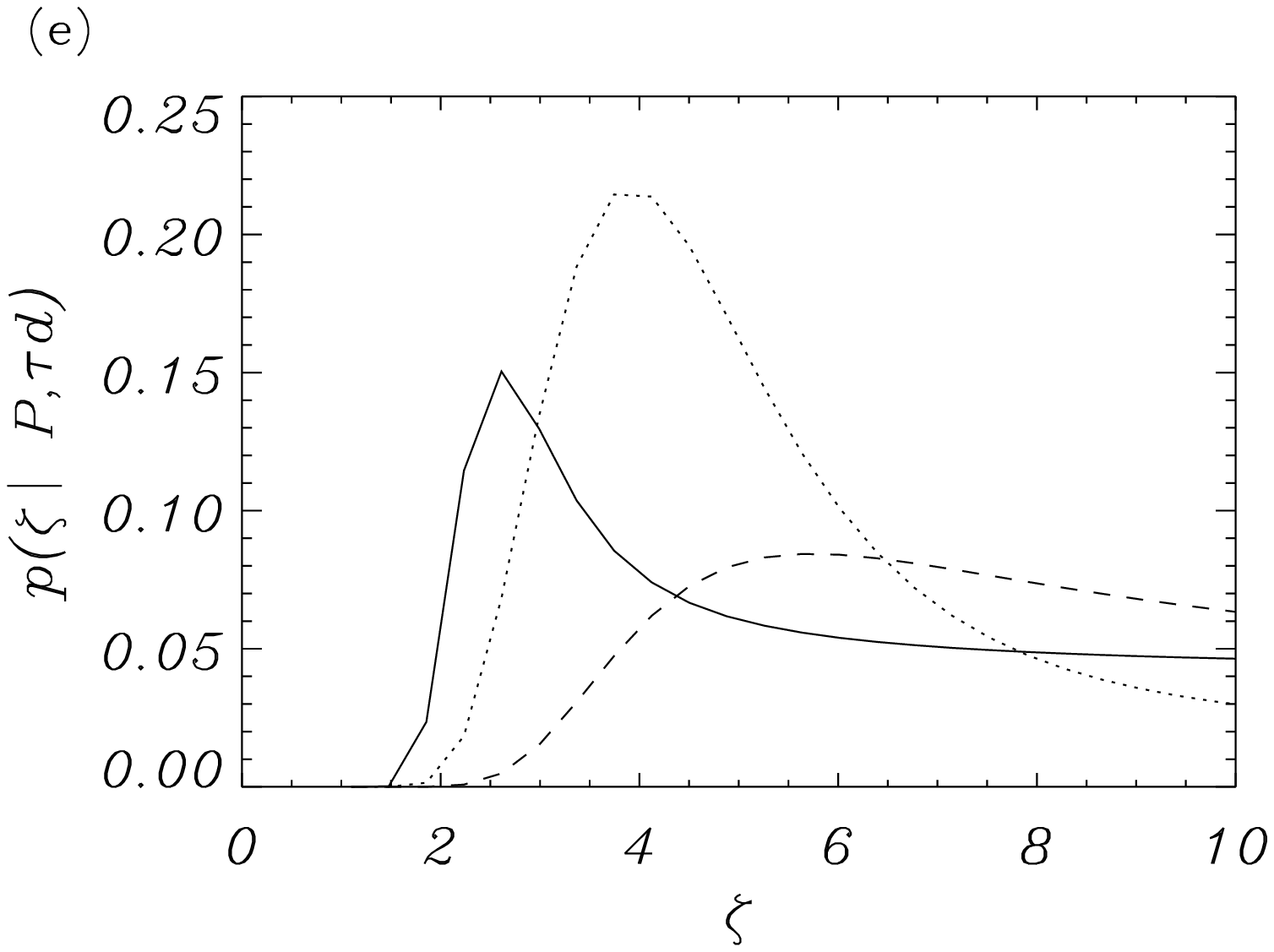}
   \includegraphics[width=0.3\textwidth]{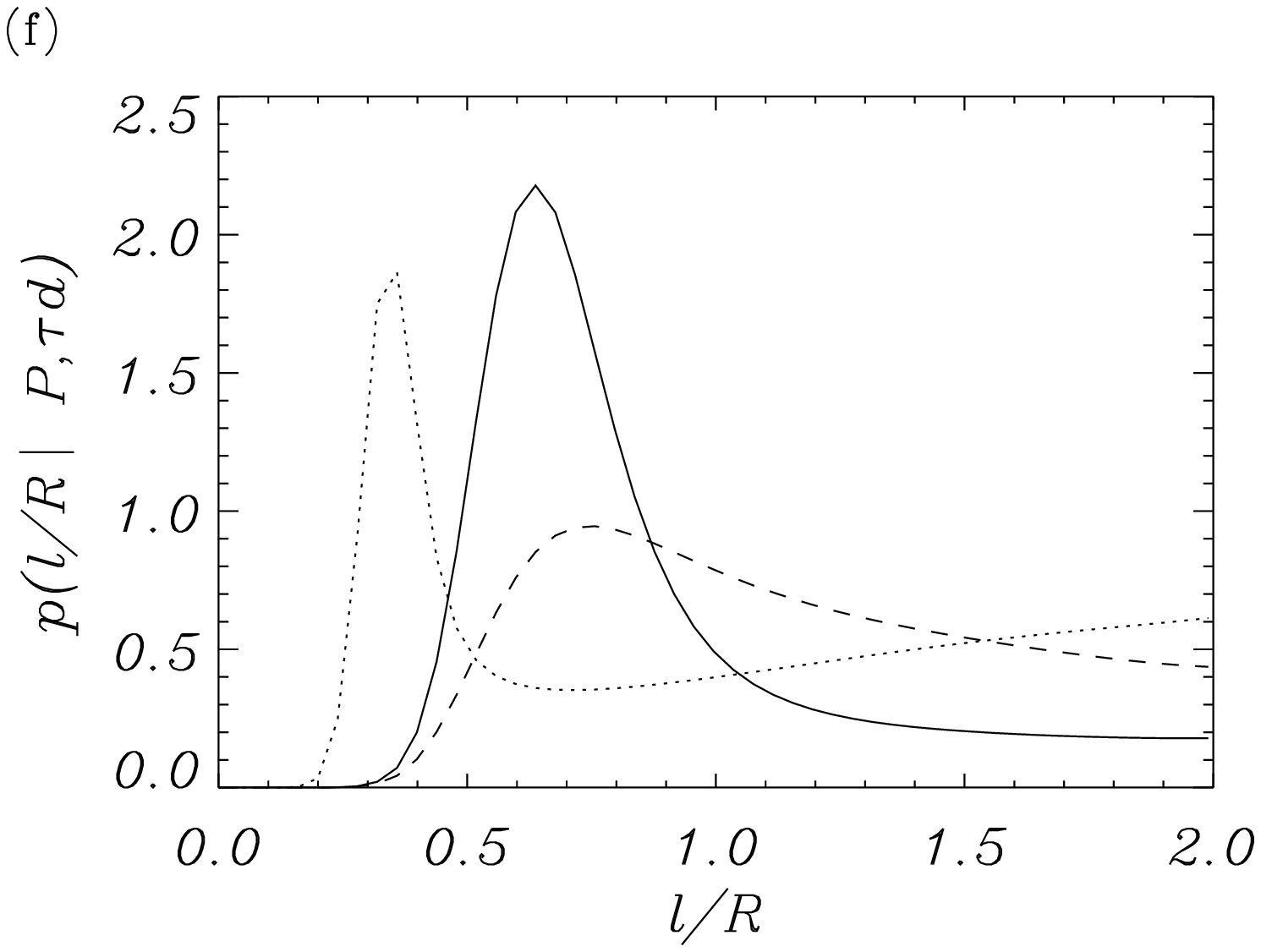}
   \caption{Marginal posteriors shown in Fig.~\ref{inf2f1} now displayed for direct comparison between results for the three alternative density models: sinusoidal (solid), linear (dotted), parabolic (dashed). Top row: TTTB results; bottom row: numerical results.}
              \label{inf2f2}%
    \end{figure*}

For the case with moderate damping, application of Bayes' rule, using the likelihood function in Eq.~(\ref{like}), uniform prior distributions, and the marginalization rules given in Eqs.~(\ref{m1}) leads to the marginal posterior density functions displayed in Figs.~\ref{inf1f1} and \ref{inf1f2}. In Fig.~\ref{inf1f1}, the results are arranged in such a way that a comparison between the inversion performed for TTTB results and fully numerical
results can be directly compared. Each of the three columns displays results for each of the three used density models. The Alfv\'en travel time can be properly constrained (see Figs.~\ref{inf1f1}a-c). There are no significant differences depending on  whether we employ the analytical approximations or the full numerical forward solutions.  The inversion suggests that low density contrast values are preferred over large contrast ones (see Figs.~\ref{inf1f1}d-f). However, the posteriors for density contrast display a rather long tail. This leads to this parameter being constrained with a large uncertainty.  More constrained marginal posteriors are obtained for the transverse inhomogeneity length scale (see Figs.~\ref{inf1f1}g-i). They show that short values of  $l/R$, below $l/R=0.5$, are more plausible than models with a fully nonuniform layer. The largest discrepancies between the TTTB approximation results and the fully numerical results are obtained for the inference of the density contrast made by employing the linear and parabolic profiles. For both the Alfv\'en travel time and the transverse inhomogeneity length scale, analytical and numerical forward solutions led to similar inversion results.

In Fig.~\ref{inf1f2}, the results are arranged in such a way that a comparison between the inversions for the three alternative density models can be directly made, with the top row showing TTTB results and the bottom row numerical results. For the inference of the Alfv\'en travel time, using the TTTB approximations for the forward problem, the inference is almost identical regardless of the employed density model (see Fig.~\ref{inf1f2}a). The density model only affects the forward solutions through the factor $F$ in Eq.~(\ref{damping}); hence the impact is minimal on the inference of the Alfv\'en travel time and only due to the coupling with Eq.~(\ref{period}) through the density contrast parameter. As the damping is moderate, slightly different marginal posteriors are obtained when fully numerical solutions for the forward problem are employed (see Fig.~\ref{inf1f2}d).  For the density contrast, outside the TTTB approximation, changing the density model slightly affects the shape and probability density functions for different density models (see Fig.~\ref{inf1f2}e). The largest differences in the computed marginal posteriors are obtained for the transverse inhomogeneity length scale (see Figs.~\ref{inf1f2}c and f). For this parameter,  differences in the marginal posteriors depending on the density model already appear when the analytical TTTB approximations are employed.

Table~\ref{inf1table} summarizes the marginal posterior density functions obtained from the inference for the three density models and the analytical and numerical forward solutions. For each parameter and case, the median of the marginal probability density function and errors at the 68\% credible interval are given. Despite the differences in the results discussed above, the results indicate that when the posteriors are summarized, almost the same results are obtained, regardless of the method employed to solve the forward problem and the adopted density model. We can therefore conclude that, for moderate damping, the use of  either TTTB or numerical forward solutions leads to very similar inversion results and that the impact of the assumed density model is small.

The analysis is now repeated for the case with strong damping. The corresponding marginal posteriors are displayed in Figs.~\ref{inf2f1} and \ref{inf2f2}.  In Fig.~\ref{inf2f1}, we compare the results between the inversion performed using TTTB and numerical forward solutions. The inference for the Alfv\'en travel time is similar in both cases, when a sinusoidal density model is employed (see Fig.~\ref{inf1f1}a), but differs when either the linear or the parabolic density models are employed (see Figs.~\ref{inf2f1}b and c). 
In these two cases, the inference with the TTTB forward solutions leads to probability density functions that are shifted toward smaller values of the parameter. The inference for the density contrast leads to similar results between the TTTB and numerical cases, when the sinusoidal density model is used (see Fig.~\ref{inf2f1}d), while it shows more marked differences when either of the other two profiles is adopted (see Figs.~\ref{inf2f1}e and f). The largest differences between the inference applied to TTTB and numerical forward solutions is obtained for the transverse inhomogeneity length scale (see Figs.~\ref{inf2f1}g-i), but in this case they are less important in the case of the parabolic density model.

In Fig.~\ref{inf2f2}, the results are arranged in such a way that a comparison between the inversions for the three alternative density models can be directly made, with the top row showing TTTB results and the bottom row numerical results. For the inference of the Alfv\'en travel time and using the TTTB forward solutions the same probability density function is obtained, regardless of the employed density model (see Fig.~\ref{inf2f2}a).  Slightly  shifted marginal posteriors are obtained, when numerical forward solutions are employed (see Fig.~\ref{inf2f2}d). We note here that the period is also a function of the thickness of the layer when no analytical approximations are used. The shifted marginal posteriors for the Alfv\'en travel time indicate that the impact of the density model is measurable in the inference of this parameter. For the density contrast, the effect of the adopted density model is also unimportant under the TTTB approximations (see Fig.~\ref{inf2f2}b), but more significant differences are visible between the marginal posteriors obtained from the numerical forward solutions (see Fig.~\ref{inf2f2}e). If we compare the inversions for the density contrast performed from numerical solutions for the moderate and strong damping cases, Figs.~\ref{inf1f2}e and \ref{inf2f2}e, we can see that the impact of the strength of the damping on the inversion of this parameter is important. Likewise, the probability density functions for the transverse inhomogeneity length scale are affected by the adopted density model, to a moderate extent under the TTTB approximations (see Fig.~\ref{inf2f2}c), but significantly for numerical forward solutions (see Fig.~\ref{inf2f2}f).

\begin{deluxetable}{cccccccc}
\tablecolumns{8} 
\tablewidth{0pc} 
\tablecaption{Summary of inference results for the moderate damping case for the three alternative models by employing the TTTB and numerical forward solutions.
$P=272$ s; $\tau_{\rm d}=849$ s; $\sigma_{\rm P}=\sigma_{\rm \tau_{\rm d}}= 30$ s. \label{inf1table} } 
\tablehead{ 
\colhead{$\theta_{\rm i}$}    &  \multicolumn{3}{c}{TTTB approximation} &   \colhead{}   & \multicolumn{3}{c}{Numerical} \\ 
\cline{1-8} \\ 
\colhead{} & \colhead{M$^{\rm S}$}   & \colhead{M$^{\rm L}$}    & \colhead{M$^{\rm P}$} & 
\colhead{}    & \colhead{M$^{\rm S}$}   & \colhead{M$^{\rm L}$}    & \colhead{M$^{\rm P}$}}
\startdata 
$\tau_{\rm Ai}$  &  171$^{+22}_{-24}$ &169$^{+24}_{-24}$&171$^{+21}_{-24}$ &&177$^{+24}_{-20}$  &169$^{+29}_{-21}$ &181$^{+21}_{-18}$\\\\
$\zeta$  &4.4$^{+3.7}_{-2.6}$ &4.1$^{+3.9}_{-2.2}$ &4.3$^{+3.8}_{-2.6}$    &&7.9$^{+8.2}_{-6.1}$  &4.6$^{+10.8}_{-2.9}$&6.7$^{+9.0}_{-4.8}$\\\\
$l/R$ &0.3$^{+0.4}_{-0.1}$ &0.2$^{+0.4}_{-0.1}$  &0.3$^{+0.4}_{-0.1}$   &&0.3$^{+0.5}_{-0.1}$  & 0.2$^{+0.8}_{-0.1}$&0.3$^{+0.4}_{-0.1}$\\
\enddata 
\end{deluxetable}

\begin{deluxetable}{cccccccc}
\tablecolumns{8} 
\tablewidth{0pc} 
\tablecaption{Summary of inference results for the strong damping case for the three alternative models by employing the TTTB and numerical forward solutions.
$P=185$ s; $\tau_{\rm d}=200$ s; $\sigma_{\rm P}=\sigma_{\rm \tau_{\rm d}}= 30$ s. \label{inf2table} } 
\tablehead{ 
\colhead{$\theta_{\rm i}$}    &  \multicolumn{3}{c}{TTTB approximation} &   \colhead{}   & \multicolumn{3}{c}{Numerical} \\ 
\cline{1-8}  \\ 
\colhead{} & \colhead{M$^{\rm S}$}   & \colhead{M$^{\rm L}$}    & \colhead{M$^{\rm P}$} & 
\colhead{}    & \colhead{M$^{\rm S}$}   & \colhead{M$^{\rm L}$}    & \colhead{M$^{\rm P}$}}
\startdata 
$\tau_{\rm Ai}$  &  121$^{+20}_{-20}$ &120$^{+20}_{-20}$&121$^{+19}_{-20}$ &&126$^{+23}_{-21}$  &147$^{+39}_{-30}$ &145$^{+27}_{-26}$\\\\
$\zeta$  &5.4$^{+3.1}_{-2.6}$ &4.9$^{+3.4}_{-2.7}$ &5.3$^{+3.2}_{-2.6}$    &&8.8$^{+7.5}_{-5.6}$  &5.4$^{+6.8}_{-1.8}$&10.1$^{+6.3}_{-4.6}$\\\\
$l/R$ &1.0$^{+0.4}_{-0.3}$ &0.7$^{+0.5}_{-0.2}$  &0.9$^{+0.4}_{-0.2}$   &&0.7$^{+0.5}_{-0.2}$  & 1.0$^{+0.5}_{-0.7}$&1.0$^{+0.6}_{-0.4}$\\
\enddata 
\end{deluxetable}

Summary data for the inferences performed for the strong damping case for the three density models and the analytical and numerical forward solutions are presented in Table~\ref{inf2table}. For the TTTB results, we obtain almost the same results when one focuses on the median of the probability density function and errors given at the 68\% credible interval. The situation is different for the case in which numerical forward solutions are used. Here, rather distinct results are obtained depending on the density model that is employed, although they are masked to a great extent when one considers the upper and lower limits of the credible intervals. We can therefore conclude that, for observations with strong damping, the inference results will significantly differ depending on whether TTTB or numerical forward solutions are used. They will also lead to different probability density functions and parameter estimates depending on the adopted density model. The differences will be more important in the case of the density contrast and the transverse inhomogeneity length scale, the two parameters that define the cross-field density variation.

Besides the particular properties of the marginal posteriors discussed above, we stress a crucial difference between our results and those obtained from the classic inversion. In the classic inversion, as emphasized by \cite{arregui07a}, \cite{goossens08a}, and \cite{soler14a}, any point along the inversion curve is equally compatible with the observations. Furthermore, the uncertainty on the measured wave properties is not taken into account. This is not true under the Bayesian formalism, where the solution is given in terms of a degree of plausibility for the different values each parameter can take on.  Given that observed data values can be obtained by different combinations of parameters and, considering uncertainty as well, some combinations are more probable than others in causing the observed values of period and damping time.
The calculation related to how many times alternative parameter combinations produce a given data realization is contained in the likelihood function. When combined with the prior, the resulting posterior gives the distribution for the grade of belief on a given parameter value conditional on observed data. 

\section{Model comparison}\label{comparison}

The second level of Bayesian inference is model comparison, which enables us to compare the plausibility of the three alternative density models to explain observed data in period and damping time. Recall that we use the names $M^i$, with $i=S,L,P$, corresponding to the sinusoidal, linear, and parabolic density profiles used in this study.  For each assumed profile, the marginal likelihood,
\begin{equation}\label{marginal}
p(d|M^i)=\int p(d,\mbox{\boldmath$\theta$}|M^i)d\mbox{\boldmath$\theta$}=\int p(d|\mbox{\boldmath$\theta$},M^i)p(\mbox{\boldmath$\theta$}|M^i)d\mbox{\boldmath$\theta$},  
\end{equation}
is a quantity that provides us with the probability of the observed data $d$, given that the model $M^i$ is true. It tells us how well the observed data are predicted by the particular model $M^i$. When two competing alternative models, $M^i$ and $M^j$, are compared one by one to explain a given set of observations, the plausibility of one model over the alternative is given by their posterior ratio. This can be computed by first rewriting 
Bayes' theorem in terms of  the probability of a given model, conditional on the observed data as

\begin{equation}
p(M|d)\propto P(d|M)p(M)
\end{equation}
and applying it to two competing models to compute their posterior ratio as

\begin{equation}
\frac{p(M^i|d)}{p(M^j|d)}=\frac{p(d|M^i)}{p(d|M^j)}\frac{p(M^i)}{p(M^j)}.
\end{equation}
Considering that both models are equally probable a priori, $p(M^i)=p(M^j)=1/2$, the posterior ratio calculation reduces to the evaluation of the ratio of marginal likelihoods for both models. The  comparison then consists of the computation of the so-called Bayes factor,

\begin{equation}\label{bf}
BF^{ij}=\frac{p(d|M^i)}{p(d|M^j)}.
\end{equation}
The magnitude of the Bayes factor is a measure of the relative plausibility between models. To translate the obtained magnitudes into levels of evidence an empirical scale is used \citep[see e.g., ][]{jeffreys61}. We use the empirical table by \cite{kass95}, which assigns evidence for model $M^i$ against model $M^j$ that is minimal evidence (ME) to values of $2\log(BF^{ij})$ in between 0 and 2, positive evidence (PE) to values in between 2 and 6,  strong evidence (SE) to values in between 6 and 10, and very strong evidence (VSE) to values of $2\log(BF^{ij})$ larger than 10.

\begin{figure}
\epsscale{1.20}
\plotone{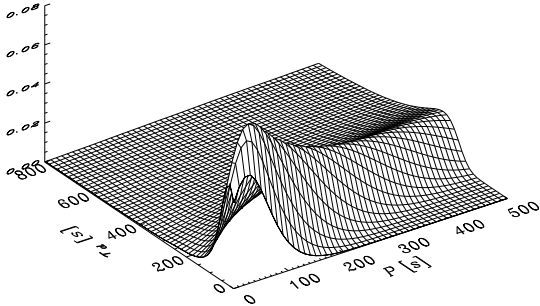}
        \caption{Two-dimensional surface plot representing the marginal likelihood for the sinusoidal density model, $p(d|M^S)$, in the observable parameter space ($P$, $\tau_{\rm d}$). The magnitude at each point is a measure of the degree of plausibility of the model for the observed data. 
        The uncertainty on period and damping time is $\sigma_{\rm P}=\sigma_{\rm \tau_{\rm d}}= 30$ s. }
              \label{surfmarginal}
    \end{figure}
    
We have first computed the marginal likelihood for the three alternative models under consideration, by applying  Eq.~(\ref{marginal}) to each of them. As an example, Figure~\ref{surfmarginal} shows the result for the sinusoidal model, $M^S$. A given pair of observed period and damping time would correspond to a point over the displayed two-dimensional surface. The magnitude of the marginal likelihood at such a point is a measure of the goodness of this particular model in explaining those observed period and damping time values. The calculation already takes into account the uncertainty on the measured wave properties. We see that certain combinations of period and damping time have a larger marginal likelihood than others. This means that when performing the integral in Eq.~(\ref{marginal}), summing up all the possible parameter combinations that could produce the observed data, those combinations have a larger plausibility. The calculation requires the use of that particular model, assumed to be true, the selection of particular values for the uncertainty on the data, and the integral of the product of the likelihood function with the prior over the assumed ranges for all parameters; hence, all the available information is used in a consistent way. 
A similar calculation was carried out for the remaining linear and parabolic density models, $M^L$ and $M^P$.  Once this is done, Bayes factors can be computed considering ratios as given by Eq.~(\ref{bf}) and the relative plausibility of one model against another can be assessed.

\begin{figure*}
   \centering
   \includegraphics[width=0.3\textwidth]{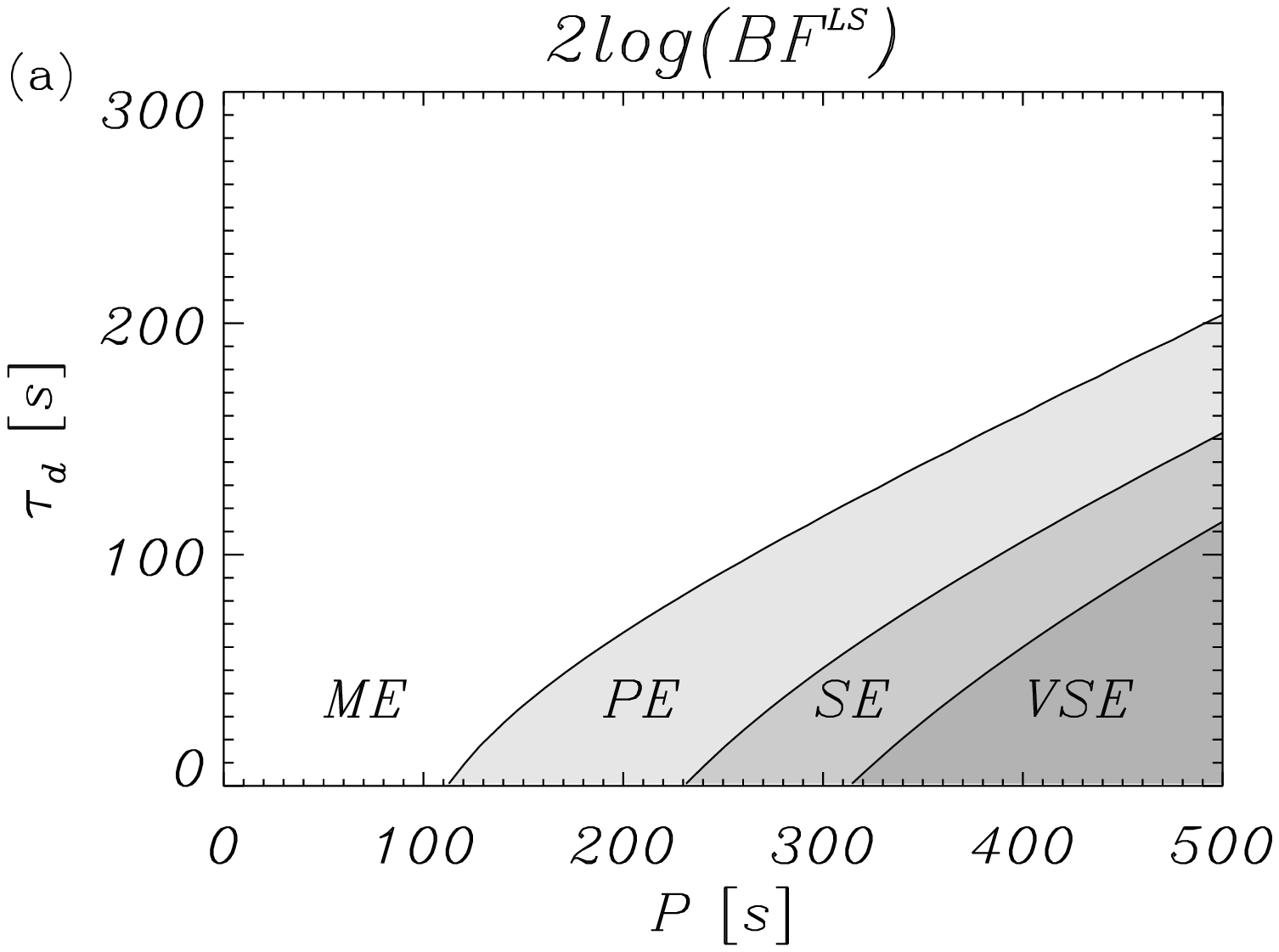}
   \includegraphics[width=0.3\textwidth]{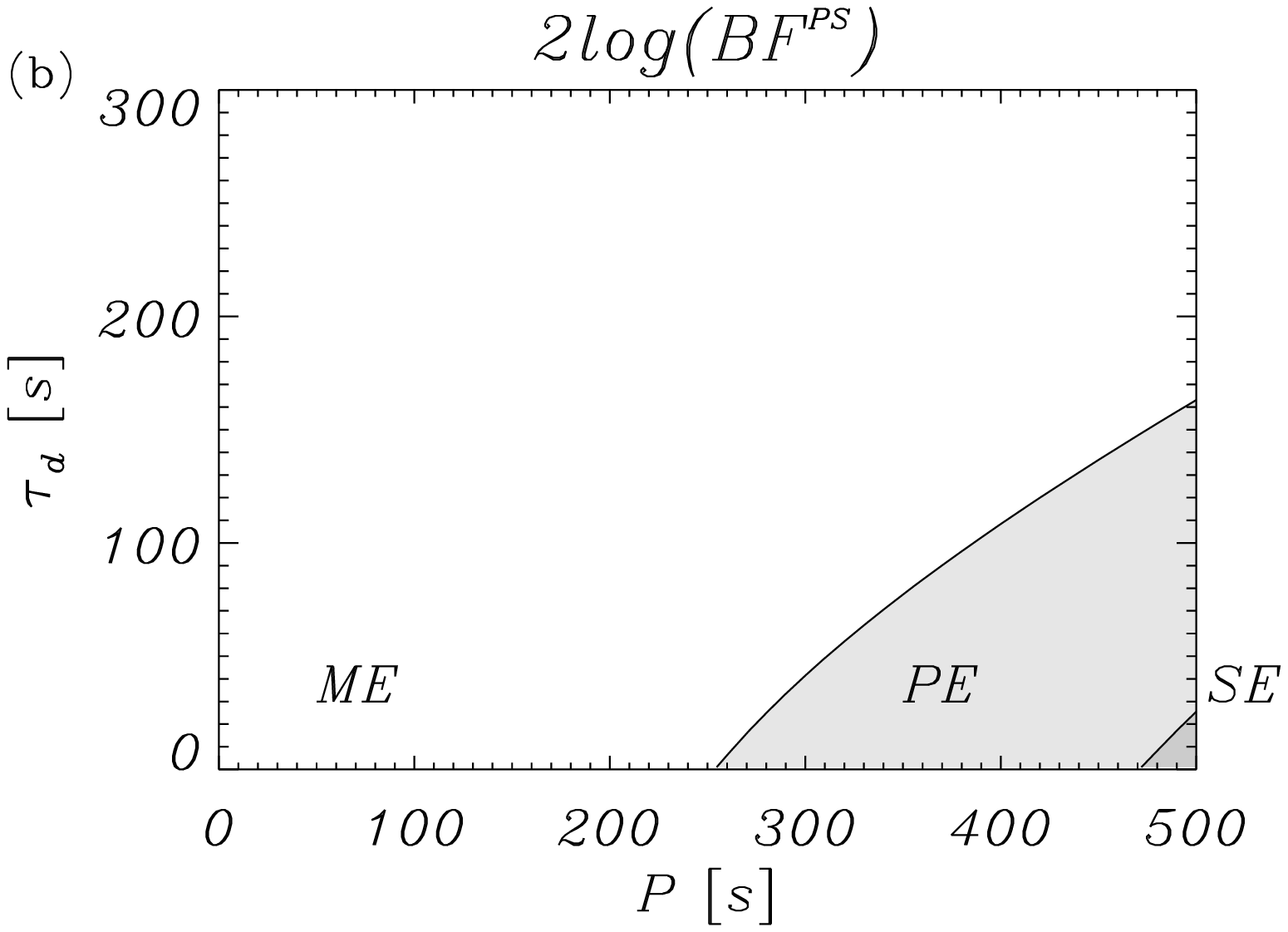}
   \includegraphics[width=0.3\textwidth]{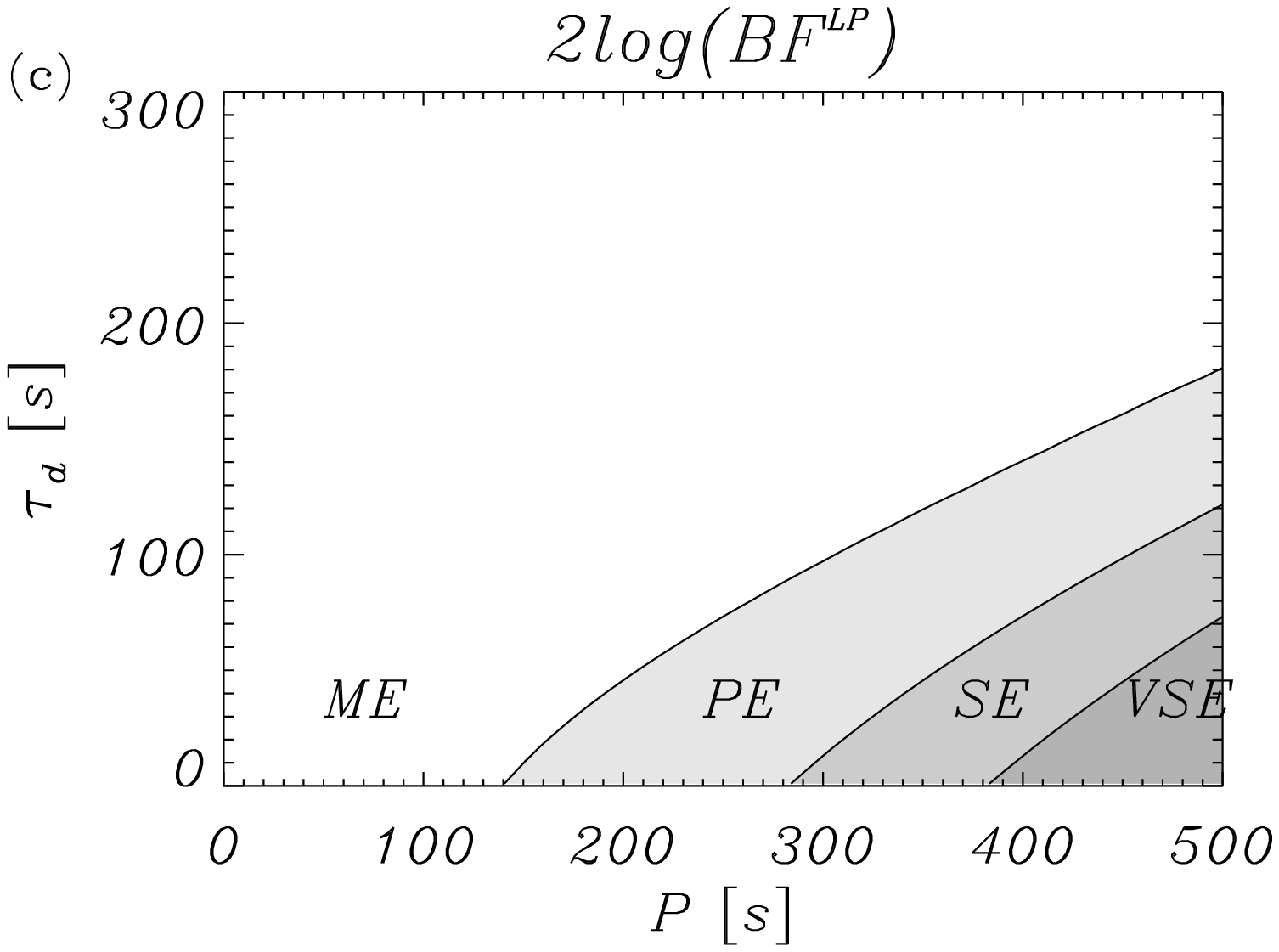}\\
   \includegraphics[width=0.3\textwidth]{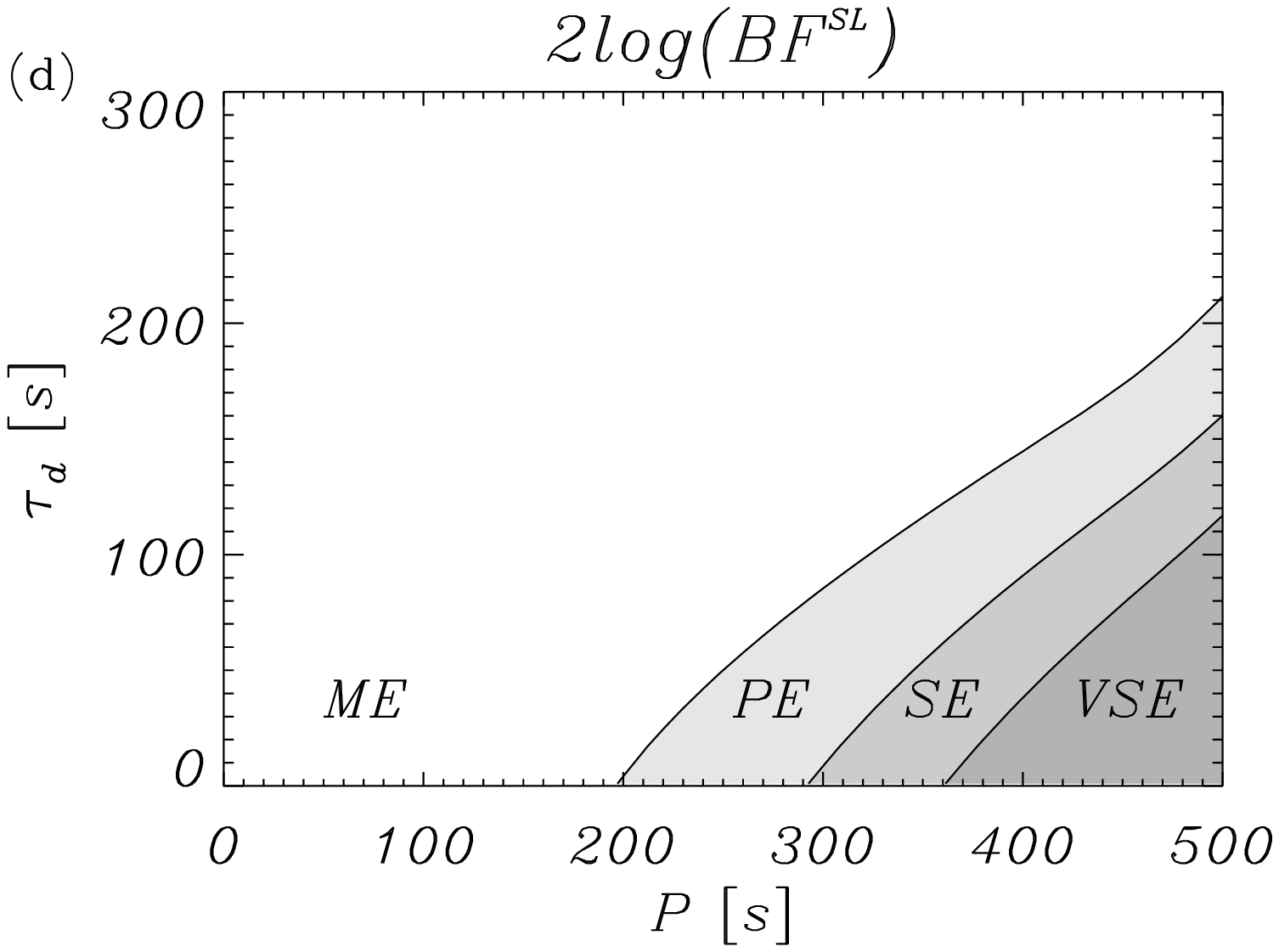}
   \includegraphics[width=0.3\textwidth]{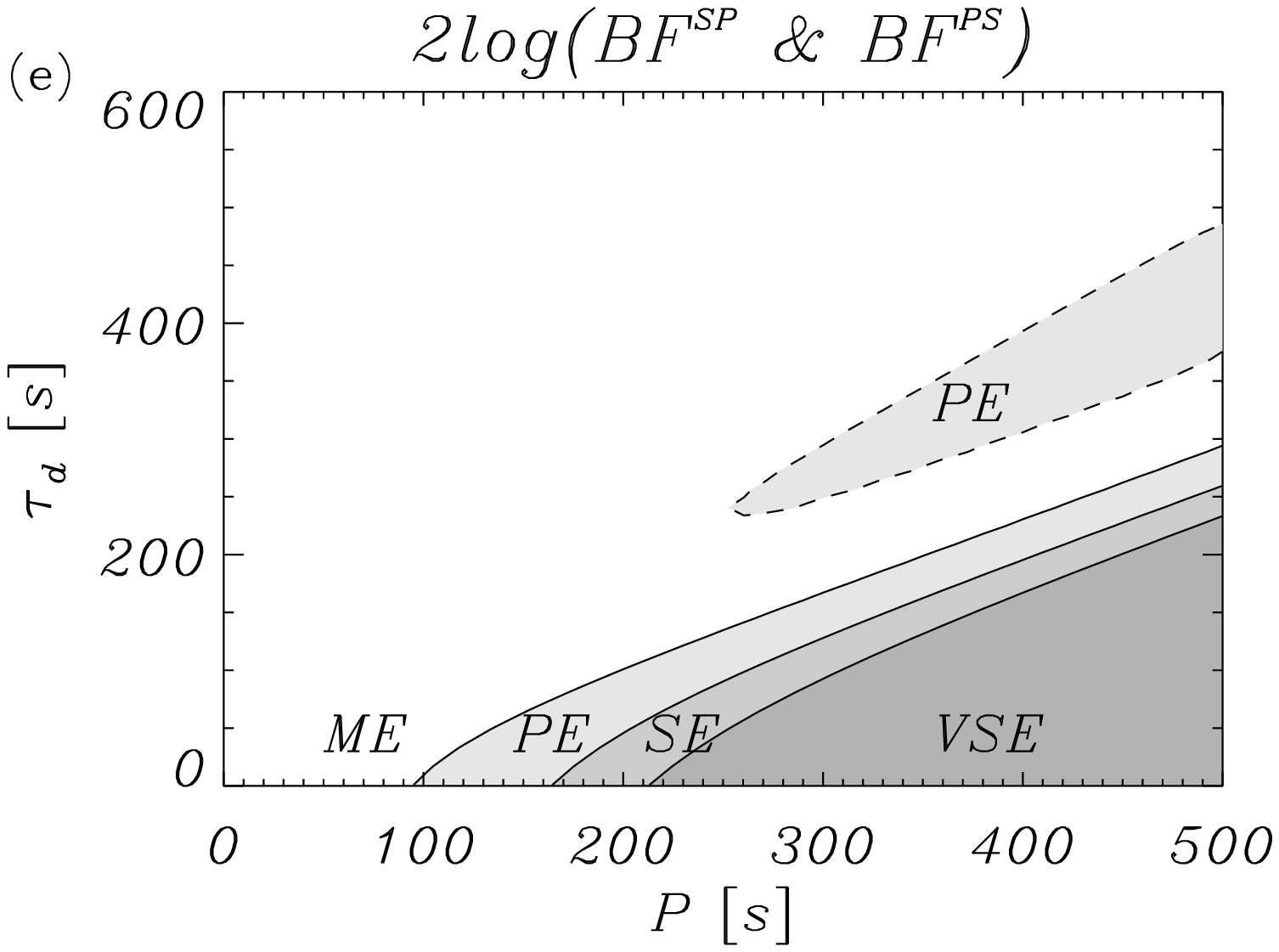}
   \includegraphics[width=0.3\textwidth]{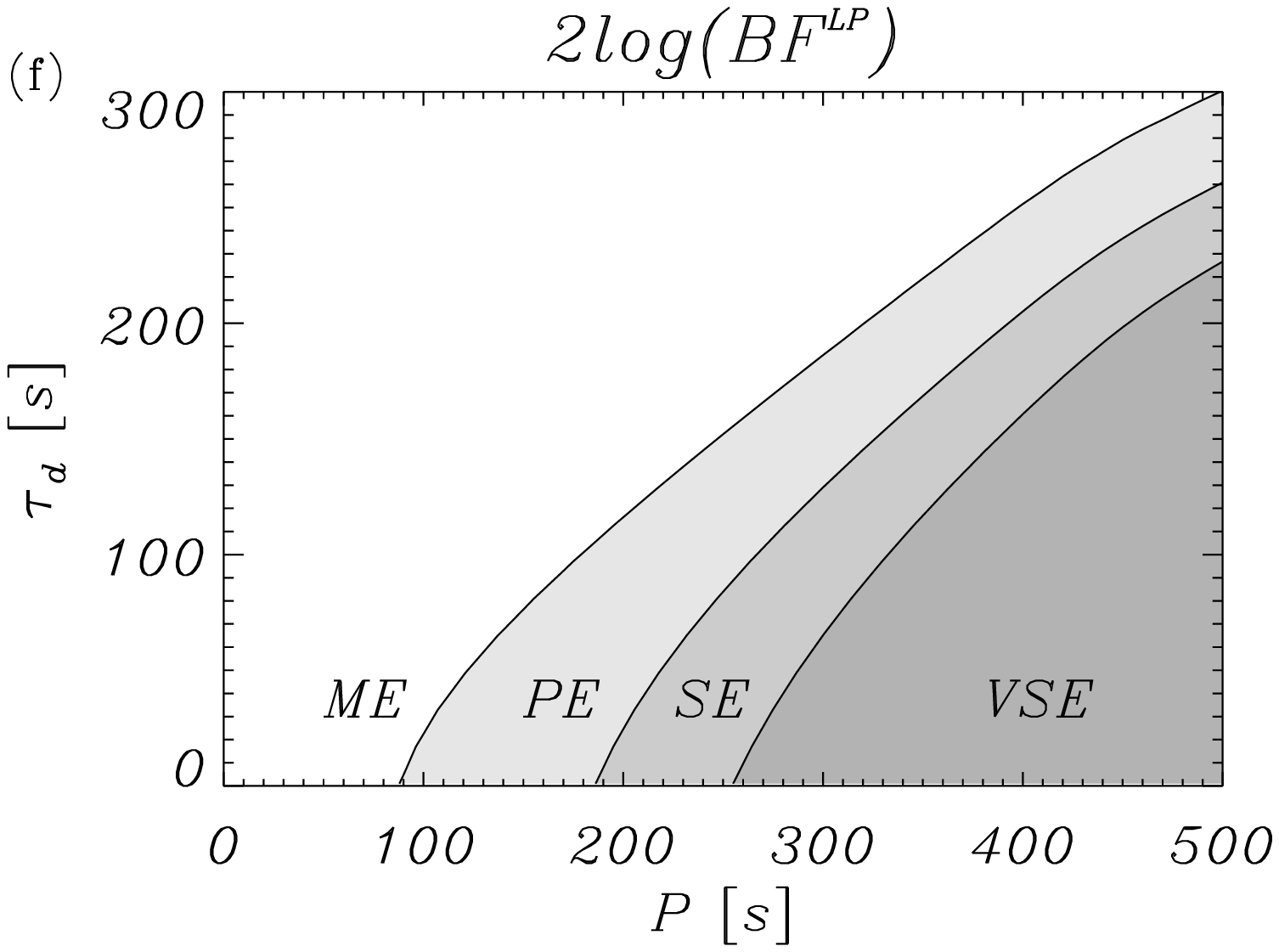}  
   \caption{Contour plots of the regions with different levels of evidence according to Bayes factor values from the comparison between alternative density models in the two-dimensional observational parameter space ($P$, $\tau_{\rm d}$). ME: minimal evidence; PE: positive evidence; SE: strong evidence; VSE: very strong evidence. Top row panels correspond to the analysis using TTTB forward solutions. Bottom row panels correspond to the analysis using numerical forward solutions. The density models being compared are: (a) and (d): sinusoidal vs. linear; (b) and (e): sinusoidal vs. parabolic; and (c) and (f): linear vs. parabolic.} 
              \label{mc}%
    \end{figure*}

Figure~\ref{mc} shows the obtained results. Each panel displays contours for the regions in period and damping time space where the evidence for one particular model against another alternative reaches a given level of evidence. These levels are based on the corresponding values of $2\log(BF^{ij})$ and the associated amount of evidence according to the \cite{kass95} table. Regions in white indicate period and damping time combinations for which minimal evidence is obtained and hence the result of the model comparison is inconclusive. Then, different levels of gray shading point to regions in which positive, strong, and very strong evidence is obtained, with the level of evidence increasing with the darkening.
The top row panels correspond to results in which the TTTB approximations to the forward solutions have been used. The bottom row panels correspond to results in which numerical forward solutions have been employed. 

Considering first model comparison results under the TTTB approximations, Fig.~\ref{mc}a shows the regions with different levels of evidence
for the comparison between the sinusoidal and the linear profiles. Note that, by construction, regions in the observed parameter space where $BF^{ij}$ and $BF^{ji}$ reach the different levels of evidence are mutually exclusive, since $\log(BF^{ij})=-\log(BF^{ji})$, and cannot overlap. In the comparison between the sinusoidal and linear profiles only $BF^{LP}$ reaches significantly positive values, with the evidence being positive, strong, and very strong. This means that the evidence supports the linear density model instead of the sinusoidal density model. However, the regions where this occurs correspond to combinations with short damping timescale, in comparison to the period (see Fig.~\ref{mc}a). A similar conclusion can be reached in the comparison between the sinusoidal and parabolic density models (see Fig.~\ref{mc}b) and the comparison between the linear and parabolic density models (see Fig.~\ref{mc}c). In the first case, we obtain positive and strong evidence for the parabolic profile. In the second case, the evidence supports the linear density model instead of the parabolic profile, with positive, strong, and very strong evidence. In both cases, this happens again for combinations of period and damping time indicative of  very strong damping. In summary, model comparison under the TTTB approximations enables us to draw conclusions about different levels of evidence among the considered models, but this evidence is only appreciable for strongly damped oscillations. If that were the case, the linear and parabolic profiles are preferred in front of the sinusoidal profile. Among them, the plausibility of the linear density model is larger than that corresponding to the parabolic density model.

Turning now to the case in which numerical solutions for the forward model are used, outside the TTTB approximations, the results are strikingly different (see Figs.~\ref{mc}d-f). In appearance, Fig.~\ref{mc}a and d, corresponding to the comparison between the sinusoidal and linear density models, look rather similar. The difference is that, in the second case, regions with significantly positive values of $\log(BF^{SL})$ are being plotted, i.e., those combinations of observed period and damping time for which the evidence supports the sinusoidal model. The model comparison result using the numerical solutions is therefore contrary to the one obtained under the TTTB approximations, where the linear density model was preferred. The situation is more involved when we compare the sinusoidal and parabolic density models. In this case, see Fig.~\ref{mc}e, there are regions in observed parameter space for which evidence supportive of both models is found. Those regions where the evidence supports the sinusoidal density model are contoured with solid lines, with the evidence being positive, strong, and very strong. The region contoured with dashed lines corresponds to period and damping time values for which positive evidence for the parabolic density model is obtained. Finally, in Fig.~\ref{mc}f, the result from the comparison between the linear and parabolic density models is shown. In this case, the result is similar to the one obtained under the TTTB approximations (compare Figs.~\ref{mc}c and f) and the evidence supports the linear profile instead of the parabolic density model. As before, model comparison using the numerical forward solutions  enables us to draw conclusions about different levels of evidence among the considered models, but this evidence is only appreciable for strongly damped oscillations.  The sinusoidal and linear density models are preferred in most of the regions in observed parameter space, with a little region where positive evidence for the parabolic density model instead of the sinusoidal profile exists.

\section{Model averaging}\label{averaging}
Let us summarize the results gathered so far. In Section~\ref{inference}, we have shown that the Bayesian solution to the inference problem enables us to obtain well-constrained marginal posteriors for the three parameters of interest. However, the obtained results depend on the particularly adopted model for the density variation at the nonuniform transitional layer. These differences are small and mainly focused on the posteriors for the transverse inhomogeneity length scale in the case with moderate damping, but can be significant and even affect the inference of the Alfv\'en travel time for cases with strong damping. The use of numerical solutions for the forward problem instead of the analytical estimates under the TTTB approximation leads to more marked differences in the corresponding inversions.  In Section~\ref{comparison}, we have shown that the application of model comparison techniques leads to different levels of evidence for one model to be preferred to another, but the regions over observable parameter space where this happens are located at combinations of period and damping time corresponding to very strong damping regimes. This means that it might be difficult to perform such assessment with well-differentiated degrees of plausibility among the competing models, because the evidence for one model to be preferred over another is not strong enough for many combinations of observed period and damping times. The question then remains as to which one of the inferences is to be preferred. 

The period and damping time values for the moderate and strong damping example cases fall in regions where the evidence is minimal in the comparisons shown in Fig.~\ref{mc}. However, the evidence for each model given a particular pair of values for period and damping time is different. For example, a calculation of the Bayes factors for the moderate damping case, using the numerical forward solution and taking the sinusoidal model as the reference model,  gives $BF^{SS}=1$; $BF^{LS}=0.67$; and $BF^{PS}=1.08$. This may lead us to chose the parabolic density model as the most plausible one in view of data and to therefore adopt the corresponding inference as the most reasonable one. However, the differences in Bayes factors are not large enough to support this conclusion.

To solve this problem, a further step can be pursued by considering a third level of Bayesian inference, namely, model averaging. Bayesian model averaging is a procedure to obtain parameter constraints that account for the uncertainty about the models under consideration.  Applied to our particular problem,  it consists of combining the posteriors obtained for each particular density models to obtain a model-averaged posterior, weighted with the evidence for each model. 

\begin{figure*}
   \centering
   \includegraphics[width=0.3\textwidth]{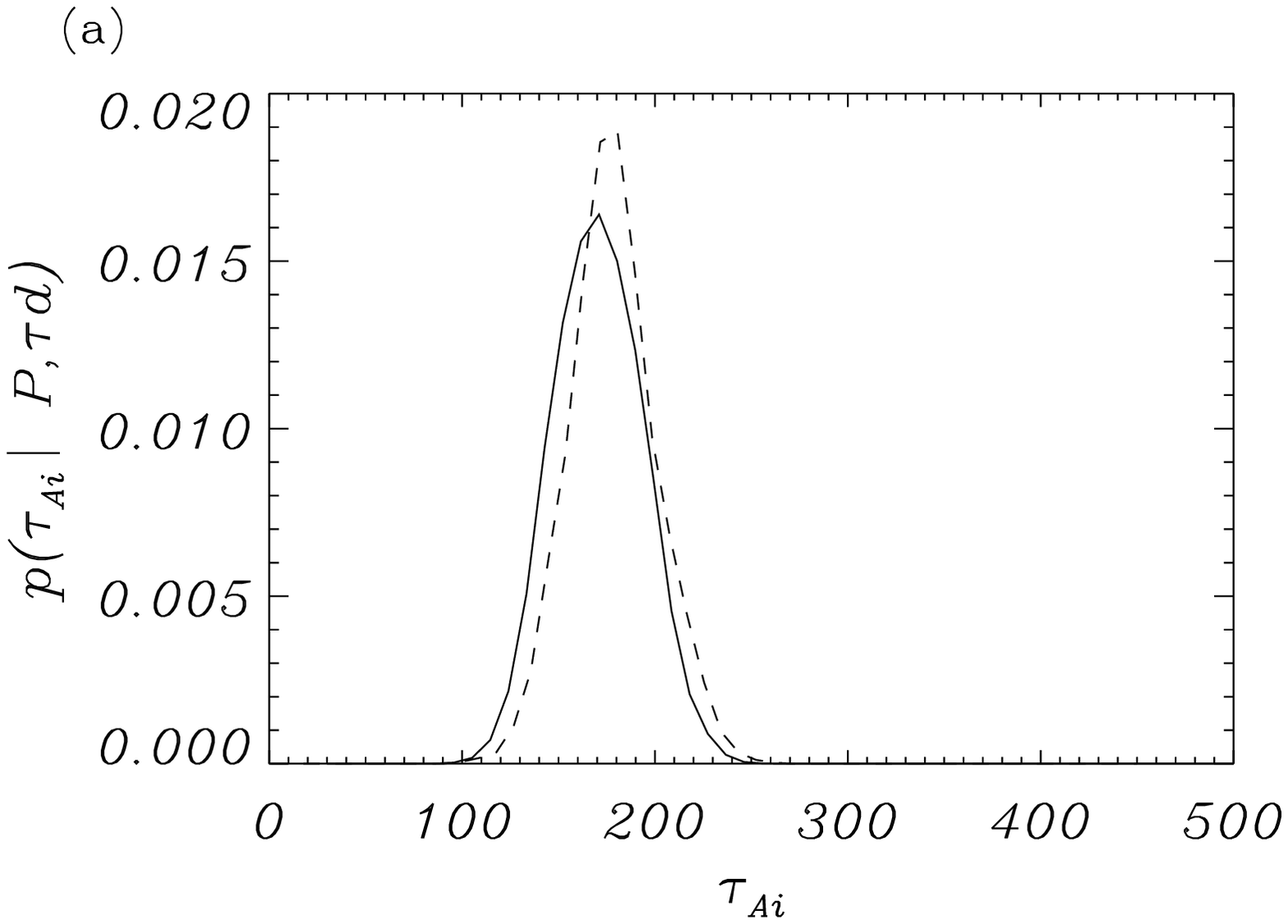}
   \includegraphics[width=0.3\textwidth]{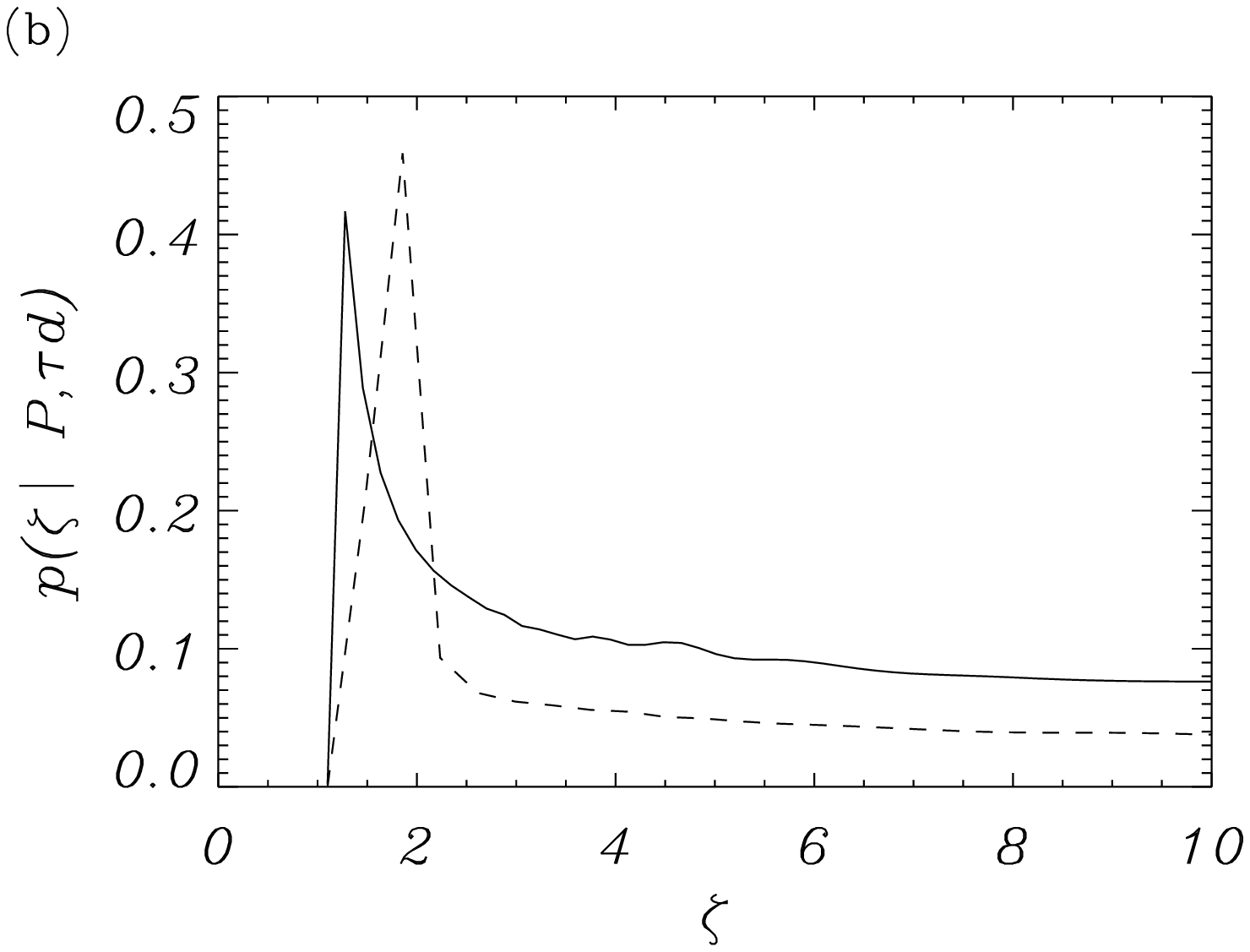}
   \includegraphics[width=0.3\textwidth]{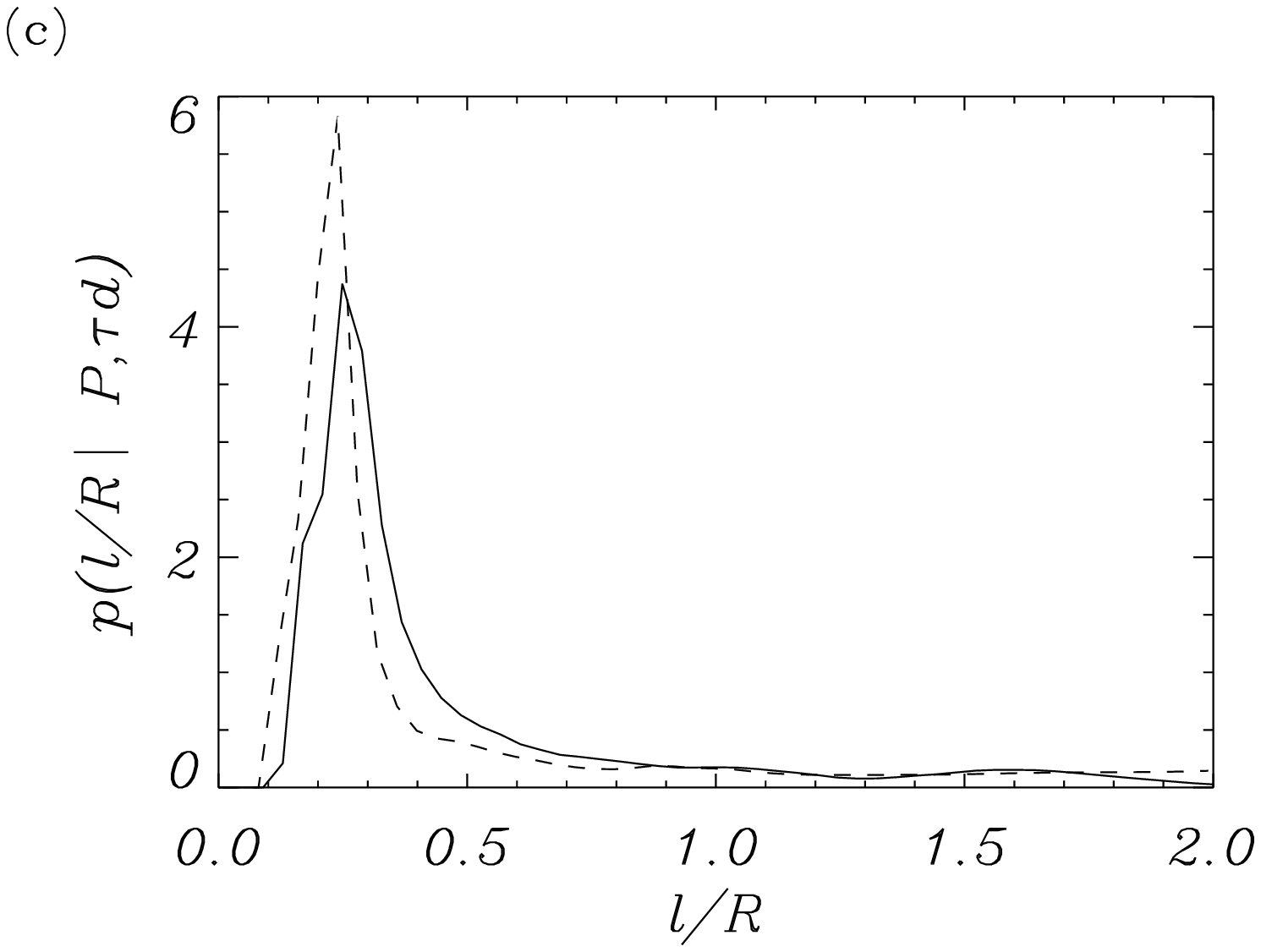}\\
   \includegraphics[width=0.3\textwidth]{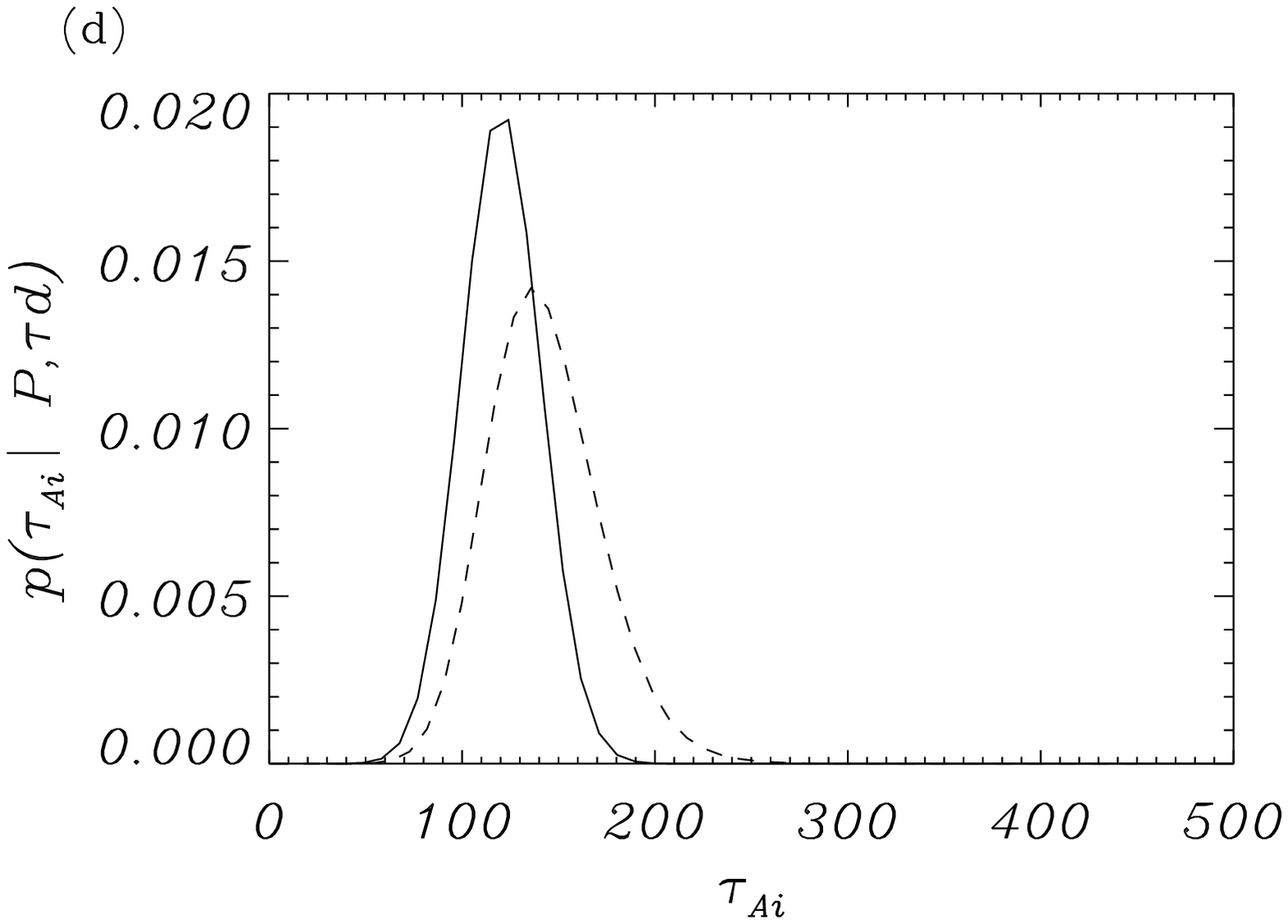}
   \includegraphics[width=0.3\textwidth]{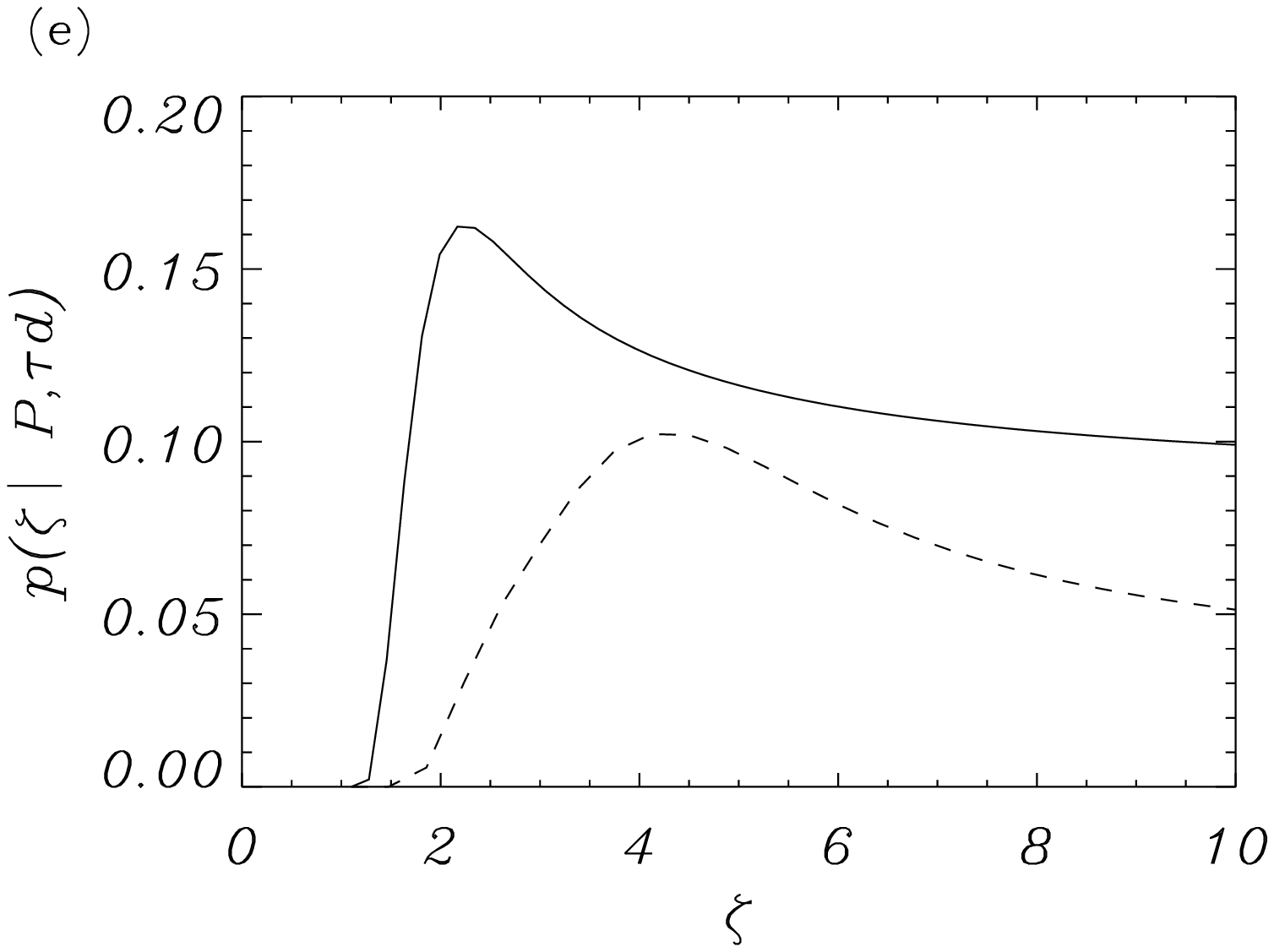}
   \includegraphics[width=0.3\textwidth]{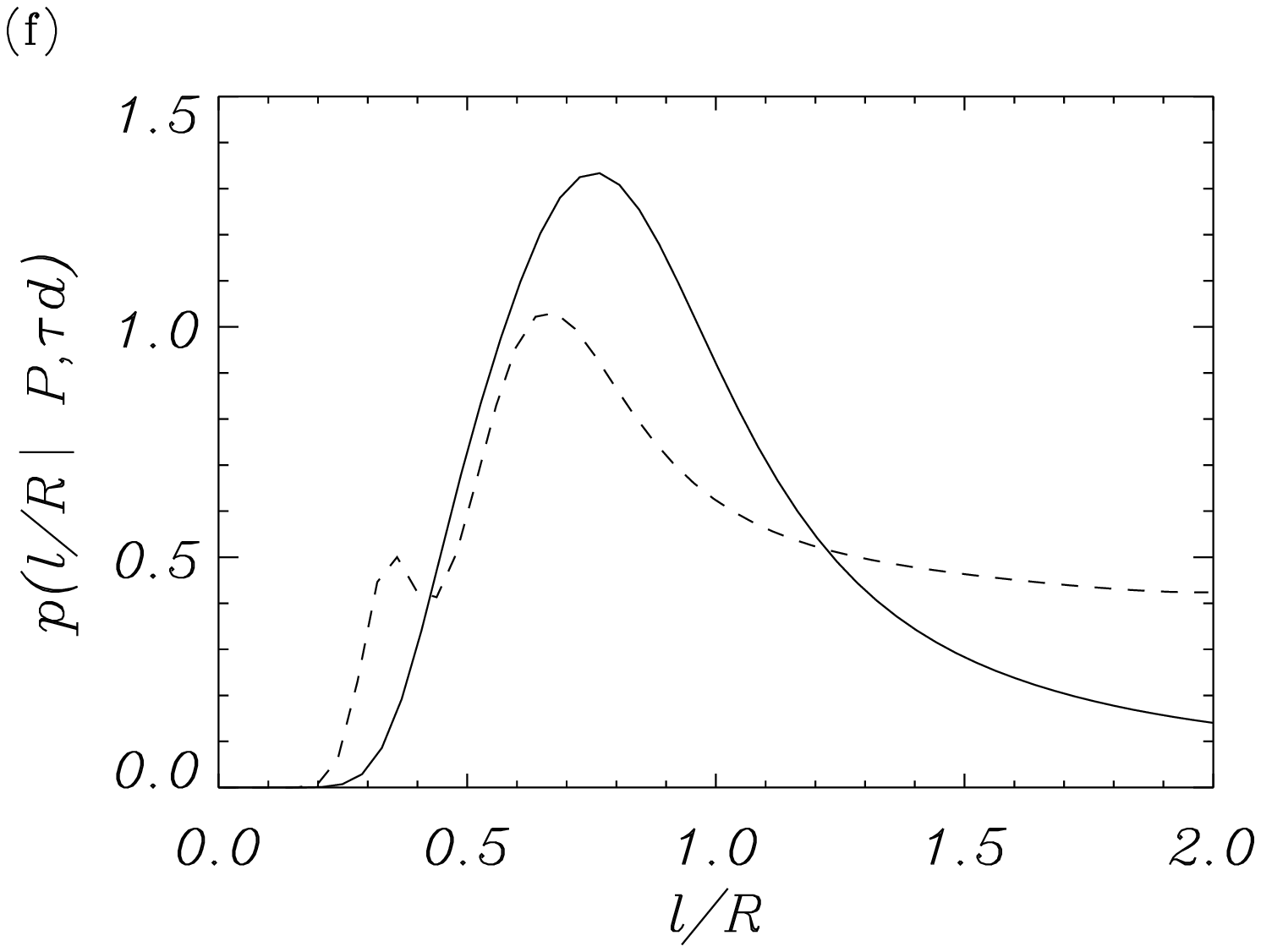}
   \caption{Model-averaged marginal posteriors for the three parameters of interest computed using expression~(\ref{average}). Top and bottom panels correspond to the moderate ($P=272$ s; $\tau_{\rm d}=849$ s) and strong ($P=185$ s; $\tau_{\rm d}=200$ s) damping cases, respectively. In all cases, $\sigma_{\rm P}=\sigma_{\rm \tau_{\rm d}}= 30$ s. In each panel, solid lines represent the inference performed using the TTTB forward solutions, while dashed lines are for the inference performed using the numerical forward solutions. Summary data for the posteriors with the median and errors at the 68\% credible region are shown in Table~\ref{infavgtable}.}
              \label{mavg}%
    \end{figure*}

The model-averaged posterior distribution for a given parameter $\theta_i$,  conditional on observed data $d$ and weighted with the probability of our set of density models $M^k$
is given by 

\begin{eqnarray}\label{average}
p(\theta_i|d)&=&\sum_{k=S, L, P}p(\theta_i|d,M^k)p(M^k|d)\nonumber\\
&=& p(M^S|d)\sum_{k=S, L, P}BF^{kS} p(\theta_i|d,M^k),
\end{eqnarray}
where in the second equality we have adopted the sinusoidal profile, $M^S$, as the reference model and replaced the models' posterior probabilities by the Bayes factors, $BF^{kS}$, with respect to the reference model. Obviously, $BF^{SS}=1$. The posterior for the reference model, $M^S$, can be calculated by considering that the sum of the probabilities for all three models must be unity, thus

\begin{equation}
p(M^S|d)=\frac{1}{1+\sum_{k=L,P}B^{kS}}.
\end{equation}
The obtained result is independent of the model chosen to be the reference.

Figure~\ref{mavg} shows the inference results for the three parameters of interest and the two considered cases, for moderate (Fig.~\ref{mavg}a-c) and strong (Fig.~\ref{mavg}d-f) damping. In all panels the model-averaged posteriors using the TTTB and the numerical forward solutions are compared.  In the moderate damping case, very similar model-averaged posteriors are obtained for the three parameters of interest. Differences are more marked in the strong damping case. 

Table~\ref{infavgtable} displays the summary statistics, with the median and errors at 68\% credible regions, for each model-averaged inference. Also, the corresponding Bayes factors are shown. For the moderate damping case and the TTTB results, the sinusoidal model is the one with the largest plausibility, with a small difference with respect to the parabolic density model. However, when numerical results are employed, the parabolic density model has the largest plausibility. Something similar occurs for the strong damping case in which the Bayes factor for the TTTB result is larger for the sinusoidal model, but the parabolic profile turns out to be the most likely when numerical results are employed.

The model-averaged posteriors and the summary values offer the most general inference result that can be obtained on the unknown parameter values. They take into account all the available information, i.e., the prior information, the observed data with their uncertainty, the modeling constraints, and the evidence for each model in view of data.

\begin{deluxetable}{cccccccc}
\tablecolumns{8} 
\tablewidth{0pc} 
\tablecaption{Summary of model averaged inference results for the moderate ($P=272$ s; $\tau_{\rm d}=849$ s) and strong ($P=185$ s; $\tau_{\rm d}=200$ s) damping cases. In both cases,  $\sigma_{\rm P}=\sigma_{\rm \tau_{\rm d}}= 30$ s.\label{infavgtable}} 
\tablehead{ 
\colhead{$\theta_{\rm i}$ \& $BF^{kS}$}    & \colhead{} & \multicolumn{2}{c}{Moderate damping} &   \colhead{}   & \multicolumn{2}{c}{Strong damping} \\ 
\cline{1-7} \\ 
\colhead{} & & \colhead{TTTB}   & \colhead{Numerical} & 
\colhead{}    & \colhead{TTTB}    & \colhead{Numerical}}
\startdata 
$\tau_{\rm Ai}$[s]& &170$^{+24}_{-24}$& 170$^{+24}_{-21}$ && 120$^{+20}_{-19}$& 141$^{+30}_{-27}$\\\\
$\zeta$&  &4.2$^{+3.8}_{-2.5}$ & 6.7$^{+9.0}_{-4.9}$ && 5.2$^{+3.2}_{-2.6}$& 8.6$^{+7.3}_{-4.3}$\\\\
$l/R$ & &0.3$^{+0.5}_{-0.1}$ & 0.3$^{+0.5}_{-0.1}$&& 0.9$^{+0.5}_{-0.3}$& 1.0$^{+0.7}_{-0.4}$\\\\
$BF^{SS}$&&1&1&&1&1\\\\
$BF^{LS}$&&0.62&0.67&&0.71&1.12\\\\
$BF^{PS}$&&0.89&1.08&&0.93&2.42\\
\enddata 
\end{deluxetable}

\section{Summary and conclusions}\label{conclusions}

The characteristic timescales and spatial scales for damping of MHD kink waves by resonant absorption and subsequent energy dissipation in coronal waveguides depend on the cross-field variation of the mass density. Observed wave damping properties can be used to perform the inference of the density profile at the nonuniform layer, upon the assumption of particular forms for the density variation in theoretically modeled density tubes. 
In a recent study, \cite{soler14a} have pointed out that the particular model adopted for the density profile at the nonuniform transitional layer
may have strong implications for coronal seismology estimates for the unknown parameters. The reason is that solving the inversion problem
in the three-dimensional space of unknown parameters leads to solution curves that are significantly different. 

In our study, we applied three levels of Bayesian inference to the problem of obtaining information on the cross-field density structuring in coronal waveguides from observed periods and damping times of transverse kink oscillations. Three alternative density models were considered as a representation of the density variation at the nonuniform transitional layer at the tube boundary, namely, sinusoidal, linear and parabolic profiles.

The application of parameter inference enables us to determine the parameters that define the density variation under the assumption of a particular model from observed oscillation properties. The inference result depends on the assumed density model. The differences are small when the damping of transverse kink oscillations is moderate and when TTTB forward solutions are employed. However, significant differences in the obtained posterior distributions were found when the damping is strong and the full numerical forward solutions are used. When the posteriors are summarized by considering, e.g., the median and errors at the 68\% credible region, the differences are concealed to a great extent. 

By applying model comparison techniques, the marginal likelihood for each model and the Bayes factors were computed. They inform us on the likelihood of obtaining given observed values for period and damping time and on the relative plausibility between the considered alternative density models in view of data. Our results indicate that although the plausibility for each density model is different and the regions in observable parameter space where the evidence for one model against an alternative is stronger can be well differentiated, those regions correspond to period and damping time values that imply very fast damping.  Therefore, a categorical assessment among models, based on positive, strong, or very strong evidence, cannot be made for typically observed damping regimes. Furthermore, as analytical TTTB and numerical forward solutions predict different period and damping times, the regions in observable parameter space for which the evidence supports a given model are different and lead in general to dissimilar conclusions.

Even if the evidence analysis does not permit to choose a particular density model, for any given observation the Bayes factors differ. This  information on the distinct plausibility of each model, in view of data, can be used to perform model averaging. The procedure combines the alternative posteriors by weighting them with the individual model evidence. Such a calculation makes use of all the available information in data and modeling in a fully consistent manner. The resulting marginal posteriors are the best inference one can obtain with the information at hand. 

\acknowledgments
We acknowledge financial support from the Spanish MICINN/MINECO through projects AYA2010-18029, AYA2011-22846, AYA2014-55456-P, AYA2014-54485-P, the Consolider-Ingenio 2010 CSD2009-00038 project, and from FEDER funds.  I.A. and A.A.R. acknowledge support by Ram\'on y Cajal Fellowships by the Spanish MINECO.  R.S. acknowledges support from MINECO through a Juan de la Cierva grant, from MECD through project CEF11-0012, and from the ``Vicerectorat d'Investigaci\'o i Postgrau'' of the Universitat de les Illes Balears.


\begin{thebibliography}{56}
\expandafter\ifx\csname natexlab\endcsname\relax\def\natexlab#1{#1}\fi

\bibitem[{{Andries} {et~al.}(2005){Andries}, {Arregui}, \&
  {Goossens}}]{andries05b}
{Andries}, J., {Arregui}, I., \& {Goossens}, M. 2005, \apjl, 624, L57

\bibitem[{{Arregui}(2012)}]{arregui12b}
{Arregui}, I. 2012, in Multi-scale Dynamical Processes in Space and
  Astrophysical Plasmas, ed. M.~P. {Leubner} \& Z.~{V{\"o}r{\"o}s}, 159

\bibitem[{Arregui(2015)}]{arregui15}
Arregui, I. 2015, Royal Society of London Philosophical Transactions Series A,
  373, 20140261; DOI: 10.1098/rsta.2014.0261

\bibitem[{Arregui {et~al.}(2007)Arregui, Andries, Van~Doorsselaere, Goossens,
  \& Poedts}]{arregui07a}
Arregui, I., Andries, J., Van~Doorsselaere, T., Goossens, M., \& Poedts, S.
  2007, Astron. Astrophys., 463, 333

\bibitem[{{Arregui} \& {Asensio Ramos}(2011)}]{arregui11b}
{Arregui}, I., \& {Asensio Ramos}, A. 2011, \apj, 740, 44

\bibitem[{{Arregui} \& {Asensio Ramos}(2014)}]{arregui14}
---. 2014, \aap, 565, A78

\bibitem[{{Arregui} {et~al.}(2013){Arregui}, {Asensio Ramos}, \&
  {Pascoe}}]{arregui13b}
{Arregui}, I., {Asensio Ramos}, A., \& {Pascoe}, D.~J. 2013, \apjl, 769, L34

\bibitem[{{Arregui} {et~al.}(2012){Arregui}, {Oliver}, \&
  {Ballester}}]{arregui12a}
{Arregui}, I., {Oliver}, R., \& {Ballester}, J.~L. 2012, Living Reviews in
  Solar Physics, 9, 2

\bibitem[{{Arregui} {et~al.}(2005){Arregui}, {Van Doorsselaere}, {Andries},
  {Goossens}, \& {Kimpe}}]{arregui05}
{Arregui}, I., {Van Doorsselaere}, T., {Andries}, J., {Goossens}, M., \&
  {Kimpe}, D. 2005, \aap, 441, 361

\bibitem[{{Bayes} \& {Price}(1763)}]{bayes63}
{Bayes}, M., \& {Price}, M. 1763, Royal Society of London Philosophical
  Transactions Series I, 53, 370

\bibitem[{{De Moortel}(2005)}]{demoortel05}
{De Moortel}, I. 2005, Royal Society of London Philosophical Transactions
  Series A, 363, 2743

\bibitem[{{De Moortel} \& {Nakariakov}(2012)}]{demoortel12}
{De Moortel}, I., \& {Nakariakov}, V.~M. 2012, Royal Society of London
  Philosophical Transactions Series A, 370, 3193

\bibitem[{{Goossens}(1991)}]{goossens91}
{Goossens}, M. 1991, in Advances in Solar System Magnetohydrodynamics, ed.
  A.~W. {Hood} \& E.~R. {Priest} (Cambridge University Press), 137

\bibitem[{{Goossens}(2008)}]{goossens08b}
{Goossens}, M. 2008, in IAU Symposium, Vol. 247, IAU Symposium, ed.
  {R.~Erd{\'e}lyi \& C.~A.~Mendoza-Brice{\~n}o}, 228--242

\bibitem[{{Goossens} {et~al.}(2002){Goossens}, {Andries}, \&
  {Aschwanden}}]{goossens02a}
{Goossens}, M., {Andries}, J., \& {Aschwanden}, M.~J. 2002, \aap, 394, L39

\bibitem[{{Goossens} {et~al.}(2012{\natexlab{a}}){Goossens}, {Andries},
  {Soler}, {Van Doorsselaere}, {Arregui}, \& {Terradas}}]{goossens12a}
{Goossens}, M., {Andries}, J., {Soler}, R., {Van Doorsselaere}, T., {Arregui},
  I., \& {Terradas}, J. 2012{\natexlab{a}}, \apj, 753, 111

\bibitem[{{Goossens} {et~al.}(2008){Goossens}, {Arregui}, {Ballester}, \&
  {Wang}}]{goossens08a}
{Goossens}, M., {Arregui}, I., {Ballester}, J.~L., \& {Wang}, T.~J. 2008, \aap,
  484, 851

\bibitem[{{Goossens} {et~al.}(2011){Goossens}, {Erd{\'e}lyi}, \&
  {Ruderman}}]{goossens11}
{Goossens}, M., {Erd{\'e}lyi}, R., \& {Ruderman}, M.~S. 2011, \ssr, 158, 289

\bibitem[{{Goossens} {et~al.}(1995){Goossens}, {Ruderman}, \&
  {Hollweg}}]{goossens95a}
{Goossens}, M., {Ruderman}, M.~S., \& {Hollweg}, J.~V. 1995, \solphys, 157, 75

\bibitem[{{Goossens} {et~al.}(2012{\natexlab{b}}){Goossens}, {Soler},
  {Arregui}, \& {Terradas}}]{goossens12b}
{Goossens}, M., {Soler}, R., {Arregui}, I., \& {Terradas}, J.
  2012{\natexlab{b}}, \apj, 760, 98

\bibitem[{{Goossens} {et~al.}(2009){Goossens}, {Terradas}, {Andries},
  {Arregui}, \& {Ballester}}]{goossens09}
{Goossens}, M., {Terradas}, J., {Andries}, J., {Arregui}, I., \& {Ballester},
  J.~L. 2009, \aap, 503, 213

\bibitem[{{Goossens} {et~al.}(2013){Goossens}, {Van Doorsselaere}, {Soler}, \&
  {Verth}}]{goossens13}
{Goossens}, M., {Van Doorsselaere}, T., {Soler}, R., \& {Verth}, G. 2013, \apj,
  768, 191

\bibitem[{{Heyvaerts} \& {Priest}(1983)}]{heyvaerts83}
{Heyvaerts}, J., \& {Priest}, E.~R. 1983, \aap, 117, 220

\bibitem[{{Hollweg} \& {Yang}(1988)}]{hollweg88}
{Hollweg}, J.~V., \& {Yang}, G. 1988, \jgr, 93, 5423

\bibitem[{{Jeffreys}(1961)}]{jeffreys61}
{Jeffreys}, H. 1961, in {Theory of Probability (3rd ed.)} (Oxford University
  Press)

\bibitem[{{Jess} {et~al.}(2015){Jess}, {Morton}, {Verth}, {Fedun}, {Grant}, \&
  {Giagkiozis}}]{jess15}
{Jess}, D.~B., {Morton}, R.~J., {Verth}, G., {Fedun}, V., {Grant}, S.~D.~T., \&
  {Giagkiozis}, I. 2015, \ssr, 190, 103

\bibitem[{{Kass} \& {Raftery}(1995)}]{kass95}
{Kass}, R.~E., \& {Raftery}, A.~E. 1995, JASA, 90, 773

\bibitem[{{Lee} \& {Roberts}(1986)}]{lee86}
{Lee}, M.~A., \& {Roberts}, B. 1986, \apj, 301, 430

\bibitem[{{Lin} {et~al.}(2009){Lin}, {Soler}, {Engvold}, {Ballester},
  {Langangen}, {Oliver}, \& {Rouppe van der Voort}}]{lin09}
{Lin}, Y., {Soler}, R., {Engvold}, O., {Ballester}, J.~L., {Langangen}, {\O}.,
  {Oliver}, R., \& {Rouppe van der Voort}, L.~H.~M. 2009, \apj, 704, 870

\bibitem[{{Nakariakov} \& {Ofman}(2001)}]{nakariakov01}
{Nakariakov}, V.~M., \& {Ofman}, L. 2001, \aap, 372, L53

\bibitem[{{Nakariakov} \& {Verwichte}(2005)}]{nakariakov05}
{Nakariakov}, V.~M., \& {Verwichte}, E. 2005, Living Reviews in Solar Physics,
  2, 3

\bibitem[{{Ofman} \& {Aschwanden}(2002)}]{ofman02b}
{Ofman}, L., \& {Aschwanden}, M.~J. 2002, \apjl, 576, L153

\bibitem[{{Parnell} \& {De Moortel}(2012)}]{parnell12}
{Parnell}, C.~E., \& {De Moortel}, I. 2012, Royal Society of London
  Philosophical Transactions Series A, 370, 3217

\bibitem[{{Pascoe} {et~al.}(2012){Pascoe}, {Hood}, {de Moortel}, \&
  {Wright}}]{pascoe12}
{Pascoe}, D.~J., {Hood}, A.~W., {de Moortel}, I., \& {Wright}, A.~N. 2012,
  \aap, 539, A37

\bibitem[{{Pascoe} {et~al.}(2010){Pascoe}, {Wright}, \& {De
  Moortel}}]{pascoe10}
{Pascoe}, D.~J., {Wright}, A.~N., \& {De Moortel}, I. 2010, \apj, 711, 990

\bibitem[{{Pascoe} {et~al.}(2011){Pascoe}, {Wright}, \& {De
  Moortel}}]{pascoe11}
---. 2011, \apj, 731, 73

\bibitem[{{Roberts} {et~al.}(1984){Roberts}, {Edwin}, \& {Benz}}]{roberts84}
{Roberts}, B., {Edwin}, P.~M., \& {Benz}, A.~O. 1984, \apj, 279, 857

\bibitem[{{Rosenberg}(1970)}]{rosenberg70}
{Rosenberg}, H. 1970, \aap, 9, 159

\bibitem[{{Ruderman} \& {Roberts}(2002)}]{ruderman02}
{Ruderman}, M.~S., \& {Roberts}, B. 2002, \apj, 577, 475

\bibitem[{{Soler} {et~al.}(2013){Soler}, {Goossens}, {Terradas}, \&
  {Oliver}}]{soler13}
{Soler}, R., {Goossens}, M., {Terradas}, J., \& {Oliver}, R. 2013, \apj, 777,
  158

\bibitem[{Soler {et~al.}(2014)Soler, Goossens, Terradas, \& Oliver}]{soler14a}
Soler, R., Goossens, M., Terradas, J., \& Oliver, R. 2014, The Astrophysical
  Journal, 781, 111

\bibitem[{{Soler} {et~al.}(2011{\natexlab{a}}){Soler}, {Oliver}, \&
  {Ballester}}]{soler11a}
{Soler}, R., {Oliver}, R., \& {Ballester}, J. 2011{\natexlab{a}}, \apj, 726, 102

\bibitem[{Soler \& Terradas(2015)}]{soler15b}
Soler, R., \& Terradas, J. 2015, The Astrophysical Journal, 803, 43

\bibitem[{{Soler} {et~al.}(2011{\natexlab{b}}){Soler}, {Terradas}, \&
  {Goossens}}]{soler11d}
{Soler}, R., {Terradas}, J., \& {Goossens}, M. 2011{\natexlab{b}}, \apj, 734,
  80

\bibitem[{{Soler} {et~al.}(2011{\natexlab{c}}){Soler}, {Terradas}, {Verth}, \&
  {Goossens}}]{soler11c}
{Soler}, R., {Terradas}, J., {Verth}, G., \& {Goossens}, M. 2011{\natexlab{c}},
  \apj, 736, 10

\bibitem[{{Terradas} {et~al.}(2008){Terradas}, {Arregui}, {Oliver},
  {Ballester}, {Andries}, \& {Goossens}}]{terradas08b}
{Terradas}, J., {Arregui}, I., {Oliver}, R., {Ballester}, J.~L., {Andries}, J.,
  \& {Goossens}, M. 2008, \apj, 679, 1611

\bibitem[{{Terradas} {et~al.}(2010){Terradas}, {Goossens}, \&
  {Verth}}]{terradas10}
{Terradas}, J., {Goossens}, M., \& {Verth}, G. 2010, \aap, 524, A23

\bibitem[{{Terradas} {et~al.}(2006{\natexlab{a}}){Terradas}, {Oliver}, \&
  {Ballester}}]{terradas06a}
{Terradas}, J., {Oliver}, R., \& {Ballester}, J.~L. 2006{\natexlab{a}}, \apj,
  642, 533

\bibitem[{{Terradas} {et~al.}(2006{\natexlab{b}}){Terradas}, {Oliver}, \&
  {Ballester}}]{terradas06c}
---. 2006{\natexlab{b}}, Royal Society of London Philosophical Transactions
  Series A, 364, 547

\bibitem[{{Uchida}(1970)}]{uchida70}
{Uchida}, Y. 1970, \pasj, 22, 341

\bibitem[{{van Ballegooijen} {et~al.}(2011){van Ballegooijen}, {Asgari-Targhi},
  {Cranmer}, \& {DeLuca}}]{vanballegooijen11}
{van Ballegooijen}, A.~A., {Asgari-Targhi}, M., {Cranmer}, S.~R., \& {DeLuca},
  E.~E. 2011, \apj, 736, 3

\bibitem[{{Van Doorsselaere} {et~al.}(2004){Van Doorsselaere}, {Andries},
  {Poedts}, \& {Goossens}}]{vandoorsselaere04a}
{Van Doorsselaere}, T., {Andries}, J., {Poedts}, S., \& {Goossens}, M. 2004,
  \apj, 606, 1223

\bibitem[{{Van Doorsselaere} {et~al.}(2014){Van Doorsselaere}, {Gijsen},
  {Andries}, \& {Verth}}]{vandoorsselaere14}
{Van Doorsselaere}, T., {Gijsen}, S.~E., {Andries}, J., \& {Verth}, G. 2014,
  \apj, 795, 18

\bibitem[{{Van Doorsselaere} {et~al.}(2008){Van Doorsselaere}, {Nakariakov},
  {Young}, \& {Verwichte}}]{vandoorsselaere08}
{Van Doorsselaere}, T., {Nakariakov}, V.~M., {Young}, P.~R., \& {Verwichte}, E.
  2008, \aap, 487, L17

\bibitem[{{Verth} {et~al.}(2008){Verth}, {Erd{\'e}lyi}, \& {Jess}}]{verth08b}
{Verth}, G., {Erd{\'e}lyi}, R., \& {Jess}, D.~B. 2008, \apjl, 687, L45

\bibitem[{{Verth} {et~al.}(2010){Verth}, {Terradas}, \& {Goossens}}]{verth10}
{Verth}, G., {Terradas}, J., \& {Goossens}, M. 2010, \apjl, 718, L102

\end{thebibliography}
\end{document}